\documentclass[a4paper, 10pt]{article}

\usepackage[utf8]{inputenc} 
\usepackage[T1]{fontenc}    
\usepackage[english]{babel} 

\usepackage{enumitem}
\usepackage{amsmath, amssymb}
\usepackage{mathtools}
\usepackage{amsfonts}
\usepackage{esvect}
\usepackage{stmaryrd}
\usepackage{esint}
\usepackage{cases}
\usepackage{cite}

\usepackage[amsthm, amsmath, thmmarks, hyperref]{ntheorem}

\newtheorem{theorem}{Theorem}
\newtheorem{lemma}{Lemma}
\newtheorem{proposition}{Proposition}

\theoremstyle{definition}
\newtheorem{definition}{Definition}

\theoremstyle{remark}
\newtheorem{remark}{Remark}

\usepackage{tikz}
\usetikzlibrary{3d}
\usepackage{pgfplots}
\pgfplotsset{compat = newest}
\usepackage{graphicx}

\usepackage{comment}
\usepackage{xcolor}

\usepackage{hyperref}
\usepackage[noabbrev]{cleveref}

\newcommand{\rmi}[1]{\mathrm{#1}}

\newcommand{\ecal}{\mathcal{E}}

\newcommand{\kcal}{\mathcal{K}}

\newcommand{\mcal}{\mathcal{M}}

\newcommand{\ocal}{\mathcal{O}}

\newcommand{\rcal}{\mathcal{R}}

\newcommand{\I}{\ensuremath{\mathbb{I}}}

\newcommand{\R}{\ensuremath{\mathbb{R}}}
\renewcommand{\S}{\ensuremath{\mathbb{S}}}
\newcommand{\T}{\ensuremath{\mathbb{T}}}
\newcommand{\U}{\ensuremath{\mathbf{U}}}

\newcommand{\gbf}{\ensuremath{\mathbf{g}}}

\newcommand{\mbf}{\ensuremath{\mathbf{m}}}
\newcommand{\nbf}{\ensuremath{\mathbf{n}}}

\newcommand{\tbf}{\ensuremath{\mathbf{t}}}
\newcommand{\ubf}{\ensuremath{\mathbf{u}}}

\newcommand{\xbf}{\ensuremath{\mathbf{x}}}
\newcommand{\ybf}{\ensuremath{\mathbf{y}}}

\renewcommand{\d}{\,\mathrm{d}}

\newcommand{\Norm}[1]{\left\| #1 \right\|}

\newcommand{\limit}[2]{\xrightarrow[#1\rightarrow #2]{}}

\renewcommand{\bar}[1]{\overline{#1}}
\newcommand{\favre}[1]{\langle {#1} \rangle}

\newcommand{\ulin}[1]{\underline{#1}}
\newcommand{\ubrace}[2]{\underset{#1}{\underbrace{#2}}}

\newcommand{\pierrick}[1]{\textcolor{green!70!blue}{[P] {#1}}}

\title{Shallow Water type Models \\ for Compressible Free-surface/Pressurized Flows: \\ Derivation, Analysis and Incompressible Limit}

\newcommand{\andline}{\end{tabular}\par\vspace{0.1ex}\begin{tabular}[t]{c}}

\author{D. Bresch,\thanks{Universit\'e Savoie Mont Blanc, UMR5127 CNRS, Laboratoire de Math\'ematiques, 73376 Le Bourget-du-Lac, France
  (\texttt{didier.bresch@univ-smb.fr}),}
  \and E.D. Fern\'andez-Nieto,\thanks{Dpto. Matem\'atica Aplicada I \& IMUS. ETS Arquitectura – Universidad de Sevilla, Avda. Reina
Mercedes S/N, 41012--Sevilla, Spain
  (\texttt{edofer@us.es, gnarbona@us.es}),}
  \and J. Garres-D\'{i}az,\thanks{Dpto. Matem\'atica Aplicada II \& IMUS. ETS Ingeniería – Universidad de Sevilla, Camino Descubrimientos S/N, 41092-Sevilla, Spain
  (\texttt{jgarres@us.es}),}  
  \andline
  P. Le Vourc'h,\thanks{IMAG, ANGUS, Univ Montpellier, Inria, CNRS, Montpellier, France
  (\texttt{pierrick.le-vourc-h@umontpellier.fr}),} 
  \and G. Narbona-Reina $^\dag$ 
}

\date{}

\begin{document}

\maketitle


\begin{abstract}
In this paper, we formally derive compressible shallow-water equations from the full compressible barotropic Navier--Stokes equations with free surface or with pressurized zones, without any weak compressibility assumption, unlike previously published works. We explain
how it is possible to obtain a closed system in the averaging process even if the pressure state law is fully nonlinear, that is, when the pressure is chosen as $p(\rho) = a \rho^\gamma$, with $a>0$ and $\gamma>1$ given constants. We discuss two situations, the free-surface and the pressurized ones. This enables us to write a system
 that allows some parts to have a free surface and some parts to be pressurized with a flow confined inside a fixed domain.
To be complete, we discuss the limit $\gamma\to + \infty$, from which we get the usual shallow-water type system derived from the incompressible Navier--Stokes free-surface/pressurized problem. We analyze this limit mathematically in the free-surface case encoding the convergence rate (by using a standard relative entropy process) and we perform some numerical simulations to illustrate the convergence. Then we characterize the steady states at rest of the pressurized system and design a well-balanced numerical scheme that preserves them. To the best of our knowledge, this is the first formal derivation of shallow-water type systems starting from the compressible barotropic Navier--Stokes system with a fully nonlinear pressure state law, in which we also characterize their formal singular limits when $\gamma \to +\infty$.
\end{abstract}

\tableofcontents

\section{Introduction}

The formal mathematical derivations of shallow water (Saint-Venant) type equations from the full incompressible Navier--Stokes equations with a free moving boundary were obtained depending on the boundary condition at the bottom: Dirichlet boundary conditions by J.--P. Vila and co-authors, (see descriptions in \cite{bresch2007}) and Navier boundary conditions by J.-F. Gerbeau -- B. Perthame (see ~\cite{gerbeau_derivation_2001} and also \cite{Ma} for the multi-dimensional case). These systems have been mathematically justified by D. Bresch -- P. Noble in \cite{bresch2007,bresch2011}.

The compressible shallow--water equations extend the classical Saint-Venant shallow--water model by incorporating fluid compressibility (density variations). The formal derivation from the full compressible Navier--Stokes equations with free surface has been recently investigated in the weakly compressible case (low Mach number) with important and interesting applications in geophysics, 
in \cite{ersoy_dimension_2016} for mixed flows in closed water pipes
and in \cite{richard_extension_2021,richard_preliminary_2023} for tsunami propagation in coastal areas. See also \cite{bonnet-bendhia2021} for weakly compressible flows with temperature, and
the recent paper \cite{zuccoli2026} on weakly compressible flows. 
\\\\
In this paper, the main goal is to obtain a closed system even if the pressure state law is fully nonlinear, therefore avoiding the weak compressibility assumption. We will consider Navier type boundary conditions at the bottom but it is important to note that our procedure could be developed also in the framework of what has been done in the incompressible case by J.--P. Vila recalled and mathematically justified in \cite{bresch2007}. Our paper will be divided into several parts: in the first part, we will explain the difficulty linked to the pressure state law $p(\rho)= a \rho^\gamma$ with $\gamma>1$ and then explain how it could be handled in the case of free surface and in the pressurized case.
Note that free surface / pressurized problems have been considered, for instance, in  \cite{BoErGe} and in  \cite{DeHe} (see also references cited therein) with linearized pressure state laws.

More importantly, we are able to write a system that allows us to have some parts with a free surface and some parts pressurized with a flow confined inside  a fixed domain defined by $\Omega = \{\xbf \in \T_L^2, \, z \in (0,h(\xbf))\}$.
More precisely, we find the following averaged system 
$$\partial_t(h\bar\rho) + \nabla_\xbf\cdot (h\bar\rho \favre{\ubf_\xbf})= 0
$$
and
$$\partial_t (h\bar\rho\favre{\ubf_\xbf}) 
+ \nabla_\xbf\cdot (h\bar\rho \favre{\ubf_\xbf}\otimes\favre{\ubf_\xbf})
+ h\bar\rho \nabla_\xbf
  \left(\frac{a\gamma}{\gamma-1} \rho_S^{\gamma-1} 
  + gh \right)= - \widehat{\kappa} \favre{\ubf_\xbf}
$$
with the relation 

$$h\bar{\rho} = \frac{a}{g}\left(\rho_0^{\gamma}-\rho_S^{\gamma}\right),\quad\mbox{with}\quad \rho_0^{\gamma-1}=\rho_S^{\gamma-1}+\frac{\gamma-1}{a\gamma}\,g\,h$$
and with the unilateral constraint
$$(h-h(\xbf))\,p(\rho_S) = 0, \qquad \rho_S\ge 0$$
where 
$$ 0 \le  h \le h(\xbf), \qquad \bar \rho = \frac{1}{h} \int_0^h \rho(t,\xbf,z) dz, \qquad  
 \favre{\ubf_\xbf} = \frac{1}{\bar\rho h(t,\xbf)}\int_0^h (\rho {\bf  u_x})(t,\xbf,z) dz$$
with $\rho$ the density, $h$ the height, $p=a\rho^\gamma$ the pressure, and ${\bf  u_x}$ is the 2D horizontal components of the velocity field of the three-dimensional flow in the shallow-water approximation. Additionally, $\rho_0$ and $\rho_S$ denotes, respectively, the density at the bottom $z=0$ and at the surface $z=h(t,\xbf)$.

Note that in the free surface case, we assume that the density at the surface is $\rho_S = 0$ then we find the explicit formula for the averaged density in terms of the unknown $h$
 $$ h\bar\rho = \frac{a}{g} \left(\frac{\gamma-1}{a\gamma} g h\right)^{\frac{\gamma}{\gamma-1}}.$$
However, in the pressurized case, the height is known, $h=h(\xbf)$ but not $\rho_S$. Thus, we have an implicit relation to express $\rho_S$ in terms of $\bar\rho$ and $h(\xbf)$. 



An interesting point is that if we pass to the limit letting $\gamma$ go to infinity in the compressible shallow-water equations written above then we get the usual shallow-water equation with possible pressurized parts for incompressible free-surface. More precisely letting $\gamma$
tend to $\infty$ and denoting $\Pi$ the limit of $a \rho_S^{\gamma-1}$ we get from the nonlinear relation between $(h,\bar\rho, \rho_S)$ the equality 
$$h\bar\rho =h$$
and therefore the limit system reads
$$\partial_t h + \nabla_x\cdot (h \favre{\ubf_\xbf})= 0
$$
and
$$\partial_t (h\favre{\ubf_\xbf}) 
+ \nabla_x\cdot (h \favre{\ubf_\xbf}\otimes\favre{\ubf_\xbf})
+ h \nabla_x
  \left(\Pi 
  + gh \right)= - \widehat{\kappa} \favre{\ubf_\xbf}
$$
with the unilateral constraint
$$(h-h(\xbf))\,\Pi = 0,\quad 0 \le  h \le h(\xbf), \qquad \Pi\ge 0.$$
It is interesting to see that such a system has been considered in \cite{GoPaSaWa}. Remark that if $h=h(x)$, then we get a system similar to the lake model, see for instance  \cite{BrMe} and the comments in our paper in the pressurized part. Readers interested by mathematical works with unilateral constraints in particular those for free surface / pressurized problems are referred to the overview paper \cite{Pe} and references cited therein such as the interesting paper \cite{La}. See also\cite{BoBrCoRi} for hierarchy of incompressible/compressible two-phase models with a derived unilateral constraint.

\bigskip

In the document, we will separate the free surface case to the pressurized case for the sake of clarity in the exposition. Note that no smallness assumption on the Mach or Oser number is required compared to papers on the subject, to the authors knowledge.  For each case, we provide interesting numerical calculations discussing the limit of $\gamma$ for the free surface shallow water equations and then discussing the use of the implicit function procedure for stationary state law in the pressurized case.

The paper is organized as follows. After this introduction, in Section 2 we present the general framework of the flow dynamics and geometry, together with the 3D equations and boundary conditions for both configurations. Section 3 is devoted to the depth-averaging process to obtain a shallow-water type model that remains to be closed depending on the physical situations. In Sections 4 and 5 we explain how the system is closed from analytic calculations for  the free surface and then for the pressurized case. In each of these sections, the energy balance of the system is analyze, and various interesting numerical results are presented with several new insights.

\section{Compressible free-surface or pressurized flows in a thin domain}

In this section, we present the geometry and dynamics of the compressible flow that we consider. Particular care will be given to the description of the two types of conditions that will be prescribed on the top boundary, namely free-surface and pressurized conditions.

\subsection{Flow configuration}

\begin{figure}[!ht]
    \centering
    \begin{tikzpicture}[scale=0.47]
        \begin{scope}[canvas is zx plane at y=-3]
            \draw (-2,-8) rectangle ++(4,16);
            \draw (0,7) node {$\T_L^2$};
        \end{scope}
        \draw (8.5,-3,2) node[right]{\large $\Omega_h$};


        \path[fill=blue!70!white, opacity=0.7](-8,0,-2) .. controls +(2,1,0) and +(-2,-.8,0).. (0,0,-2)
         .. controls +(2,1,0) and +(-2,-.6,0) .. (8,0,-2)
         .. controls +(0,-.8,2) and +(0,1.5,0.01) ..  (8,0,2)
         ..controls +(-2,-.6,0) and +(2,1,0) ..(0,0,2)
         .. controls +(-2,-.8,0) and +(2,1,0) ..(-8,0,2)
         ..controls +(0,1.5,0.01) and +(0,-.8,2) ..(-8,0,-2);
 
         \draw[thick, color=blue](-8,0,-2) .. controls +(2,1,0) and +(-2,-.8,0).. (0,0,-2)
         .. controls +(2,1,0) and +(-2,-.6,0) .. (8,0,-2)
         .. controls +(0,-.8,2) and +(0,1.5,0.01) ..  (8,0,2)
         ..controls +(-2,-.6,0) and +(2,1,0) ..(0,0,2)
         .. controls +(-2,-.8,0) and +(2,1,0) ..(-8,0,2)
         ..controls +(0,1.5,0.01) and +(0,-.8,2) ..(-8,0,-2);
 
        \draw [thick,->,color=blue!80!black] (0.3,0,-0.5) -- (-0.1,1, -0.5) ; 
        \draw [color=blue!80!black] (-0.1,1,-0.5) node[left] {$\nbf_h$};
        \draw [color=blue!80!black] (0.2,0.23,-0.5)--(0.45,0.35,-0.5)--(0.55,0.15,-0.5);

        \draw [thick,<->] (-8,1,-2) -- (8,1,-2) ;
        \draw[thick,<->] (-8.3,0.4,2.3)-- (-8.3,0.4,-1.8);
        \draw (0,1,-2) node[above] {$L\gg h$}; 
        \draw (-8.3,0.6,0.1) node[left] {$L$};

        \draw[thick, ->] (-9,-3.5,2.5)--(-8,-3.5,2.5);
        \draw (-8,-3.5,2.5) node[right] {$x$};
        \draw[thick, ->] (-9,-3.5,2.5)--(-9,-2.5,2.5);
        \draw (-9,-2.5,2.5) node[above] {$z$};
        \draw[thick, ->] (-9,-3.5,2.5)--(-9,-3.5,3.5);
        \draw (-9,-3.5,4) node {$y$};

        \draw [->,color=blue!50!red,very thick] (-4,-1.5,0) -- (-2.5,-2,1.5);
        \draw (-3,-1.5,1) node [right] {$\ubf=(\ubf_\xbf,w)=(u,v,w)$};

        \draw[thick,<->] (8.3,0.15,0)--(8.3,-2.6,0);
        \draw (8.3,-1.5,0) node[right] {$h$};
    \end{tikzpicture}
    \caption{Flow configuration}
    \label{fig1:flow_config}
    \end{figure}
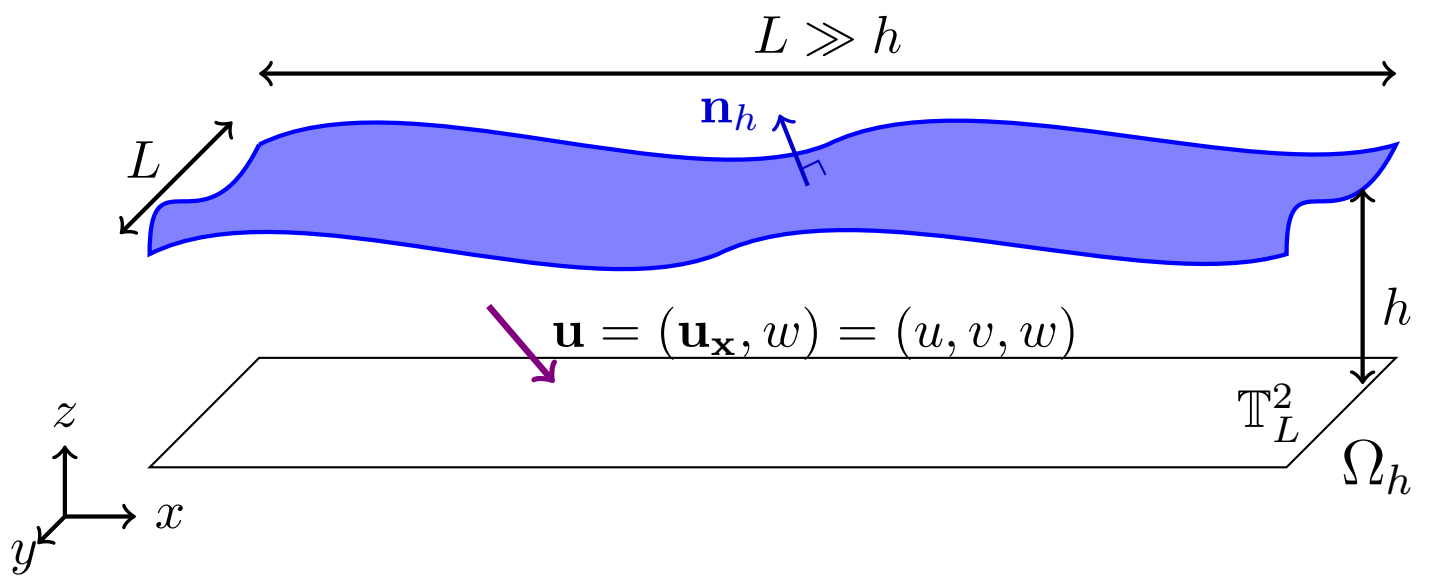

    We consider a three-dimensional compressible flow in a thin domain which might depend on time. The region $\Omega_h(t)$ filled by the fluid at time $t \in \R_+$ is characterized by a smooth function $h(t,x,y)$ representing the height of the fluid: 

    \[ \Omega_h(t) = \{(\xbf,z)=(x,y,z) \mid \xbf \in \T_L^2, \, z \in (0,h(t,\xbf))\},\]
    
    \noindent where $\T_L^2$ is the two-dimensional torus of period $L$ (see~\cref{fig1:flow_config}). We also define the top boundary $\Gamma_h$:
    \begin{equation}
        \forall t \in \R_+, \qquad \Gamma_h(t)=\{(\xbf,h(t,\xbf)), \; \xbf \in \T_L^2\}.
    \label{eq:def_free_surface}
    \end{equation}

    In this work, we assume that the domain $\Omega_h$ is thin, which means that the characteristic vertical dimension of the flow is much smaller that the horizontal dimensions of the domain (see~\cref{sec:averaging} for a more detailed presentation of the thin-domain assumption and its implications).

\subsection{The compressible barotropic Navier--Stokes system}

    Let $\mathcal{D}_h=\{(t,\xbf,z) \in \R_+ \times \T_L^2 \times \R \mid (\xbf,z) \in \Omega_h(t)\}$. Let $\rho: \mathcal{D}_h \to \R_+$ denote the density of the fluid, $\ubf=(u,v,w):\mathcal{D}_h \to\R^3$ its velocity and $p: \mathcal{D}_h \to \R_+$ its pressure. The flow is described by the following compressible Navier--Stokes system on $\R_+ \times \Omega_h$:

     \begin{subequations}
        \begin{align}
            \partial_t \rho + \nabla \cdot (\rho \ubf) &= 0, \label{eq:Navier_Stokes_mass}\\
            \partial_t (\rho \ubf) + \nabla \cdot (\rho \ubf \otimes \ubf) + \nabla p &= \nabla \cdot \S + \rho \gbf,
        \label{eq:Navier_Stokes_momentum}
        \end{align}
        \label{eq:Navier_Stokes}
     \end{subequations}
    
    \noindent where the viscous tensor $\S$ and the gravity force $\gbf$ are defined as follows:
    \begin{align}
        \S &= \mu((\nabla \otimes \ubf) + (\nabla \otimes \ubf)^{\rm T}) + \lambda (\nabla \cdot \ubf) \I,\label{eq:def_viscous_tensor}\\
        \gbf&= \begin{pmatrix}
        0 & 0 & -g
        \end{pmatrix}^{\rm T}
    \end{align}
     and $\mu$ and $\lambda$ are positive constants which denote the shear and bulk viscosity of the fluid respectively.
    
    \noindent Since we consider a barotropic flow, we assume that the fluid's pressure only depends on its density. More precisely, we prescribe the following ideal gas equation of state:
    \begin{equation}
        p(\rho)=a\rho^\gamma,
        \label{eq:barotropy_law}
    \end{equation}
    where $a>0$ and $\gamma>1$ are two constants. 
    To close the model made of \cref{eq:Navier_Stokes}, we need to prescribe boundary conditions at the bottom wall and at the surface $\Gamma_h$. First, we prescribe a slip condition at the bottom of the domain. We assume that the bottom boundary is horizontal and there is no topography, which means that this boundary is described by the equation $z=0$. The conditions read
    \begin{equation}
        \left\{ \begin{array}{rcl}
        w &=& 0,\\
        -\mu (\partial_z \ubf_\xbf+\nabla_\xbf w)+\kappa \ubf_\xbf &=&0,
        \end{array}\right.\qquad \text{at } z=0,
        \label{eq:bottom_boundary_condition}
    \end{equation}
    where $\nabla_\xbf$ is the horizontal gradient operator.
    \noindent The first equation of this condition represents the impermeability of the bottom boundary. The second one describes the horizontal friction of the fluid against the wall. The quantity $\kappa$ is the friction coefficient at the boundary.
    
    Now, we present the two kinds of top conditions that will be studied in this paper.
    
    \noindent {\bf Free surface conditions.} In the case of free-surface flows, the fluid height $h$ is an unknown of the problem. For all $t \in \R_+$, the free surface $\Gamma_h(t)$ corresponds to the level set $\psi(t,\xbf,z)=0$ where $\psi(t,\xbf,z)=z-h(t,\xbf)$. The surface $\Gamma_h$ being transported by the velocity field $\ubf(h)=\ubf(t,\xbf,h(t,\xbf))$, 
    \[ \partial_t \psi + \ubf\cdot \nabla \psi=0,\]
    which yields the evolution equation for $h$:
    \begin{equation}
        \partial_t h +\ubf_\xbf\cdot \nabla_\xbf h=w,\qquad \text{at } z=h.
        \label{eq:advection_h}
    \end{equation}
    
    \noindent In addition to this equation, we introduce the total stress tensor $\sigma =-p\I + \S$ and the unitary normal vector to the interface pointing upwards (see~\cref{fig1:flow_config})
    \begin{equation}
        \nbf_{h}=
        \frac{1}{\sqrt{1+\Norm{\nabla_\xbf h}^2}}
        \begin{pmatrix}
            -\nabla_\xbf h \\ 1
        \end{pmatrix}.
    \label{eq:def_normal_free_surface}
    \end{equation}
    We prescribe the following free-surface condition at $\Gamma_h$:
    \begin{equation}
        \sigma \nbf_h=0,\qquad \text{at } z=h.
        \label{eq:free_surface_condition}
    \end{equation}
    This condition translates the equilibrium of the boundary, or in other words the continuity of the stress applied to the normal vector to $\Gamma_h$.
        
    \noindent {\bf Pressurized conditions.} In the pressurized case, the height $h$ is prescribed depending on ${\bf x}$ and therefore is not an unknown anymore. In this case, the following conditions hold:
    \begin{subequations}\label{eq:pressurized_bc}
        \begin{numcases}{}
            \ubf_\xbf\cdot \nabla_\xbf h = w,\label{eq:pressurized_impermeability}\\
            (\sigma \nbf_h)_{\text{tang}}=0, \label{eq:pressurized_tensor}
        \qquad \text{at } z=h,\end{numcases}
    \end{subequations}
    where $(\cdot)_{\text{tang}}$ denotes the component of a vector which is tangential to the surface $\Gamma_h$.
    
    In this paper, we use the thin-domain assumption to derive a two-dimensional compressible flow model. To do so, we perform an asymptotic expansion of the model developed above with respect to the vertical variable and average it with respect to the vertical space variable. The presentation of this process is given in the next section. 

\section{Averaging the equations}\label{sec:averaging}

\subsection{The thin domain assumption and associated scalings}

    As mentioned in the previous section, this paper studies a flow in a thin domain. More specifically, we study the asymptotic regime $\varepsilon \to 0$ where $\varepsilon$ is a parameter representing the ratio between the vertical and horizontal dimensions of the flow. As the parameter $\varepsilon$ becomes small, some terms in the equations and boundary conditions become negligible. To make the influence of the parameter $\varepsilon$ clear, we introduce below the different scalings taking place in our configuration and rescale the model accordingly. 

    Let us introduce some characteristic dimensions. Denote $H$ the characteristic height of the domain, $L$ the characteristic horizontal period of the flow and $T$ its characteristic time. Let $\varepsilon=H/L \ll 1$ be the thickness parameter, that is assumed to be small. Let us temporarily denote the dimensionless variables $\Tilde{x}=x/L$, $\Tilde{y}=y/L$, $\Tilde{z}=z/H=z/(\varepsilon L)$ and $\Tilde{t}=t/T$. 
    
    Now, we introduce characteristic dimensions relatively to the different unknowns and parameters of the configuration: $\varrho$ for the density, $U=L/T$ for the horizontal velocity, $W=h/T$ for the vertical velocity, $P=\varrho U^2$ for the pressure, $M=RUL$ for the viscosity coefficients, and $K=RU$ for the friction. We also introduce the Froude number $\text{Fr}=U/\sqrt{gH}$ which compares the flow inertia to the gravity force, the Reynolds number $\text{Re}=M/\mu$ which compares inertia to the viscosity, and the Mach number $\text{Ma}=U/\sqrt{p'(\varrho)}=U/\sqrt{a\gamma\varrho^{\gamma-1}}$ which compares the flow velocity to the sound velocity in the medium. The dimensionless functions are denoted $\Tilde{h}=h/H$, $\Tilde{\rho}=\rho/\varrho$, $\widetilde{\ubf_\xbf}=\ubf_\xbf/U$, $\Tilde{w}=w/W$ and $\Tilde{p}=p/P$. For compatibility, the consistency factor $a$ in the pressure law is $\tilde a = a \varrho^\gamma/P$. Denote $\Tilde{\mu}=1/\rm{Re}$, $\Tilde{\lambda}=\lambda/M$, $\Tilde{\kappa}=\kappa/K$ and $\tilde{g}=1/\text{Fr}^2$.

    Note that in virtue of the thin domain assumption, $W=\varepsilon U$. Other assumptions regarding the dependency of the viscosity and friction parameters with respect to $\varepsilon$ will be given later on. 
    
    The flow is now studied on the rescaled domain
    \begin{equation}
        \Omega=\T^2 \times (0,\tilde{h})
        \label{eq:def_domain_rescaled}
    \end{equation}
    where $\T=\T_1$ is the torus of period $1$. The advection equation~\eqref{eq:advection_h} reads the same once rescaled. Denote $\Gamma$ the rescaled version of $\Gamma_h$ and $\Delta_\xbf$ the horizontal Laplace operator. Removing the $\Tilde{\cdot}$ for the sake of clarity, the Navier--Stokes system~\eqref{eq:Navier_Stokes} is rescaled as 

    \begin{subequations}
        \begin{align}
            \partial_t \rho + \nabla \cdot (\rho \ubf) &= 0,
        \label{eq:Navier_Stokes_mass_rescaled}\\
            \partial_t (\rho \ubf_\xbf) + \nabla_\xbf \cdot (\rho \ubf_\xbf \otimes \ubf_\xbf) + \partial_z(\rho w \ubf_\xbf) + \nabla_\xbf p &= (\lambda+\mu) \nabla_\xbf(\nabla \cdot \ubf) \nonumber \\
            & \quad + \mu\Delta_\xbf \ubf_\xbf +  \frac{\mu}{\varepsilon^2} \partial_{zz}^2 \ubf_\xbf,
        \label{eq:Navier_Stokes_momentum_horizontal_rescaled}\\
            \varepsilon^2(\partial_t(\rho w)+\nabla_\xbf \cdot (\rho w \ubf_\xbf)+\partial_z(\rho w^2)) +\partial_z p &=-\rho g + (\lambda+\mu)\partial_z (\nabla \cdot \ubf) \nonumber \\
            & \quad + \mu \partial_{zz}^2 w + \varepsilon^2 \mu \partial_{zz}^2 w.
        \label{eq:Navier_Stokes_momentum_vertical_rescaled}
        \end{align}
    \end{subequations}

    \noindent We separated the horizontal and vertical components of the momentum equations in the system above since the rescaling affects them differently. The equation of state~\eqref{eq:barotropy_law} reads the same after rescaling. 
    
    Next, we take care of the bottom boundary condition~\eqref{eq:bottom_boundary_condition}:
    \begin{equation}
        \left\{ \begin{array}{rcl}
        w &=& 0,\\
        -\frac{\mu}{\varepsilon} (\partial_z\ubf_\xbf+\varepsilon^2 \nabla_\xbf w) + \kappa \ubf_\xbf &=& 0,
        \end{array}\right.\qquad \text{at } z=0.
        \label{eq:bottom_boundary_rescaled}
    \end{equation}

    There remains to rescale the top boundary conditions. For the boundary conditions \eqref{eq:free_surface_condition} and \eqref{eq:pressurized_bc}, we first rescale the normal unitary vector to the free surface
    \[\nbf=\frac{1}{\sqrt{1+\varepsilon^2\Norm{\nabla_\xbf h}^2}} \begin{pmatrix}
        -\varepsilon \nabla_\xbf h\\
        1
    \end{pmatrix}.
    \]
    We also introduce two tangential vectors to the interface $\Gamma$ in order to express the boundary conditions with respect to its tangential and normal components:
    \[ \tbf^x = \begin{pmatrix} 1\\0\\ \varepsilon \partial_x h \end{pmatrix} \quad \text{and} \quad \tbf^y=\begin{pmatrix} 0\\1\\ \varepsilon \partial_y h \end{pmatrix}.\]
    The rescaling of the free surface and the pressurized conditions need the tangent and normal component of the normal stress tensor $\sigma \nbf$, which we collect in the following lemma.
    \begin{lemma}
        The decomposition of the rescaled version of $\sigma \nbf$ at $z=h$ in the directions $\tbf^x$, $\tbf^y$ and $\nbf$ reads: 
        \begin{subequations}
            \begin{align}
                \sigma \nbf \cdot \tbf^x|_{z=h}&=\frac{\mu}{\sqrt{1+\varepsilon^2|\nabla_\xbf h|^2}} \left[ \frac{\partial_z u}{\varepsilon} + \varepsilon (-2\partial_x h(\partial_x u -\partial_z w) -\partial_yh(\partial_x v + \partial_y u)+\partial_x w-(\partial_xh)^2 \partial_z u\right.\nonumber \\
                & \left.\quad -\partial_x h \partial_y h \partial_z v)-\varepsilon^3\Big((\partial_x h)^2 \partial_x w + \partial_x h \partial_y h \partial_y w\Big)\right]|_{z=h} \label{eq:rescaled_constraint_tx}\\
                \sigma \nbf \cdot \tbf^y|_{z=h}&=\frac{\mu}{\sqrt{1+\varepsilon^2|\nabla_\xbf h|^2}} \left[ \frac{\partial_z v}{\varepsilon} + \varepsilon (-2\partial_y h(\partial_y v-\partial_z w)  -\partial_xh(\partial_x v + \partial_y u)+\partial_y w-(\partial_yh)^2 \partial_z v \right.\nonumber \\
                &\left. \quad -\partial_x h \partial_y h \partial_z u)-\varepsilon^3((\partial_y h)^2 \partial_y w + \partial_x h \partial_y h \partial_x w)\right]|_{z=h}\label{eq:rescaled_constraint_ty}\\
                \sigma \nbf \cdot \nbf|_{z=h} &= (-p+\lambda) (\nabla \cdot \ubf)|_{z=h} +\frac{2\mu}{1+\varepsilon^2|\nabla_\xbf h|^2}  \left[ \partial_z w -\Big(\partial_x h\partial_z u  + \partial_y h \partial_z v\Big) \right.\nonumber \\
                & \left.\quad +\varepsilon^2\Big( (\partial_xh)^2 \partial_x u + (\partial_y h)^2\partial_y v +\partial_x h \partial_x w+\partial_y h \partial_y w+ \partial_x h \partial_y h (\partial_y u+\partial_x v)\Big)\right]|_{z=h}\label{eq:rescaled_constraint_normal}
            \end{align}
            \label{eq:rescaled_constraint}
        \end{subequations}
    \end{lemma}
    
    \noindent Note that the pressure does only appear in the normal component of $\sigma \nbf|_{z=h}$. We are now able to rescale both types of boundary conditions. 
    
    \noindent \textbf{Free surface conditions.} The kinematic condition \eqref{eq:advection_h} does not change.  The rescaled free surface condition \eqref{eq:free_surface_condition} is 
    \begin{equation}\label{eq:free_surface_condition_tensor}
        \sigma \nbf \cdot \tbf^x|_{z=h}=\sigma \nbf \cdot \tbf^y|_{z=h}=\sigma \nbf \cdot \nbf|_{z=h}=0,
    \end{equation}
    where each component is given in~\eqref{eq:rescaled_constraint}.

    \noindent \textbf{Pressurized.} The rescaling process keeps \eqref{eq:pressurized_impermeability} unchanged and the tensor condition is
    \begin{equation}\label{eq:pressurized_tensor2}
        \sigma \nbf \cdot \tbf^x|_{z=h}=\sigma \nbf \cdot \tbf^y|_{z=h}=0,
    \end{equation}
    where the components are given by \eqref{eq:rescaled_constraint_tx} and \eqref{eq:rescaled_constraint_ty} for a given $h=h(\xbf)$.

\subsection{First-order approximation}

    In a first simplification of the model, we prescribe the following scalings for the viscosity and friction coefficients:
    \begin{equation}
        \mu=\widehat{\mu} \varepsilon, \quad \lambda=\widehat{\lambda} \varepsilon \quad \text{and} \quad \kappa = \widehat{\kappa} \varepsilon.
        \label{eq:scaling_viscosity_friction}
    \end{equation}
    
    \noindent We integrate these scalings into the rescaled model and neglect the terms of order $1$ in $\varepsilon$ in the equations. We write all equations for the sake of clarity
    \begin{subequations}
        \begin{align}
        \partial_t \rho + \nabla_\xbf \cdot (\rho \ubf_\xbf)+\partial_z(\rho w) &= 0,
        \label{eq:Navier_Stokes_mass_asymptotic}\\
            \partial_t(\rho \ubf_\xbf)+\nabla_\xbf \cdot(\rho \ubf_\xbf \otimes \ubf_\xbf) + \partial_z(\rho w \ubf_\xbf)+ \nabla_\xbf p &= \frac{\widehat{\mu}}{\varepsilon}\partial_{zz} \ubf_\xbf + O(\varepsilon),
        \label{eq:horizontal_momentum_asymptotic}\\
            \partial_z p &=-\rho g+O(\varepsilon).
        \label{eq:vertical_momentum_asymptotic}
        \end{align}
    \end{subequations}

    \noindent We also write the boundary conditions with the new scalings for the viscosity and friction coefficients, neglecting also terms of order $\varepsilon$. 
    \\
    The bottom boundary conditions are as follows
    \begin{subequations}\label{eq:bc_bottom}
        \begin{align}
        w|_{z=0}=0, \label{eq:bottom_impermeability}\\
        \left(-\frac{\widehat{\mu}}{\varepsilon}\partial_z \ubf_\xbf+\widehat{\kappa} \ubf_\xbf\right)\Big|_{z=0}=0.
    \label{eq:horizontal_friction_asymptotic}
    \end{align}
    \end{subequations}

    \noindent \textbf{Free surface conditions.} From \eqref{eq:advection_h} and \eqref{eq:free_surface_condition_tensor},
    \begin{subequations}\label{eq:asymptotic_free_surface}
        \begin{align}
            \partial_t h +\ubf_\xbf|_{z=h}\cdot\nabla_\xbf h &= w|_{z=h}\label{eq:advection_h2}\\
            \widehat{\mu} (\partial_z \ubf_\xbf)|_{z=h}&=O(\varepsilon^2) \label{eq:asymptotic_constraint_tang}\\
            p|_{z=h}&=O(\varepsilon).\label{eq:asymptotic_constraint_normal}
        \end{align}
    \end{subequations}
    The second equation corresponds to the tangential component of the free-surface condition and the third one to its normal component. 
    
    \noindent \textbf{Pressurized conditions.} The pressurized conditions \eqref{eq:pressurized_tensor2} become
    \begin{subequations}
        \begin{align}
            \ubf_\xbf|_{z=h}\cdot \nabla_\xbf h(\xbf) &=w|_{z=h}, \label{eq:asymptotic_pressurized_normal}\\
            \widehat{\mu} (\partial_z \ubf_\xbf)|_{z=h}&=O(\varepsilon^2). \label{eq:asymptotic_pressurized_tang}
        \end{align}
        \label{eq:asymptotic_pressurized}
    \end{subequations}
Notice that since in the pressurized case $h=h(\xbf)$ does not depend on time, conditions \eqref{eq:advection_h2} and \eqref{eq:asymptotic_pressurized_normal} are equivalent.

    \begin{remark}
        The horizontal momentum equation cannot be further simplified. Indeed, it tells us that $\widehat{\mu} \partial_{zz}^2 \ubf_\xbf$ is of order $\varepsilon$.
        \hfill$\square$
    \end{remark}
    
    \begin{remark}
        Combining the asymptotic horizontal momentum equation~\eqref{eq:horizontal_momentum_asymptotic} and the bottom boundary condition~\eqref{eq:horizontal_friction_asymptotic}, we can show that
        \begin{equation}
           \forall (t,\ybf) \in \R_+ \times \Omega, \qquad 
           \widehat{\mu} \partial_z\ubf_\xbf (t,\ybf)=O(\varepsilon),
        \label{eq:estimate_derivative_u}
        \end{equation}
        thanks to a Taylor expansion.
        \hfill$\square$
    \end{remark}
\subsection{Averaging the equations}

    In this section, we average the system composed of the rescaled mass conservation equation~\eqref{eq:Navier_Stokes_mass_asymptotic} and the approximated horizontal momentum equation~\eqref{eq:horizontal_momentum_asymptotic} by integrating them in the vertical direction. Thanks to this procedure, we obtain a two-dimensional system. The boundary terms appearing in the integration process are closed thanks to the boundary conditions. We will also approximate some pointwise values of the variables by their average to have a model expressed fully in terms of averaged unknowns.
    
    Before proceeding to integration, we need to introduce two different notions of averages. Indeed, since we work in a compressible setting, many terms in the equations are nonlinear, and the notion of average must be chosen accordingly. 
    
    \begin{definition}[Averages]
        Let $f$ be a function defined in $\R_+\times \Omega$. Its average $\Bar{f}$ is defined by 
        \begin{equation}
            \forall (t,\xbf) \in \R_+ \times \T^2, \qquad
            \Bar{f}(t,\xbf)=\frac{1}{h(t,\xbf)}\int_0^{h(t,\xbf)} f(t,\xbf,z) \d z.
        \label{eq:definition_average}
        \end{equation}
        The Favre average of $f$ with respect to $\rho$, denoted $\favre{f}$, is defined by 
        \begin{equation}
            \forall (t,\xbf) \in \R_+ \times \T^2, \qquad
            \favre{f}(t,\xbf)=\frac{\Bar{\rho f}}{\Bar{\rho}}=\frac{1}{\Bar{\rho} h(t,\xbf)}\int_0^{h(t,\xbf)} (\rho f)(t,\xbf,z) \d z.
        \label{eq:definition_favre_average}
        \end{equation}
        \hfill$\square$
    \end{definition}
    
    Let us start by integrating the mass conservation equation~\eqref{eq:Navier_Stokes_mass_asymptotic} with respect to $z \in (0,h)$. Thanks to the Leibniz integral rule, for all smooth functions $f$, 
        \begin{equation}
            h \bar{\partial_t f} = \partial_t (h\bar{f})-f(h)\partial_t h, \quad h \bar{\nabla_\xbf f} = \nabla_\xbf (h\bar{f})-f(h)\nabla_\xbf h.
        \label{eq:interversion_derivative_integral}
        \end{equation}
    
    \noindent Therefore, integrating~\eqref{eq:Navier_Stokes_mass_asymptotic} yields
    \begin{equation}\label{eq:averaged_mass_interm}
        \partial_t(h\Bar{\rho}) +\nabla_\xbf \cdot(h \Bar{\rho} \favre{\ubf_\xbf}) -(\rho\partial_t h+ \rho \ubf_\xbf \cdot \nabla_\xbf h - \rho w)|_{z=h}-(\rho w)|_{z=0}=0,
    \end{equation}
where the last term vanishes due to the impermeability condition at the bottom \eqref{eq:bottom_impermeability}.
    Next, we integrate the horizontal momentum equation~\eqref{eq:horizontal_momentum_asymptotic} to write
    \begin{equation*}
    \begin{array}{l}
         \partial_t (h\Bar{\rho}\favre{\ubf_\xbf})+\nabla_\xbf \cdot (h\Bar{\rho}\favre{\ubf_\xbf \otimes \ubf_\xbf})+\nabla_\xbf (h\Bar{p})\\
         \qquad -p|_{z=h}\nabla_\xbf h-\Big(\rho \ubf_\xbf(\partial_t h +\ubf_\xbf\cdot\nabla_\xbf h - w)\Big)|_{z=h}  -(\rho  \ubf_\xbf w)|_{z=0}\\
         =\frac{\widehat{\mu}}{\varepsilon} (\partial_z \ubf_\xbf)|_{z=h}-\frac{\widehat{\mu}}{\varepsilon} (\partial_z \ubf_\xbf)|_{z=0}+O(\varepsilon).
    \end{array}
    \end{equation*}
We use conditions at the bottom \eqref{eq:bottom_impermeability} and \eqref{eq:horizontal_friction_asymptotic} to get
\begin{equation}
\begin{array}{l}
         \partial_t (h\Bar{\rho}\favre{\ubf_\xbf})+\nabla_\xbf \cdot (h\Bar{\rho}\favre{\ubf_\xbf \otimes \ubf_\xbf})+\nabla_\xbf (h\Bar{p})
         \\\qquad 
         -p|_{z=h}\nabla_\xbf h-\Big(\rho \ubf_\xbf(\partial_t h +\ubf_\xbf\cdot\nabla_\xbf h - w)\Big)|_{z=h}  \\
         =\frac{\widehat{\mu}}{\varepsilon} (\partial_z \ubf_\xbf)|_{z=h}-\widehat{\kappa}\ubf_\xbf|_{z=0}+O(\varepsilon).
    \end{array}    \label{eq:intermediate_avg_momentum}
    \end{equation}

    Next, we want to express the average $\favre{\ubf_\xbf \otimes \ubf_\xbf}$ in terms of the Favre average of the velocity. Using the estimate \eqref{eq:estimate_derivative_u}, any pointwise value of $\ubf_\xbf$ can be approximated by its Favre average up to an error of order $\varepsilon$. Thus, 
    \[\favre{\ubf_\xbf \otimes \ubf_\xbf}=\favre{(\favre{\ubf_\xbf}+O(\varepsilon))\otimes \ubf_\xbf}=\favre{\ubf_\xbf}\otimes \favre{\ubf_\xbf}+O(\varepsilon).\]
    Injecting this relation in~\eqref{eq:intermediate_avg_momentum} yields 
    \begin{equation}
    \begin{array}{l}
         \partial_t (h\Bar{\rho}\favre{\ubf_\xbf})+\nabla_\xbf \cdot (h\Bar{\rho}\favre{\ubf_\xbf} \otimes \favre{\ubf_\xbf})+\nabla_\xbf (h\Bar{p})
         \\\qquad 
         -p|_{z=h}\nabla_\xbf h-\Big(\rho \ubf_\xbf(\partial_t h +\ubf_\xbf\cdot\nabla_\xbf h - w)\Big)|_{z=h}  \\
         =\frac{\widehat{\mu}}{\varepsilon} (\partial_z \ubf_\xbf)|_{z=h}-\widehat{\kappa}\ubf_\xbf|_{z=0}+O(\varepsilon).
    \end{array} \label{eq:generic_avg_momentum}
    \end{equation}
Finally, we also integrate the vertical momentum equation \eqref{eq:vertical_momentum_asymptotic}. Taking into account the pressure law \eqref{eq:barotropy_law} we write
    \[ \partial_z p=\partial_z (a\rho^\gamma)=a\gamma \rho^{\gamma-1}\partial_z \rho=a\frac{\gamma}{\gamma-1} \rho\, \partial_z (\rho^{\gamma-1}),\]
    and then, the equation \eqref{eq:vertical_momentum_asymptotic} reads
    \[ a\frac{\gamma}{\gamma-1} \rho\, \partial_z (\rho^{\gamma-1})=-\rho g +O(\varepsilon).\]
    We divide by $\rho$ and neglect the terms of order $\varepsilon$ obtaining a $z$-derivative that is a constant, 
    \[ \partial_z (\rho^{\gamma-1})=-\frac{\gamma-1}{a\gamma} g.\]
    We integrate this equation in $[z,h]$ to obtain 
    \begin{equation}\label{eq:vertical_momentum_integrated}
        \rho^{\gamma-1}=(\rho|_{z=h})^{\gamma-1}+\frac{\gamma-1}{a\gamma}g(h-z).
    \end{equation}

  There remains to take care of the top boundary terms, depending on the configuration we choose. 

    \noindent \textbf{Free surface conditions.} They are given in ~\eqref{eq:asymptotic_free_surface}. We use the kinematic condition \eqref{eq:advection_h2} to write the mass conservation equation \eqref{eq:averaged_mass_interm} as
    \begin{equation}
        \partial_t(h\Bar{\rho})+\nabla_\xbf \cdot(h \Bar{\rho} \favre{\ubf_\xbf})=0.
    \label{eq:averaged_mass}
    \end{equation}
    We also use this condition to simplify the corresponding term in the horizontal momentum equation \eqref{eq:generic_avg_momentum}. Additionally, taking into account \eqref{eq:asymptotic_constraint_tang} and \eqref{eq:asymptotic_constraint_normal} we finally write the averaged momentum equation
    \begin{equation}
        \partial_t (h\Bar{\rho}\favre{\ubf_\xbf})+\nabla_\xbf \cdot (h\Bar{\rho}\favre{\ubf_\xbf} \otimes \favre{\ubf_\xbf})+\nabla_\xbf (h\Bar{p})=-\widehat{\kappa} \ubf_\xbf|_{z=0}+O(\varepsilon).
    \label{eq:momentum_free_surface_general}
    \end{equation}
Thanks to condition \eqref{eq:asymptotic_constraint_normal} and the pressure law \eqref{eq:barotropy_law}, we deduce $\rho|_{z=h}=O(\varepsilon)$, and the integrated vertical equation \eqref{eq:vertical_momentum_integrated} becomes
\begin{equation}\label{eq:vertical_free_surface_general}
        \rho^{\gamma-1}=\frac{\gamma-1}{a\gamma}g(h-z).
    \end{equation}
 At this stage, we have four unknowns : $h$, $\Bar{\rho}$, $\favre{\ubf_\xbf}$, $\Bar{p}$, and two equations \eqref{eq:averaged_mass}, \eqref{eq:momentum_free_surface_general} together with the constraints for density and pressure given by equations \eqref{eq:barotropy_law}, \eqref{eq:asymptotic_constraint_normal} and \eqref{eq:vertical_free_surface_general}.

 In~\cref{sec:closure_free_surface}, we use these equations  to exhibit relations between $\bar{\rho}$, $\bar{p}$ and $h$, which reduce the number of unknowns to two ($h$ and $\favre{\ubf_\xbf}$) and therefore close the system. \\\\
    
    \noindent \textbf{Pressurized conditions.} Recall that in this case the height $h$ is given as $h=h(\xbf)$ and is not an unknown of the system. This fact, together with condition \eqref{eq:asymptotic_pressurized_normal}, allows us to write the mass conservation equation \eqref{eq:averaged_mass_interm} as in the free surface case, given by \eqref{eq:averaged_mass}.

    Condition \eqref{eq:asymptotic_pressurized_tang} enables us to neglect the shear boundary term and condition \eqref{eq:asymptotic_pressurized_normal} the kinematic term. The momentum equation then reads 
    \begin{equation}
         \partial_t (h\Bar{\rho}\favre{\ubf_\xbf})+\nabla_\xbf \cdot (h\Bar{\rho}\favre{\ubf_\xbf} \otimes \favre{\ubf_\xbf})+\nabla_\xbf (h\Bar{p})-p|_{z=h}\nabla_\xbf h =-\widehat{\kappa} \ubf_\xbf|_{z=0}+O(\varepsilon).
    \label{eq:momentum_pressurized_general}
    \end{equation}
    In this configuration, we have three unknowns $\Bar\rho$, $\favre{\ubf_\xbf}$, $\Bar{p}$, for equations \eqref{eq:averaged_mass} and \eqref{eq:momentum_pressurized_general}, together with density and pressure constraints given by equations \eqref{eq:barotropy_law} and \eqref{eq:vertical_momentum_integrated}. However, we have no further information on the pressure at the surface, so we also must determine $p|_{z=h}$ depending on $h$, $\bar{\rho}$ and $\bar{p}$ to achieve the closure. This procedure is detailed in~\cref{sec:closure_pressurized}.

\medskip 
    \noindent Thanks to \eqref{eq:estimate_derivative_u} we can approximate $\ubf_\xbf|_{z=0}$ by $\favre{\ubf_\xbf}$, and we denote $\rho_S(t,\xbf)=\rho(t,\xbf,h(t,\xbf))$ and $p_S(t,\xbf)=p(t,\xbf,h(t,\xbf))$. Thus, we obtain the following system of first order in $\varepsilon$
    \begin{subequations}\label{eq:GeneralDepthAveragedSystem}
    \begin{equation} \label{eq:GeneralDepthAveragedSystem1}
	    	\left\{\begin{array}{l}
	 \partial_t(h\Bar{\rho})+\nabla_\xbf \cdot(h \Bar{\rho} \favre{\ubf_\xbf})=0,\\[2mm]
		\partial_t (h\Bar{\rho}\favre{\ubf_\xbf})+\nabla_\xbf \cdot (h\Bar{\rho}\favre{\ubf_\xbf} \otimes \favre{\ubf_\xbf})+\nabla_\xbf (h\Bar{p})
       - p_S \nabla_\xbf h =-\widehat{\kappa} \favre{\ubf_\xbf},
			    	\end{array}\right.
	\end{equation}
    subject to the additional constraints \eqref{eq:barotropy_law}, $p=a\rho^\gamma$, and \eqref{eq:vertical_momentum_integrated} that reads
    \begin{equation}\label{eq:GeneralDepthAveragedSystem2}
    \rho^{\gamma-1}=\rho_S^{\gamma-1}+\frac{\gamma-1}{a\gamma}g(h-z).
    \end{equation}
    \end{subequations}
	    
    As we explained before, this system must still be closed. Depending on the two considered cases, we resume the particularities to the system \eqref{eq:GeneralDepthAveragedSystem} as follows:

\medskip 
    
\noindent {\bf Free surface case.}\\ 
- $p_S=\rho_S=0$ due to condition~\eqref{eq:asymptotic_constraint_normal}.
\\
- Unknowns: $h$, $\Bar{\rho}$, $\favre{\ubf_\xbf}$ and $\Bar{p}$.
\\

\noindent{\bf Pressurized case.}\\ 
- $h=h(\xbf)$ known.
\\
- Unknowns: $\Bar{\rho}$, $\favre{\ubf_\xbf}$, $\Bar{p}$ and $p_S$.
\\\\
The objective in the next sections \ref{sec:closure_free_surface} and \ref{sec:closure_pressurized} is to find a relation between $h$, $\Bar\rho$ and $\Bar p$ and thus reduce the number of unknowns to have a closed system. Notice that in the linear case $\gamma=1$ we have $\Bar p = a \Bar \rho$ and the system is close. In the fully compressible environment considered here for $\gamma>1$, we do not have $\Bar{p(\rho)}=p(\Bar{\rho})$. This prevents us from obtaining a direct relation between the average density and the average pressure. This difficulty was overcome in \cite{ersoy_dimension_2016} by assuming a low Oser number $C=\frac{\rm Ma}{\rm Fr}$, which added to the asymptotic assumptions in \eqref{eq:scaling_viscosity_friction} gives $\Bar{p(\rho)}=p(\Bar{\rho})+O(C^2)$ and then the system is closed at order $C^2$, avoiding nonlinearity (see Remark \ref{rem:oser} for more details).


\begin{remark}\label{rem:oser}
The Oser number $C=\frac{Ma}{Fr}$ determines the importance of the gravity in front of the compressibility effects. A low Oser number assumption is only valid when gravity has no influence in the flow and in this case $\Bar{\rho^\gamma} = \bar\rho^\gamma$. In contrast, when the fluid flow is affected by gravity, this assumption is wrong, and the averaged pressure cannot be directly obtained through the averaged density. In \cite{ersoy_dimension_2016} a low Oser number assumption is considered to close the system for the strictly nonlinear pressure law, when $\gamma>1$. Namely, it is  considered $C=O(\varepsilon^l)$ for $l\geq 1/2$ which gives $\Bar{p(\rho)}=a\Bar{\rho^\gamma} = a\bar\rho^\gamma+O(\varepsilon^{2l})$ and the proposed model is closed on the order of $C^2$.
\\
Note that the vertical equation \eqref{eq:Navier_Stokes_momentum_vertical_rescaled} 
can be equivalently written in terms of the Oser number.
Taking into account the scaling in \eqref{eq:scaling_viscosity_friction} it gives
$$\partial_z p =-\gamma \frac{\text{Ma}^2}{\text{Fr}^2}\rho+O(\varepsilon)=-\gamma \,C^2\rho+O(\varepsilon).
$$
Under a small Oser number $C$, the pressure does not depend on $z$, so neither the density due to the state law. The system is then closed. The development performed here shows how to obtain the compressible system when gravity effects are important in flow dynamics, being a significant extension of the model presented in \cite{ersoy_dimension_2016}.
\end{remark}
        
\section{Closing the system in the free surface configuration}
\label{sec:closure_free_surface}
In this section, we focus on the closure of system \eqref{eq:GeneralDepthAveragedSystem} for the free surface case. First, we will study the particular case $\gamma=2$ to show the strategy for the simplest non-linear pressure law, then we will analyze the general case.

\subsection[Quadratic barotropy case]{Case $\gamma=2$}

    In this section, we find new relations to close the averaged system~\eqref{eq:GeneralDepthAveragedSystem}. Using the vertical momentum equation~\eqref{eq:vertical_momentum_asymptotic} and the pressure law \eqref{eq:barotropy_law} for $\gamma=2$, we will express $\bar{\rho}$ and $\bar{p}$ in terms of $h$, resulting in a two-equation system with two unknowns. 
    
    \subsubsection{Determining the averaged density and pressure}
    
Let us focus on the first-order approximation of the integrated vertical momentum equation~\eqref{eq:GeneralDepthAveragedSystem2}. Due to condition~\eqref{eq:asymptotic_constraint_normal}, the pressure on the free surface is zero at the main order, which gives $p_S(t,\xbf)=p(t,\xbf,h(t,\xbf))=0$. Using the equation of state \eqref{eq:barotropy_law}, we have $\rho_S=0$.  Embedding this value in equation \eqref{eq:GeneralDepthAveragedSystem2} yields  
    \begin{equation}\label{eq:rho_gamma2}
        \rho  = \frac{g}{2a}(h - z).
    \end{equation} 
    Therefore, taking the average of the previous equation yields the relation between $\Bar \rho$ and $h$,
    \begin{equation}
        \Bar{\rho} = \frac{gh}{4a}.
    \label{eq:rho_avg_free_surface_quad}
    \end{equation}
Next, we determine $\Bar{p}$ using again the equation of state, 
    \[p=a\rho^2 = a\frac{g^2}{4a^2}(h-z)^2.\]

    \noindent To find a relation between $\bar p$ and $h$, we take the average of the previous relation,
    \begin{equation}
        \Bar{p} = \frac{g^2h^2}{12a}.
    \label{eq:p_avg_free_surface_quad}
    \end{equation}
   
    \begin{remark}
        We have confirmation that $\Bar{\rho^2}\neq \Bar{\rho}^2$.

        \hfill $\square$
    \end{remark}
    Equations \eqref{eq:rho_avg_free_surface_quad} and \eqref{eq:p_avg_free_surface_quad} relate the unknowns $h$, $\bar p$ and $\bar\rho$, allowing to keep just one of them as an unknown of the system.
    Thus, embedding these expressions of $\bar{\rho}$ and $\bar{p}$ in the averaged mass conservation and momentum equations in \eqref{eq:GeneralDepthAveragedSystem}, we have the following system for unknowns $h$ and $\favre{\ubf_\xbf}$:
    
    \begin{align}
        \partial_t (h^2)+\nabla_\xbf \cdot (h^2 \favre{\ubf_\xbf})&=0,
        \label{eq:mass_free_surface}\\
        \partial_t (h^2\favre{\ubf_\xbf}) +\nabla_\xbf \cdot (h^2 \favre{\ubf_\xbf} \otimes \favre{\ubf_\xbf}) + \nabla_\xbf \left( \frac{gh^3}{3}\right)&=-\frac{4a}{g}\widehat{\kappa}\favre{\ubf_\xbf}.
        \label{eq:momentum_free_surface}
    \end{align}
    This system is complete since we have two equations and two unknowns $h$ and $\favre{\ubf_\xbf}$. In the following subsection, we are going to derive the averaged energy equation of the system and compare it to the energy equation of the averaged model. 

    \subsubsection{Averaged (energy equation) vs. (averaged energy) equation}

    In the first part of this section, we introduce the energy equation of the compressible Navier--Stokes system and average it in the vertical direction. Then, we compute the energy equation associated to the averaged model~\eqref{eq:mass_free_surface}-\eqref{eq:momentum_free_surface}, to finally compare both equations and energies to discuss the validity of our derivation. 
    
    \paragraph{Averaging the energy equation of the Navier--Stokes system.}

    Let us introduce the specific energy of the fluid 
    \begin{equation}
        E=\frac{1}{2} \Norm{\ubf}^2 + e(\rho) + gz
        \label{eq:energy_definition}
    \end{equation}
    where $e(\rho)$ is the internal energy of the fluid given for iso-entropic processes as 
    \begin{equation}\label{eq:internal_energy_definition}
        e(\rho)=\frac{p}{\rho(\gamma-1)}.
    \end{equation}
    The energy equation associated with~\eqref{eq:Navier_Stokes_mass}-\eqref{eq:Navier_Stokes_momentum} is 
    \begin{equation}
        \partial_t (\rho E) + \nabla\cdot ((\rho E+p) \ubf)= \nabla \cdot (\S \ubf).
        \label{eq:Navier_Stokes_energy}
    \end{equation}
    Rescaling the equation and neglecting the terms of order $1$ in $\varepsilon$ yields
    \begin{equation}
        \partial_t (\rho E_\xbf)+\nabla \cdot ((\rho E_\xbf+p)\ubf)= \frac{\mu}{\varepsilon^2} \partial_z (\ubf_\xbf\cdot \partial_z \ubf_\xbf)+O(\varepsilon),
    \end{equation}
    where $E_\xbf=\Norm{\ubf_\xbf}^2/2+e(\rho)+gz$. 

    Let us integrate the previous equation in the vertical direction. We use the kinematic condition \eqref{eq:advection_h2} for the free surface case and \eqref{eq:asymptotic_pressurized_normal} with $h=h(\xbf)$ for the pressurized case. We additionally use the impermeability condition on the bottom of the domain \eqref{eq:bottom_impermeability}, to write
    \begin{multline*}
        \partial_t (h\Bar{\rho}\favre{E_\xbf}) + \nabla_\xbf \cdot (h\Bar{\rho} \favre{E_\xbf \ubf_\xbf}+h\Bar{p\ubf_\xbf})-p_S(\ubf_\xbf\cdot \nabla_\xbf h-w)|_{z=h}\\
        =\frac{\widehat{\mu}}{\varepsilon}\partial_z \ubf_\xbf \cdot \ubf_\xbf(h)-\frac{\widehat{\mu}}{\varepsilon}\partial_z \ubf_\xbf \cdot \ubf_\xbf(0) +O(\varepsilon).
    \end{multline*}
The boundary conditions \eqref{eq:asymptotic_constraint_tang} and \eqref{eq:asymptotic_pressurized_tang} allow to determine the right-hand side, 
\begin{equation*}
        \partial_t (h\Bar{\rho}\favre{E_\xbf}) + \nabla_\xbf \cdot (h\Bar{\rho} \favre{E_\xbf \ubf_\xbf}+h\Bar{p\ubf_\xbf})-p_S(\ubf_\xbf\cdot \nabla_\xbf h-w)|_{z=h}\\
        =-\widehat{\kappa} \Norm{\ubf_\xbf(0)}^2+O(\varepsilon).
    \end{equation*}
Notice that the term evaluated at the free surface, $p_S(\ubf_\xbf\cdot \nabla_\xbf h-w)|_{z=h}$ vanishes in the two configurations due, respectively, to conditions  \eqref{eq:asymptotic_constraint_normal} and \eqref{eq:asymptotic_pressurized_normal}.
As before, we approximate $\ubf_\xbf$ by $\favre{\ubf_\xbf}$ thanks to 
\eqref{eq:estimate_derivative_u} to finally write the general equation of energy for the Navier-Stokes system
\begin{subequations}\label{eq:energyNS_general}
\begin{equation}\label{eq:energyNS_general_eq}
        \partial_t (h\Bar{\rho}\favre{E_\xbf}) + \nabla_\xbf \cdot (h(\Bar{\rho} \favre{E_\xbf}+\Bar{p}) \favre{\ubf_\xbf})
        =-\widehat{\kappa} \Norm{\favre{\ubf_\xbf}}^2+O(\varepsilon),
    \end{equation}
with
\begin{equation}\label{eq:definition_favre_energy}
    \favre{E_\xbf}=\frac{\Norm{\favre{\ubf_\xbf}^2}}{2}+\favre{e(\rho)+gz}+O(\varepsilon)=\frac{\Norm{\favre{\ubf_\xbf}^2}}{2}+\frac{1}{\gamma-1}\frac{\Bar{p}}{\Bar{\rho}}+\frac{\bar{\rho g z}}{\bar{\rho}}+O(\varepsilon).
\end{equation}
where we used \eqref{eq:internal_energy_definition} to write $\favre{e(\rho)}=\frac{1}{\gamma-1}\favre{\frac{p}{\rho}}=\frac{1}{\gamma-1}\frac{\bar p}{\bar \rho}$. Notice that this energy equation is the same for both configurations, free surface or pressurized.
\end{subequations}
\bigskip

Let us analyze the free surface case when $\gamma=2$. Thanks to \eqref{eq:asymptotic_constraint_normal} we have $p_S=0$ and then, neglecting terms of order $\varepsilon$, the energy equation becomes
    
    \begin{equation}
        \partial_t (h\Bar{\rho}\favre{E_\xbf}) + \nabla_\xbf \cdot (h(\Bar{\rho} \favre{E_\xbf}+\Bar{p}) \favre{\ubf_\xbf})
        =-\widehat{\kappa} \Norm{\favre{\ubf_\xbf}}^2,
    \end{equation}
and
    \begin{equation}
        \favre{E_\xbf}
        =\frac{\Norm{\favre{\ubf_\xbf}^2}}{2}+\frac{\Bar{p}}{\Bar{\rho}}+\frac{\bar{\rho g z}}{\bar{\rho}}+O(\varepsilon),
\end{equation}
where $\bar p$ and $\bar\rho$ are given in \eqref{eq:p_avg_free_surface_quad} and \eqref{eq:rho_avg_free_surface_quad} respectively.
\\
To calculate the average $\bar{\rho g z}$, we use the expression of the density \eqref{eq:rho_gamma2}. We obtain
$$
\favre{e(\rho)}=\frac{\Bar{p}}{\Bar{\rho}}=\frac{gh}{3}, \qquad \favre{gz}=\frac{\bar{\rho g z}}{\bar{\rho}}=\frac{gh}{3}.
$$
The averaged energy equation then reads
    \begin{multline}\label{eq:averaged_energy_free_surf_gamma2}
        \partial_t \left( h^2 \left( \frac{\Norm{\favre{\ubf_\xbf}}^2}{2}+\frac{2gh}{3}\right)\right)+\nabla_\xbf \cdot \left( h^2 \left( \frac{\Norm{\favre{\ubf_\xbf}}^2}{2}+gh\right)\favre{\ubf_\xbf}\right) 
        = -4\frac{a}{g}\widehat{\kappa} \Norm{\favre{\ubf_\xbf}}^2.
    \end{multline}

    \paragraph{The energy equation of the averaged model.}
    
    In this section, we derive the energy equation linked to the system~\eqref{eq:mass_free_surface}-\eqref{eq:momentum_free_surface}. Let us introduce $\ecal(h^2)$ a function that satisfies
    \begin{equation}
        \ecal'(h^2)h^4=\frac{gh^3}{3}.
    \label{eq:internal_energy_ecal}
    \end{equation}
    We multiply the averaged mass conservation equation~\eqref{eq:mass_free_surface} by $\ecal'(h^2)$ and it yields
    \begin{equation*}
        \partial_t (\ecal(h^2))+(\favre{\ubf_\xbf} \cdot \nabla_\xbf) \ecal(h^2) + \ecal'(h^2)h^2 \nabla_\xbf \cdot \favre{\ubf_\xbf}=0
    \end{equation*}
    
    \noindent Multiply the previous equation by $h^2$ and use the averaged mass conservation equation~\eqref{eq:mass_free_surface} to obtain 
    
    \begin{equation}
        \partial_t(h^2\ecal(h^2))+\nabla_\xbf \cdot (h^2\ecal(h^2)\favre{\ubf_\xbf}) + \ecal'(h^2)h^4 \nabla_\xbf \cdot \favre{\ubf_\xbf}=0.
    \label{eq:avg_equation_internal_energy}
    \end{equation}
    
    Then, we take the scalar product of the momentum equation~\eqref{eq:momentum_free_surface} with $\favre{\ubf_\xbf}$. Using the mass conservation equation~\eqref{eq:mass_free_surface},
    \begin{multline}
        \partial_t \left( h^2\frac{\Norm{\favre{\ubf_\xbf}}^2}{2}\right)+\nabla_\xbf \cdot \left( h^2\left(  \frac{\Norm{\favre{\ubf_\xbf}}^2}{2}+\frac{gh}{3}\right) \favre{\ubf_\xbf}\right)-\left( \frac{gh^3}{3}\right) \nabla_\xbf \cdot \favre{\ubf_\xbf}\\
        =-4\frac{a}{g} \widehat{\kappa}\Norm{\favre{\ubf_\xbf}}^2.
    \label{eq:avg_equation_kinetic_energy}
    \end{multline}
    
    Summing equations~\eqref{eq:avg_equation_internal_energy} and~\eqref{eq:avg_equation_kinetic_energy}, we get
    \begin{multline*}
        \partial_t \left( h^2\left(\frac{\Norm{\favre{\ubf_\xbf}}^2}{2}+\ecal(h^2)\right)\right)+\nabla_\xbf \cdot \left( h^2\left(  \frac{\Norm{\favre{\ubf_\xbf}}^2}{2}+\ecal(h^2)+\frac{gh}{3}\right) \favre{\ubf_\xbf}\right)\\
        +\left(\ecal'(h^2)h^4-\frac{gh^3}{3}\right) \nabla_\xbf \cdot \favre{\ubf_\xbf}=-4\frac{a}{g}\widehat{\kappa} \Norm{\favre{\ubf_\xbf}}^2.
    \end{multline*}
    
    We finally use \eqref{eq:internal_energy_ecal} to obtain the energy equation 
    \begin{multline}
        \partial_t \left( h^2\left(\frac{\Norm{\favre{\ubf_\xbf}}^2}{2}+\ecal(h^2)\right)\right)+\nabla_\xbf \cdot \left( h^2\left(  \frac{\Norm{\favre{\ubf_\xbf}}^2}{2}+\ecal(h^2)+\frac{gh}{3}\right) \favre{\ubf_\xbf}\right)\\
        =-4\frac{a}{g}\widehat{\kappa} \Norm{\favre{\ubf_\xbf}}^2.
    \end{multline}
    
    \begin{remark}
        The relation~\eqref{eq:internal_energy_ecal} gives us 
        \[ \ecal'(h^2)=\frac{g}{3h}=\frac{2g}{3} \frac{1}{2\sqrt{h^2}}.\]
        Therefore, up to a constant that we choose equal to $0$, 
        \[ \ecal(h^2)=\frac{2g}{3} \sqrt{h^2}=\frac{2gh}{3}.\]
        Therefore, we recover the same energy equation as~\eqref{eq:averaged_energy_free_surf_gamma2}.
    \hfill $\square$
\end{remark}

\subsection{General case}

    In this section, we consider a barotropy law of the form in \eqref{eq:barotropy_law} $p(\rho)=a\rho^\gamma$, with $a>0$ and $\gamma>1$. Note that the linear case $\gamma=1$ has been considered by various authors such as \cite{ersoy_dimension_2016}.
    The precise form of the barotropy law has no influence on the computations carried in sections 2 and 3. Let us see the impact it has when we close the system for the free surface configuration. 
    
    \subsubsection{Closing the system}

        We start from the integrated vertical equation \eqref{eq:GeneralDepthAveragedSystem2}.
    The density at the free surface still being zero at the main order due to condition~\eqref{eq:asymptotic_constraint_normal}, we have
    \begin{equation}
        \rho=\left( \frac{\gamma-1}{a\gamma} g(h-z)\right)^{\frac{1}{\gamma-1}}.
    \label{eq:density_free_surface}
    \end{equation}
    Following \eqref{eq:barotropy_law} the pressure law can be expressed as 
    \begin{equation}
        p(\rho)=a\left( \frac{\gamma-1}{a\gamma} g(h-z)\right)^{\frac{\gamma}{\gamma-1}}.
    \label{eq:pressure_free_surface}
    \end{equation}

    \noindent Averaging the previous expressions, we get, respectively
    \begin{equation}
        \bar{\rho}=\frac{1}{h}\int_0^h \left( \frac{\gamma-1}{a\gamma} g(h-z)\right)^{\frac{1}{\gamma-1}} \d z = \left( \frac{\gamma-1}{\gamma}\right)^{\frac{\gamma}{\gamma-1}} \left( \frac{gh}{a}\right)^{\frac{1}{\gamma-1}}
    \label{eq:avg_density_gamma}
    \end{equation}
    \begin{equation}
                \bar{p}= \frac{\gamma-1}{2\gamma-1} a \left(\frac{\gamma-1}{a\gamma}gh\right)^{\frac{\gamma}{\gamma-1}}
    \label{eq:avg_pressure_gamma}
    \end{equation}
    In particular, we have the following relation between averaged density and pressure
    \begin{equation}
        \bar{p}=  \frac{\gamma-1}{2\gamma-1} g h\ \bar\rho.
    \end{equation}

    \begin{remark}
        For more general barotropy laws, we introduce the function
        \begin{equation}
            F(\rho)=\int_0^\rho \frac{p'(s)}{s}\d s
        \label{eq:def_F_integral_form}
        \end{equation}
        This function satisfies the equation
        \[\partial_z F(\rho)=-g.\]
        Integrating the equation yields $F(\rho)=-gz+\varphi(t,\xbf)$, where $\varphi$ is a function independent of $z$. Computing the expression of $F$ thanks to its integral definition, we can express $\rho$ and $p$ in terms of $h$ and $\varphi$. Using the condition $p(\rho_S)=0$ at the free surface, we can determine $\varphi$ and then compute the averages of $\rho$ and $p$.\hfill $\square$
    \hfill $\square$
\end{remark}
    
    We then obtain the following averaged model with unknowns $h$ and $\favre{\ubf_\xbf}$:
    
    \begin{subequations}
        \begin{equation}
            \partial_t (h^{\frac{\gamma}{\gamma-1}})+\nabla_\xbf \cdot (h^{\frac{\gamma}{\gamma-1}} \favre{\ubf_\xbf})=0,
        \label{eq:mass_free_surface_gamma}\\
        \end{equation}
        \begin{multline}
            \partial_t (h^{\frac{\gamma}{\gamma-1}}\favre{\ubf_\xbf}) +\nabla_\xbf \cdot (h^{\frac{\gamma}{\gamma-1}} \favre{\ubf_\xbf} \otimes \favre{\ubf_\xbf}) + \nabla_\xbf \left( \frac{\gamma-1}{2\gamma-1}gh^{\frac{2\gamma-1}{\gamma-1}}\right)\\
            =-\left( \frac{\gamma}{\gamma-1}\right)^{\frac{\gamma}{\gamma-1}}\left( \frac{a}{g}\right)^{\frac{1}{\gamma-1}} \widehat{\kappa} \favre{\ubf_\xbf}.
        \label{eq:momentum_free_surface_gamma}
        \end{multline}
    \label{eq:avg_free_surface_gamma}
    \end{subequations}
    
    \subsubsection{The averaged energy of the compressible model}
    
    We consider the general equation \eqref{eq:energyNS_general} with $p_S=0$, and we calculate the average of the internal energy and the gravity term. With a general barotropy law, $e(\rho)=\frac{p}{(\gamma-1)\rho}=\frac{a\rho^{\gamma-1}}{\gamma-1}$ and its Favre average can be computed using~\eqref{eq:avg_density_gamma} and~\eqref{eq:avg_pressure_gamma}:
    \begin{equation*}
        \favre{e(\rho)}=\frac{\bar{\rho e(\rho)}}{\bar{\rho}}=\frac{1}{\gamma-1} \frac{\bar{p}}{\bar{\rho}}=\frac{1}{2\gamma-1}gh.
    \end{equation*}

    \noindent Moreover, this time the average of the gravity contribution using \eqref{eq:density_free_surface} reads 
    \[ \bar{\rho g z}
        =\frac{\gamma-1}{2\gamma-1}\left(\frac{1}{a}\right) ^\frac{1}{\gamma-1}\left(\frac{\gamma-1}{\gamma}gh\right)^\frac{\gamma}{\gamma-1} \]
    
    \noindent Therefore, its Favre average is
    \[\favre{g z}=\frac{\bar{\rho g z}}{\bar{\rho}}=\frac{\gamma-1}{2\gamma-1}gh.\]

    We embed these values together with $p_S=0$ in \eqref{eq:energyNS_general} to get the averaged energy equation,
    \begin{multline}\label{eq:averaged_energy_free_surf_gamma_gral}
        \partial_t \left( h^{\frac{\gamma}{\gamma-1}} \left( \frac{\Norm{\favre{\ubf_\xbf}}^2}{2}+\frac{\gamma}{2\gamma-1}gh\right)\right)
        +\nabla_\xbf \cdot \left( h^{\frac{\gamma}{\gamma-1}} \left( \frac{\Norm{\favre{\ubf_\xbf}}^2}{2}+gh\right)\favre{\ubf_\xbf}\right) \\
        = -\left( \frac{\gamma}{\gamma-1}\right)^{\frac{\gamma}{\gamma-1}}\left( \frac{a}{g}\right)^{\frac{1}{\gamma-1}} \widehat{\kappa} \Norm{\favre{\ubf_\xbf}}^2.
    \end{multline}

\subsubsection{The energy equation of the averaged model}
    
    Let us derive the energy equation linked to the system~\eqref{eq:mass_free_surface_gamma}-\eqref{eq:momentum_free_surface_gamma}. In the general case, we consider $\ecal(h^{\frac{\gamma}{\gamma-1}})$ a function that satisfies
    \begin{equation}
        \ecal'(h^{\frac{\gamma}{\gamma-1}})h^{\frac{2\gamma}{\gamma-1}}=\frac{\gamma-1}{2\gamma-1}gh^{\frac{2\gamma-1}{\gamma-1}}.
    \label{eq:internal_energy_ecal_gamma}
    \end{equation}
    We first multiply the averaged mass conservation equation~\eqref{eq:mass_free_surface_gamma} by $\ecal'(h^{\frac{\gamma}{\gamma-1}})$ to get
    \[\partial_t (\ecal(h^{\frac{\gamma}{\gamma-1}}))+\favre{\ubf_\xbf}\cdot \nabla_{\xbf} (\ecal(h^{\frac{\gamma}{\gamma-1}})) + \ecal'(h^{\frac{\gamma}{\gamma-1}})h^{\frac{\gamma}{\gamma-1}} \nabla_{\xbf}\cdot \favre{\ubf_\xbf}=0.\]
    Then, we multiply this equation by $h^{\frac{\gamma}{\gamma-1}}$ and use again \eqref{eq:mass_free_surface_gamma} to obtain
   \begin{equation}
        \partial_t(h^{\frac{\gamma}{\gamma-1}}\ecal(h^{\frac{\gamma}{\gamma-1}}))+\nabla_\xbf \cdot (h^{\frac{\gamma}{\gamma-1}}\ecal(h^{\frac{\gamma}{\gamma-1}})\favre{\ubf_\xbf}) + \ecal'(h^{\frac{\gamma}{\gamma-1}})h^{\frac{2\gamma}{\gamma-1}} \nabla_\xbf \cdot \favre{\ubf_\xbf}=0.
    \label{eq:avg_equation_internal_energy_gamma}
    \end{equation}
    
    Then, take the scalar product of the momentum equation~\eqref{eq:momentum_free_surface_gamma} with $\favre{\ubf_\xbf}$. Using the mass conservation equation~\eqref{eq:mass_free_surface_gamma},
    \begin{multline}
        \partial_t \left( h^{\frac{\gamma}{\gamma-1}}\frac{\Norm{\favre{\ubf_\xbf}}^2}{2}\right)+\nabla_\xbf \cdot \left(h^{\frac{\gamma}{\gamma-1}}\left(   \frac{\Norm{\favre{\ubf_\xbf}}^2}{2}+\frac{\gamma-1}{2\gamma-1}gh\right) \favre{\ubf_\xbf}\right)\\
        -\left(\frac{\gamma-1}{2\gamma-1}gh^{\frac{2\gamma-1}{\gamma-1}}\right) \nabla_\xbf \cdot \favre{\ubf_\xbf}
        =-\left( \frac{\gamma}{\gamma-1}\right)^{\frac{\gamma}{\gamma-1}}\left( \frac{a}{g}\right)^{\frac{1}{\gamma-1}} \widehat{\kappa} \Norm{\favre{\ubf_\xbf}}^2.
    \label{eq:avg_equation_kinetic_energy_gamma}
    \end{multline}
    
    Summing the equations~\eqref{eq:avg_equation_internal_energy_gamma},~\eqref{eq:avg_equation_kinetic_energy_gamma} and taking into account \eqref{eq:internal_energy_ecal_gamma}, we get

    \begin{multline}
        \partial_t \left( h^{\frac{\gamma}{\gamma-1}}\left(\frac{\Norm{\favre{\ubf_\xbf}}^2}{2}+\ecal(h^{\frac{\gamma}{\gamma-1}})\right)\right)\\
        +\nabla_\xbf \cdot \left(h^{\frac{\gamma}{\gamma-1}}\left(   \frac{\Norm{\favre{\ubf_\xbf}}^2}{2}+\ecal(h^{\frac{\gamma}{\gamma-1}})+\frac{\gamma-1}{2\gamma-1}gh\right) \favre{\ubf_\xbf}\right)\\
        =-\left( \frac{\gamma}{\gamma-1}\right)^{\frac{\gamma}{\gamma-1}}\left( \frac{a}{g}\right)^{\frac{1}{\gamma-1}} \widehat{\kappa} \Norm{\favre{\ubf_\xbf}}^2.
    \end{multline}
    
    \begin{remark}
        The relation~\eqref{eq:internal_energy_ecal_gamma} gives us 
        \[ \ecal'(h^{\frac{\gamma}{\gamma-1}})=\frac{\gamma-1}{2\gamma-1}gh^{-\frac{1}{\gamma-1}}=\frac{\gamma-1}{2\gamma-1}g\left(h^{\frac{\gamma}{\gamma-1}}\right)^{\frac{-1}{\gamma}}.\]
        Therefore, up to a constant, 
        \[ \ecal(h^{\frac{\gamma}{\gamma-1}})=\frac{\gamma}{2\gamma-1}g\left(h^{\frac{\gamma}{\gamma-1}}\right)^{\frac{\gamma-1}{\gamma}}=\frac{\gamma}{2\gamma-1}gh.\]
        We thus recover the expression of the averaged energy equation in \eqref{eq:averaged_energy_free_surf_gamma_gral}. 
    \hfill $\square$
\end{remark}
    
    \subsubsection{Some basic properties of the  averaged  model}
    
    In this section, we show that the averaged model is strictly hyperbolic and symetrizable. The model being invariant by Galilean transformation and frame rotation, we make the analysis for a the system in one space dimension. We consider the averaged model obtained for free surface compressible flow with a general barotropic law~\eqref{eq:avg_free_surface_gamma}. Therefore, to lighten the notations, we denote $u$ the averaged velocity $\favre{\ubf_\xbf}$ and the model reads
    $$
    \begin{array}{l}
    \partial_t (h^\frac{\gamma}{\gamma-1}) + \partial_x(h^\frac{\gamma}{\gamma-1}u)=0  \\
    \partial_t u + u\partial_x u +g\partial_x h = -\left( \frac{\gamma}{\gamma-1} \right)^\frac{\gamma}{\gamma-1} \left(\frac{a}{g}\right)^{\frac{1}{\gamma-1}}\frac{1}{h^\frac{\gamma}{\gamma-1}} \widehat{\kappa} u
    \end{array}
    $$
Denoting $H=h^\frac{\gamma}{\gamma-1}$, and $U=\begin{pmatrix}
        H & u
    \end{pmatrix}^{\rm T}$, we can write the model as follows:
    \begin{equation}
        \partial_t U + A(U) \partial_x U +B(U)=0,
    \label{eq:free_surface_gamma_matrix}
    \end{equation}
    \begin{flalign}
        \text{where} && \left\{
        \begin{array}{rcl}
        A(U)&=&\begin{pmatrix}
            u & H\\
            \frac{\gamma-1}{\gamma}g\frac{1}{H^{1/\gamma}} & u
        \end{pmatrix},\\
        B(U)&=&\begin{pmatrix}
            0,& \left( \frac{\gamma}{\gamma-1} \right)^\frac{\gamma}{\gamma-1} \left(\frac{a}{g}\right)^{\frac{1}{\gamma-1}}\frac{\widehat{\kappa}u}{H}
        \end{pmatrix}^{\rm T}.
        \end{array}\right. &&
    \end{flalign}

    \noindent The characteristic polynomial of $A(U)$ is 
    \[\chi_A=(X-u)^2-\frac{\gamma-1}{\gamma}g H^{\frac{\gamma-1}{\gamma}}=(X-u)^2-\frac{\gamma-1}{\gamma}gh = (X-(u-c))(X-(u+c))\]
    with $c= \sqrt{\frac{\gamma-1}{\gamma}gh}$.
    The system~\eqref{eq:free_surface_gamma_matrix} is therefore strictly hyperbolic. 
    
    Next, we introduce the following symmetric positive definite matrix:
    \begin{equation}
        S(U)=\begin{pmatrix}
            \frac{\gamma-1}{\gamma}g & 0\\
            0 & H^{1+1/\gamma}
        \end{pmatrix}.
    \end{equation}

    \noindent This matrix is positive definite since $H>0$. Moreover, since 
    \[ S(U)A(U)=\begin{pmatrix}
        \frac{\gamma-1}{\gamma}g u & \frac{\gamma-1}{\gamma}gH\\
        \frac{\gamma-1}{\gamma}gH & u H^{1+1/\gamma}
    \end{pmatrix}\]
    is symmetric, the system~\eqref{eq:avg_free_surface_gamma} is symmetrizable. Therefore, Kato's theorem provides us with the existence and uniqueness of smooth, local-in-time solutions to the Cauchy problem for the averaged model with smooth initial conditions.

    \subsection[The limit gamma to infinity]{The limit $\gamma \to \infty$}

    In this section, we prove the convergence of solutions of the compressible free surface model~\eqref{eq:avg_free_surface_gamma} to solutions of the incompressible shallow water equations when $\gamma \to \infty$. Such asymptotics have been studied in~\cite{lions_free_1999} to prove the convergence of a compressible free-boundary flow to a congestion model. In our case, we use a relative entropy technique to prove the convergence and its associated speed with respect to the parameter $\gamma$. 

    \noindent Denoting
    \[ \left\{ \begin{array}{rcl}
        r &=& h^{\frac{\gamma}{\gamma-1}},\\
        p_\gamma(r)&=& \frac{\gamma-1}{2\gamma-1}gr^{2-\frac{1}{\gamma}}\\
        \kappa_\gamma &=& \left( \frac{\gamma}{\gamma-1}\right)^{\frac{\gamma}{\gamma-1}}\left( \frac{a}{g}\right)^{\frac{1}{\gamma-1}} \widehat{\kappa},
    \end{array}\right.\]
    the averaged compressible model reads

    \begin{subequations}
        \begin{align}
            \partial_t r+\nabla_\xbf \cdot (r \ubf)&=0,\label{eq:AvgFreeSurf_mass}\\
            \partial_t (r \ubf) +\nabla_\xbf \cdot (r \ubf \otimes \ubf) + \nabla_\xbf p_\gamma 
            &=-\kappa_\gamma \ubf. \label{eq:AvgFreeSurf_momentum}
        \end{align}
        \label{eq:AvgFreeSurf}
    \end{subequations}

    Let us take the formal limit $\gamma \to \infty$ in the previous system. Denoting $h$ the limit of $r$ and $\U$ the limit of $\ubf$, we recover the shallow-water equations:
    \begin{subequations}
        \begin{align}
            \partial_t h + \nabla_\xbf \cdot (h \U)&=0,
        \label{eq:incompressible_mass}\\
            \partial_t (h \U) + \nabla_\xbf \cdot (h \U \otimes \U) + \nabla_\xbf \left(g\frac{h^2}{2}\right) &= -\widehat{\kappa} \U.
        \label{eq:incompressible_momentum}
        \end{align}
    \label{eq:incompressible_shallow_water}
    \end{subequations}
    Note that the limit quantities $h$ and $\U$ correspond to the usual quantities studied in the shallow-water equations, that is the fluid height and its horizontal velocity.
    
    In this section, we focus on studying the convergence of strong solutions of the compressible free-surface system~\eqref{eq:AvgFreeSurf} to strong solutions of the shallow-water model~\eqref{eq:incompressible_shallow_water} when $\gamma$ tends to $+\infty$. A result establishing the existence of globally strong solutions in time for the shallow-water equations is given in~\cite{yong_entropy_2004, lin_strong_2013}. These results also apply to the averaged compressible model~\eqref{eq:AvgFreeSurf} for all $\gamma>1$, but it is not certain that the estimate used to obtain the result is uniform with respect to $\gamma$. We therefore assume the existence of such solutions on a time interval $[0,T]$ with $0<T<+\infty$ independent from $\gamma$.
    However we mention recent results on such system (see for instance \cite{CrDa} and references cited therein) that could provide such results. 
    
    To perform a rate of convergence, we use a relative entropy method which we introduce just below.

    \subsubsection{Relative entropy: definition and convergence result}

    Let us start by introducing an entropy function $\eta$ for the compressible free-surface model~\eqref{eq:AvgFreeSurf}. Denote $(r,\mbf=r\ubf)$ the conserved quantities of the system. We define $\eta$ as 
    \begin{equation}
        \eta(r,\mbf)=\frac{|\mbf|^2}{2r}+H_\gamma(r),
    \label{eq:eta_energy_functional}
    \end{equation}
    where $H_\gamma(r)$ is defined by
    \begin{equation}
        H_\gamma(r)=r\int \frac{p(z)}{z^2} \d z = \frac{\gamma}{2\gamma-1}g r^{2-\frac{1}{\gamma}}.
    \label{eq:potential_energy_functional}
    \end{equation}

    \begin{remark}
        Since we work in a barotropic setting, the energy is an entropy. That is why $\eta$ corresponds to the energy of the system~\eqref{eq:AvgFreeSurf}.
    \hfill $\square$
\end{remark}

    \noindent The function $\eta$ is strictly convex and satisfies the following equation: 
    \begin{equation}
        \partial_t \eta(r,\mbf) + \nabla_\xbf \cdot \left(\left(\eta(r,\mbf)+p_\gamma(r)\right)\ubf \right)=-\kappa_\gamma|\ubf|^2.
    \label{eq:entropy_equation}
    \end{equation}
    In other words, $(\eta,(\eta+p_\gamma)\ubf)$ is an entropy couple in the sense of Lax~\cite{lax_hyperbolic_1973} for the system~\eqref{eq:AvgFreeSurf}.

    Now, we introduce the relative entropy associated with $\eta$. It corresponds to the linearization of $\eta$ in a neighborhood of a solution to the shallow-water equations~\eqref{eq:incompressible_shallow_water} and enables us to compare strong solutions of both models. It reads as follows
    \begin{equation}
        E_\gamma(r,\ubf|h,\U)=\frac{r}{2} |\ubf-\U|^2+H_\gamma(r|h),
    \label{eq:relative_entropy_explicite}
    \end{equation}
    \begin{flalign}
        \text{where} && H_\gamma(r|h)&=H_\gamma(r)-H_\gamma(h)-H_\gamma'(h)(r-h).&&
    \label{eq:relative_potential_energy}
    \end{flalign}

    In order to set the relative entropy method, let us rewrite the shallow-water equations~\eqref{eq:incompressible_shallow_water} in a more suitable way:
    \begin{subequations}
        \begin{align}
            \partial_t h + \nabla_\xbf \cdot (h\U)&=0,\\
            \partial_t (h\U) + \nabla_\xbf \cdot (h\U \otimes \U) + \nabla_\xbf p_\gamma(h) &= \nabla_\xbf \pi(h)-\widehat{\kappa} \U,
        \end{align}
        \label{eq:shallow_water_pi}
    \end{subequations}
    where $\pi(h)=p_\gamma(h)-gh^2/2$. We thus compare two systems with similar left-hand sides.

    Let $(r,\ubf$ be a strong solution of the compressible free-surface model~\eqref{eq:AvgFreeSurf} and $(h,\U)$ be a strong solution of~\eqref{eq:shallow_water_pi}. The relative entropy $E_\gamma$ satisfies the following equation:
    \begin{multline}
        \partial_t E_\gamma(r,\ubf|h,\U) + \nabla_\xbf \cdot (E_\gamma(r,\ubf|h,\U)\ubf+(p_\gamma(r)-p_\gamma(h))(\ubf-\U)) \\
        =-r((\ubf-\U)\cdot \nabla_\xbf) \U \cdot (\ubf-\U) -(p_\gamma(r)-p_\gamma(h)-(r-h)p_\gamma'(h))\nabla_\xbf \cdot \U\\
        -\kappa_\gamma \ubf\cdot(\ubf-\U)-\widehat{\kappa} \frac{r}{h} \U\cdot(\ubf-\U)- \frac{r}{h} \nabla_\xbf \pi(h) \cdot (\ubf-\U).
    \label{eq:relative_entropy_equation}
    \end{multline}

    Finally, let us introduce the quantity
    \begin{equation}
        \ecal_\gamma(r,\ubf|h,\U)=\int_{\T^2} E_\gamma(r,\ubf|h,\U) \d \xbf. 
    \label{eq:relative_entropy_integral}
    \end{equation}
    This new quantity enables us to compare the solutions to both models. Indeed, we have the following proposition~\cite{sueur_inviscid_2014}:
    \begin{proposition}\label{swprop.equiv_Hgamma_norm}
        Let $h$ be a smooth bounded function. 
        
        \noindent There exist two positive constants $C_1$, $C_2>0$ independent of $\gamma$ such that for all nonnegative functions $r$ and all $\gamma \geqslant \ulin{\gamma}>1$,
        \begin{equation}
            C_1\left(|r-h|_{r\leqslant 2h}^2+|r-h|_{r>2h}^{2-\frac{1}{\gamma}}\right) \leqslant H(r|h) \leqslant C_2|r-h|^2.
        \label{eq:equivalence_potential_norm}
        \end{equation}
    \end{proposition}

    Controlling $\ecal_\gamma(r,\ubf|h,\U)$ is then equivalent to controlling the distance between $r$ and $h$ in a $L^p(\T^2)$ space, with $1<p<2$, and the $L^2(\T^2)$ norm of $\sqrt{r}(\ubf-\U)$. More precisely, we have the following result:
    \begin{theorem}\label{thm:convergence_shallow_water}
        Let $T \in \R_+^*$ be independent of $\gamma$. Let $(h,\U)$ be the strong solution to the shallow-water equations~\eqref{eq:shallow_water_pi} associated with smooth initial conditions $(h_0,\U_0)$ such that
        \begin{flalign}
            \forall (t,\xbf) \in [0,T] \times \T^2, && 0 < \check{h} \leqslant h(t,\xbf) &\leqslant \hat{h},&&
        \label{eq:boundedness_h_sv}
        \end{flalign}
        where $\check{h}$, $\hat{h}$ are two positive constants independent of $\gamma$.

        Let $(r,\ubf)$ be the strong solution of the compressible free-surface model~\eqref{eq:AvgFreeSurf} on $[0,T] \times \T^2$ associated with the smooth initial conditions $(r_0,\ubf_0)$.

        Then, there exist $\ulin{\gamma}>1$ and a constant $C>0$ independent of $\gamma$ such that for all $\tau \in [0,T]$ and all $\gamma >\ulin{\gamma}$, we have
        \begin{equation}
            \ecal_\gamma(r,\ubf|h,\U)(\tau) \leqslant \ecal_\gamma(r_0,\ubf_0|h_0,\U_0) \exp(C\tau) + \ocal\left(\frac{1}{\gamma^2}\right).
        \label{eq:entropy_estimate}
        \end{equation}

        \noindent In particular, if we assume that
        \begin{equation}
            \Norm{r_0-h_0}_{L^2(\T^2)} + \Norm{\sqrt{r_0}(\ubf_0-\U_0)}_{L^2(\T^2)} \limit{\gamma}{+\infty} 0,
        \label{eq:convergence_initial_conditions}
        \end{equation}
        then
        \begin{equation}
            \sup_{\tau \in [0,T]} \Norm{r-h}_{L^1(\T^2)}(\tau) + \Norm{\sqrt{r}(\ubf-\U)}_{L^2(\T^2)}(\tau) \limit{\gamma}{\infty} 0.
        \label{eq:convergence_euler_to_sv}
        \end{equation}
        Moreover, if we fix a convergence rate at $0$ of the order $\ocal\left({1}/{\gamma^2}\right)$ for the initial data in 
        \eqref{eq:convergence_initial_conditions}, then we get the same convergence order in \eqref{eq:convergence_euler_to_sv}.
    \end{theorem}

    \begin{remark}
        As detailed in~\cite{sueur_inviscid_2014}, convergence in the sense of~\eqref{eq:convergence_euler_to_sv} implies that
        \begin{equation}
            \sup_{\tau \in [0,T]} \Norm{r\ubf-h\U}_{L^1(\T^2)} \limit{\gamma}{+\infty} 0.
        \end{equation}
        \begin{flalign*}
            \text{Indeed,} && r\ubf-h\U &= (r-h)\U +\sqrt{r} \sqrt{r}(\ubf-\U), &&
        \end{flalign*}
        and thanks to a triangle inequality and the boundedness of $\U$ on $[0,T] \times \T^2$, we obtain for all $\tau \in [0,T]$ that
        \begin{equation*}
            \Norm{r\ubf-h\U}_{L^1(\T^2)}(\tau) \leqslant \Norm{\U}_{L^\infty([0,T] \times \T^2)} \Norm{r-h}_{L^1(\T^2)}(\tau)  + \Norm{\sqrt{r}\sqrt{r}(\ubf-\U)}_{L^1(\T^2)}(\tau).
        \end{equation*}
        Thanks to Young's inequality on the second term of the right-hand side of the inequality, we have for all $\tau \in [0,T]$:
        \begin{equation*}
            \Norm{r\ubf-h\U}_{L^1(\T^2)} (\tau)\leqslant C\left( \Norm{r-h}_{L^1(\T^2)} (\tau) + \Norm{\sqrt{r}}_{L^2(\T^2)}(\tau)\Norm{\sqrt{r}(\ubf-\U)}_{L^2(\T^2)}(\tau)\right),
        \end{equation*}
        where $C>0$ is independent of $\gamma$.
    \hfill $\square$
\end{remark}

    The rest of this section is dedicated to proving~\cref{thm:convergence_shallow_water}.

\subsubsection{Proving the convergence}

    The idea of the proof is to exhibit an inequality satisfied by $\ecal_\gamma$ that we can apply Grönwall's lemma on to recover~\eqref{eq:entropy_estimate}. The rest of the theorem is deduced from this estimate. 

    Let us start by integrating the relative entropy equation~\eqref{eq:relative_entropy_equation} on $[0,\tau] \times \T^2$ for all $\tau \in [0,T]$. It yields 
    \begin{equation}
        \ecal_\gamma(r,\ubf|h,\U)(\tau) = \ecal_\gamma(r,\ubf|h,\U)(0) + \int_0^\tau \rcal_\gamma(r,\ubf|h,\U)(t) \d t,
    \label{eq:relative_entropy_inequality}
    \end{equation}
    where the remainder term $\rcal_\gamma(r,\ubf|h,\U)$ is defined by
    \begin{multline}
        \rcal_\gamma(r,\ubf|h,\U) = -\int_{\T^2} r((\ubf-\U)\cdot \nabla_\xbf) \U \cdot (\ubf-\U) \d \xbf - \int_{\T^2} \frac{r}{h} \nabla_\xbf \pi(h)  \cdot (\ubf-\U) \d \xbf\\
        -\int_{\T^2} (p_\gamma(r)-p_\gamma(h)-(r-h)p_\gamma'(h))\nabla_\xbf \cdot \U \d \xbf \\
        -\kappa_\gamma \int_{\T^2} \ubf\cdot(\ubf-\U) \d \xbf +\widehat{\kappa} \int_{\T^2} \frac{r}{h} \U\cdot(\ubf-\U) \d \xbf .
    \label{eq:def_remainder}
    \end{multline}
    The goal here is thus to control this remainder term by $\ecal_\gamma$ and some other terms that tend to $0$ as $\gamma \to +\infty$. More precisely, we prove the following proposition:
    \begin{proposition}\label{swprop.gronwall_relative_entropy}
        Under the assumptions of~\cref{thm:convergence_shallow_water}, there exists a constant $C>0$ independent from $\gamma$ such that for all $\tau \in [0,T]$, 
        \begin{equation}
            \ecal_\gamma(r,\ubf|h,\U)(\tau) \leqslant \ecal_\gamma(r,\ubf|h,\U)(0) + C \int_0^\tau \ecal_\gamma(r,\ubf|h,\U)(t) \d t +\ocal\left(\frac{1}{\gamma^2}\right).
        \label{eq:gronwall_relative_entropy}
        \end{equation}
    \end{proposition}

        \begin{proof}
        The remainder $\rcal_\gamma(r,\ubf|h,\U)$ is composed of four terms that will be estimated separately:
        \begin{itemize}
            \item a convection term:
            \begin{equation*}
                \rcal_1 = -\int_{\T^2} r((\ubf-\U)\cdot \nabla_\xbf) \U \cdot (\ubf-\U) \d \xbf;
            \end{equation*}
            \item the linearized pressure on a neighbourhood of $h$: 
            \begin{equation*}
                \rcal_2 = -\int_{\T^2} (p_\gamma(r)-p_\gamma(h)-(r-h)p_\gamma'(h))\nabla_\xbf \cdot \U \d \xbf;
            \end{equation*}
            \item the friction term:
            \begin{equation*}
                \rcal_3 = -\kappa_\gamma \int_{\T^2} \ubf\cdot(\ubf-\U) \d \xbf +\widehat{\kappa} \int_{\T^2} \frac{r}{h} \U\cdot(\ubf-\U) \d \xbf;
            \end{equation*}
            \item the difference between the pressures of both models:
            \begin{equation*}
                \rcal_4 = - \int_{\T^2} \frac{r}{h} \nabla_\xbf \pi(h) \cdot (\ubf-\U)\d \xbf.
            \end{equation*}
        \end{itemize}

        In the following, we estimate every one of these terms. We denote by $C$ any positive constant independent from $\gamma$.

        \noindent \paragraph{Estimate of $\rcal_1$ (convection).}  Since $\U$ is smooth on $[0,T]\times \T^2$, 
        \begin{equation}
            |\rcal_{1}| \leqslant \Norm{\nabla_\xbf \U}_{L^\infty([0,T]\times \T^2)} \int_{\T^2} r |\ubf -\U|^2 \d \xbf \leqslant C \ecal_\gamma(r,\ubf|h,\U).
        \label{eq:estimate_r1}
        \end{equation}

        \noindent \paragraph{Estimate of $\rcal_2$ (linearized pressure).}
        We use the fact that for all $z>0$, $H_\gamma(z)=\frac{\gamma}{\gamma-1}p_\gamma(z)$. Therefore,
        \begin{align*}
            p_\gamma(r)-p_\gamma(h)-(r-h)p_\gamma'(h) &= \frac{\gamma-1}{\gamma} (H_\gamma(r)-H_\gamma(h)-(r-h)H_\gamma'(h)),\\
            &= \frac{\gamma-1}{\gamma} H_\gamma(r|h).
        \end{align*}
        Since $\frac{\gamma-1}{\gamma}\leqslant 1$ and since $\U$ is smooth, we have
        \begin{equation}
            |\rcal_2| \leqslant \Norm{\nabla_\xbf \cdot \U}_{L^\infty([0,T]\times \T^2)} \int_{\T^2} H_\gamma(r|h) \d \xbf \leqslant C \ecal_\gamma(r,\ubf|h,\U).
        \label{eq:estimate_r2}
        \end{equation}

        \noindent \paragraph{Estimate of $\rcal_3$ (friction).} First, we rewrite $\rcal_3$ as
        \begin{equation*}
            \rcal_3 = -\kappa_\gamma \int_{\T^2} |\ubf-\U|^2 \d \xbf + (\widehat{\kappa}-\kappa_\gamma)\int_{\T^2} \U \cdot (\ubf-\U) \d \xbf
            + \widehat{\kappa} \int_{\T^2} \frac{r-h}{h}\U \cdot (\ubf-\U)\d \xbf.
        \end{equation*}

        \noindent The first term being negative, is will be used to compensate some other terms afterwards. Thanks to Young's inequality, the second term satisfies the following inequality:
        \[ \U \cdot (\ubf-\U) \leqslant \frac{1}{2\kappa_\gamma} |\U|^2 + \frac{\kappa_\gamma}{2}|\ubf-\U|^2.\]
        Then,
        \begin{equation}
            (\widehat{\kappa}-\kappa_\gamma)\int_{\T^2} \U \cdot (\ubf-\U) \d \xbf \leqslant \frac{|\widehat{\kappa}-\kappa_\gamma|^2}{2\kappa_\gamma} \int_{\T^2} |\U|^2 \d \xbf + \frac{\kappa_\gamma}{2} \int_{\T^2} |\ubf-\U|^2 \d \xbf.
        \label{eq:estimate_r3,2}
        \end{equation}
        Thanks to a series expansion applied to the definition of $\kappa_\gamma$,
        \begin{equation*}
            \widehat{\kappa}-\kappa_\gamma = \ocal\left(\frac{1}{\gamma}\right).
        \end{equation*}
        This term then goes to $0$ as $\gamma$ goes to $+\infty$. There remains to deal with the last term of $\rcal_3$. Denote $\mcal=\{ \xbf \in \T^2 \mid r\leqslant 2h \}$ and $\mcal^c=\T^2 \setminus \mcal$. 
        Write the integral as the sum of an integral over $\mcal$ and one over $\mcal^c$: 
        \begin{equation*}
            \int_{\T^2} \frac{r-h}{h}\U \cdot (\ubf-\U)\d \xbf =  \int_\mcal \frac{r-h}{h}\U \cdot (\ubf-\U)\d \xbf
            +\int_{\mcal^c} \frac{r-h}{h}\U \cdot (\ubf-\U)\d \xbf.
        \end{equation*}

        \noindent On $\mcal$, thanks to Young's inequality,
        \begin{equation*}
            \int_\mcal \frac{r-h}{h}\U \cdot (\ubf-\U)\d \xbf \leqslant \frac{1}{\delta} \int_\mcal (r-h)^2|\U|^2 \d \xbf + \delta \int_\mcal \frac{1}{h^2}|\ubf-\U|^2 \d \xbf.
        \end{equation*}
        where $\delta>0$ is a parameter that will be adjusted later. Since $h$ and $\U$ are bounded,
        \begin{equation*}
            \int_\mcal \frac{r-h}{h}\U \cdot (\ubf-\U)\d \xbf \leqslant C \left(\frac{1}{\delta} \int_\mcal (r-h)^2 \d \xbf + \delta \int_\mcal |\ubf-\U|^2 \d \xbf\right).
        \end{equation*}
        According to ~\eqref{eq:equivalence_potential_norm}, we can bound $(r-h)^2$ by $H_\gamma(r|h)$ on $\mcal$, independently from $\gamma$:
        \begin{equation*}
            (r - h)^2 \leqslant C_1^{-1} H_\gamma(r|h).
        \end{equation*} 
        \begin{flalign*}
            \text{Consequently,} && \int_\mcal (r-h)^2 \d \xbf &\leqslant C_1^{-1} \int_\mcal H_\gamma(r|h) \d \xbf \leqslant C \ecal_\gamma(r,\ubf|h,\U).&&
        \end{flalign*}
        
        \noindent On the set $\mcal^c$ ($r > 2h$), we use the following lemma from~\cite{bresch_navier-stokes-korteweg_2018}:
        \begin{lemma}\label{swlemma.didier_marguerite}
            If $r > 2h$, there exists $C > 0$ independent from $\gamma$ such that
            \begin{equation}
                H_\gamma(r|h) \geqslant C(1+r)^{2-\frac{1}{\gamma}}.
            \label{eq:lemma_relative_entropy}
            \end{equation}
        \end{lemma}

        \begin{proof}
            We start by writing
            \begin{equation*}
                |r-h|^{2-\frac{1}{\gamma}} = (1+r)^{2-\frac{1}{\gamma}} \left| 1-\frac{1+h}{1+r}\right|^{2-\frac{1}{\gamma}}.
            \end{equation*}
            On $\mcal^c$, since $x \mapsto x^{2-\frac{1}{\gamma}}$ is nondecreasing $\R_+$, we have
            \begin{equation*}
                \left| 1-\frac{1+h}{1+r}\right|^{2-\frac{1}{\gamma}} \geqslant \left| 1-\frac{1+h}{1+2h}\right|^{2-\frac{1}{\gamma}} = \left( \frac{h}{1+2h}\right)^{2-\frac{1}{\gamma}}.
            \end{equation*}
            Since $h/(1+2h) \in (0,1)$,
            \begin{equation*}
                |r-h|^{2-\frac{1}{\gamma}} \geqslant (1+r)^{2-\frac{1}{\gamma}}\left( \frac{h}{1+2h}\right)^{2},
            \end{equation*}
            and the boundedness of $h$ enables us to bound the factor $( h/(1+2h))^{2}$. We then inject the resulting inequality in~\eqref{eq:equivalence_potential_norm} to get the result.
        \end{proof}
        
        Let us go back to the integral on  $\mcal^c$. Using Young's inequality and the boundedness of $h$ and $\U$, we get
        \begin{align*}
            \int_{\mcal^c} \frac{r-h}{h}\U \cdot (\ubf-\U)\d \xbf &\leqslant \int_{\mcal^c} \frac{r-h}{\sqrt{r}}\frac{1}{h} \U\cdot (\sqrt{r}(\ubf-\U)) \d \xbf,\\
            &\leqslant C \left( \delta \int_{\mcal^c} \frac{(r-h)^2}{r} \d \xbf + \frac{1}{\delta} \int_{\mcal^c} r |\ubf-\U|^2 \d \xbf\right),\\
            & \leqslant C \left(\delta \int_{\mcal^c} \frac{(r-h)^2}{r} \d \xbf + \frac{1}{\delta} \ecal_\gamma(r,\ubf|h,\U)\right).
        \end{align*}
       The triangle inequality then yields
        \begin{equation*}
            \frac{(r-h)^2}{r} \leqslant (h+r)\left| \frac{h}{r}-1\right|.
        \end{equation*}
        Since $r>2h$ on $\mcal^c$, we have $|h/r-1|<1/2$. Consequently,
        \begin{equation*}
            \frac{(r-h)^2}{r}
             \leqslant C(1+r).
        \end{equation*}
        Since $1+r \geqslant 1$ and $2-\frac{1}{\gamma}>1$,
        \begin{equation*}
            (1+r) \leqslant (1+r)^{2-\frac{1}{\gamma}}.
        \end{equation*}
        We then use estimate~\eqref{eq:lemma_relative_entropy} of~\cref{swlemma.didier_marguerite} to obtain
        \begin{equation}
            \int_{\mcal^c} \frac{(r-h)^2}{r} \d \xbf \leqslant C \int_{\mcal^c} H_\gamma(r|h) \d \xbf \leqslant C \ecal_\gamma(r,\ubf|h,\U).
        \label{eq:majoration_complementaire_mcal}
        \end{equation}

        \noindent Finally, the last term of $\rcal_3$ is bounded from above as follows:
        \begin{equation}
            \widehat{\kappa}\int_{\T^2} \frac{r-h}{h}\U \cdot (\ubf-\U)\d \xbf \leqslant C\left(\delta + \frac{1}{\delta}\right) \ecal_\gamma(r,\ubf|h,\U) + C\delta \int_{\T^2} |\ubf-\U|^2 \d \xbf.
        \label{eq:estimate_r3,3}
        \end{equation}

        \noindent Gathering estimates~\eqref{eq:estimate_r3,2} and~\eqref{eq:estimate_r3,3} yields
        \begin{multline}
            \rcal_3 \leqslant C\left(\delta + \frac{1}{\delta}\right) \ecal_\gamma(r,\ubf|h,\U) +  (-\kappa_\gamma + \frac{\kappa_\gamma }{2}+C\delta) \int_{\T^2} |\ubf-\U|^2 \d \xbf \\
            + \frac{|\widehat{\kappa}-\kappa_\gamma|^2}{2\kappa_\gamma} \int_{\T^2} |\U|^2 \d \xbf.
        \label{eq:estimate_r3}
        \end{multline}

        \noindent \paragraph{Estimate of $\rcal_4$.} 
        We first write $\rcal_4$ as follows:
        \begin{equation*}
            \rcal_4 = \int_{\T^2} \frac{h-r}{h}\nabla_\xbf \pi(h) \cdot (\ubf -\U) \d \xbf - \int_{\T^2} \nabla_\xbf \pi(h) \cdot (\ubf -\U) \d \xbf.
        \end{equation*}

        \noindent The function $\pi(s)=\frac{\gamma-1}{2\gamma-1}gs^{2-\frac{1}{\gamma}}-\frac{gs^2}{2}$ is smooth and its derivative reads
        \begin{equation*}
            \pi'(s)=gs\left( \frac{\gamma-1}{\gamma}s^{-\frac{1}{\gamma}}-1\right).
        \end{equation*}
        Since $h$ is bounded, the Lipschitz constant $0<C_\pi(\gamma)$ of $\pi$ on $[\check{h}, \hat{h}]$ goes to zero when $\gamma$ goes to infinity. More precisely, thanks to a series expansion of $\pi'(s)$ when $\gamma$ is in a neighbourhood of infinity and $s$ is bounded, we can show that $C_\pi(\gamma)=\ocal(1/\gamma)$. Therefore, the second term composing $\rcal_4$ is bounded thanks to Young's inequality:
        \begin{equation*}
            \left|\int_{\T^2} \nabla_\xbf \pi(h) \cdot (\ubf-\U) \d \xbf\right| \leqslant \frac{1}{\delta}\left(\int_{\T^2} |\nabla_\xbf \pi(h)|^2 \d \xbf\right)+\delta \left(\int_{\T^2} |\ubf-\U|^2 \d \xbf\right).
        \end{equation*}
        Since $h$ is smooth, $|\nabla_\xbf \pi(h)|\leqslant C_\pi(\gamma) C$, hence
        \begin{equation}
            \left|\int_{\T^2} \nabla_\xbf \pi(h) \cdot (\ubf-\U) \d \xbf\right|\leqslant \frac{1}{\delta} C_\pi(\gamma)^2 +\delta \int_{\T^2} |\ubf-\U|^2 \d \xbf.
        \label{eq:estimate_r4,1}
        \end{equation}
        
        Let us decompose the first term of $\rcal_4$ according to the sets $\mcal$ and $\mcal^c$.
        \begin{multline*}
            \int_{\T^2} \frac{h-r}{h}\nabla_\xbf \pi(h) \cdot (\ubf -\U) \d \xbf =\int_\mcal \frac{h-r}{h}\nabla_\xbf \pi(h) \cdot (\ubf -\U) \d \xbf \\
            + \int_{\mcal^c} \frac{h-r}{h}\nabla_\xbf \pi(h) \cdot (\ubf -\U) \d \xbf.
        \end{multline*}
        On $\mcal$, we use Young's inequality and the boundedness of $h$ and $\nabla_\xbf \pi(h)$ to obtain
        \begin{equation*}
            \int_\mcal \frac{h-r}{h}\nabla_\xbf \pi(h) \cdot (\ubf -\U) \d \xbf \leqslant C_\pi(\gamma) C \left( \frac{1}{\delta} \int_\mcal (r-h)^2 \d \xbf + \delta \int_\mcal |\ubf -\U|^2 \d \xbf\right).
        \end{equation*}
        As we did for $\rcal_3$, we use~\eqref{eq:equivalence_potential_norm} to bound the first term of the right-hand side by $\ecal_\gamma$ (up to a multiplicative factor) and get
        \begin{equation}
            \int_\mcal \frac{h-r}{h}\nabla_\xbf \pi(h) \cdot (\ubf -\U) \d \xbf \leqslant C_\pi(\gamma) \frac{C}{\delta} \ecal_\gamma(r,\ubf|h,\U)+C_\pi(\gamma) C \delta \int_\mcal |\ubf -\U|^2 \d \xbf.
        \label{eq:estimate_r4,2}
        \end{equation}
        
        On $\mcal^c$, we use Young's inequality and the boundedness of $h$ and $\nabla_\xbf \pi(h)$ to get
        \begin{align*}
            \int_{\mcal^c} \frac{h-r}{h}\nabla_\xbf \pi(h) \cdot (\ubf -\U) \d \xbf &\leqslant \int_{\mcal^c} \frac{|h-r|}{\sqrt{r}} \frac{1}{h} \nabla_\xbf \pi(h) \cdot (\sqrt{r}(\ubf -\U)) \d \xbf,\\
            &\leqslant C_\pi(\gamma) C \left( \delta \int_{\mcal^c} \frac{(h-r)^2}{r} \d \xbf + \frac{1}{\delta} \int_{\mcal^c} r |\ubf -\U|^2 \d \xbf\right),\\ 
            & \leqslant C_\pi(\gamma) C \left(\delta \int_{\mcal^c} \frac{(h-r)^2}{r} \d \xbf + \frac{1}{\delta} \ecal_\gamma(r,\ubf|h,\U)\right).
        \end{align*}
        The remaining term composing $\rcal_4$ has already been estimated when we studied $\rcal_3$ (see~\cref{eq:majoration_complementaire_mcal}). This term is controlled by $\ecal_\gamma(r,\ubf|h,\U)$ up to a factor $C$. Consequently,
        \begin{equation}
            \int_{\mcal^c} \frac{h-r}{h}\nabla_\xbf \pi(h) \cdot (\ubf -\U) \d \xbf \leqslant C_\pi(\gamma) C \left(\delta + \frac{1}{\delta}\right)\ecal_\gamma(r,\ubf|h,\U) .
        \label{eq:estimate_r4,3}
        \end{equation}
        Gathering estimates~\eqref{eq:estimate_r4,1},~\eqref{eq:estimate_r4,2} et~\eqref{eq:estimate_r4,3}, we have the following estimate for $\rcal_4$:
        \begin{equation}
            \rcal_4 \leqslant C_\pi(\gamma)C\left(\delta + \frac{1}{\delta}\right) \ecal_\gamma(r,\ubf|h,\U) + C(\delta+C_\pi(\gamma)\delta) \int_{\T^2} |\ubf-\U|^2 \d \xbf+\frac{1}{\delta} C_\pi(\gamma)^2.
        \label{eq:estimate_r4}
        \end{equation}

        Now that we have estimated every partial remainder, we assemble the different estimates~\eqref{eq:estimate_r1},~\eqref{eq:estimate_r2},~\eqref{eq:estimate_r3} and~\eqref{eq:estimate_r4}. It yields
        \begin{multline}
            \rcal_\gamma(r,\ubf|h,\U) \leqslant C\left(1+\left(\delta + \frac{1}{\delta}\right)(1+C_\pi(\gamma))\right) \ecal_\gamma(r,\ubf|h,\U) \\
            + \left(-\kappa_\gamma + \frac{\kappa_\gamma}{2}+C(2\delta + C_\pi(\gamma)\delta) \right) \int_{\T^2} |\ubf-\U|^2 \d \xbf\\
             + \frac{|\kappa_\gamma-\widehat{\kappa}|^2}{2\kappa_\gamma} \int_{\T^2} |\U|^2 \d \xbf+\frac{1}{\delta}C_\pi(\gamma)^2.
        \label{eq:estimate_remainder}
        \end{multline}

        \noindent Since $C_\pi(\gamma)$ tends to $0$ when $\gamma$ tends to $+\infty$, there exists $\ulin{\gamma}>1$ such that we can choose $\delta>0$ independent from $\gamma$ that satisfies
        \[ C(2\delta + C_\pi(\ulin{\gamma})\delta)\leqslant \frac{\widehat{\kappa}}{4}.\]
        Therefore, for all $\gamma \geqslant \ulin{\gamma}$,
        \begin{equation*}
            \rcal_\gamma(r,\ubf|h,\U) \leqslant C \ecal_\gamma(r,\ubf|h,\U) + \frac{|\widehat{\kappa}-\kappa_\gamma|^2}{2\kappa_\gamma}\int_{\T^2} |\U|^2 \d \xbf+\frac{1}{\delta}C_\pi(\gamma)^2.
        \end{equation*}

        \noindent By applying this bound in~\eqref{eq:relative_entropy_inequality}, we recover inequality~\eqref{eq:gronwall_relative_entropy} of~\cref{swprop.gronwall_relative_entropy}: 
        \begin{equation*}
            \ecal_\gamma(r,\ubf|h,\U)(\tau) \leqslant \ecal_\gamma(r,\ubf|h,\U)(0) + C \int_0^\tau \ecal_\gamma(r,\ubf|h,\U)(t) \d t +\ocal\left(\frac{1}{\gamma^2}\right).
        \end{equation*}
    \end{proof}

    We can now conclude the proof of~\cref{thm:convergence_shallow_water}. We apply Grönwall's Lemma to inequality~\eqref{eq:gronwall_relative_entropy} to get that for all $\tau \in [0,T]$:
    \begin{equation*}
        \ecal_\gamma(r,\ubf|h,\U)(\tau) \leqslant \left(\ecal_\gamma(r,\ubf|h,\U)(0)+\ocal\left(\frac{1}{\gamma^2}\right)\right) \exp(C\tau).
    \end{equation*}

    \noindent Since $\tau \mapsto \exp(C \tau)$ is bounded on $[0,T]$, we recover inequality~\eqref{eq:entropy_estimate} of~\cref{thm:convergence_shallow_water}:
    \begin{flalign*}
        \forall \tau \in [0,T], && \ecal_\gamma(r,\ubf|h,\U)(\tau) &\leqslant \ecal_\gamma(r,\ubf|h,\U)(0) \exp(C\tau)+\ocal\left(\frac{1}{\gamma^2}\right).&&
    \end{flalign*}

    Now assume that the initial conditions $(r_0,\ubf_0)$ tend to $(h_0,\U_0)$ in the following sense:
    \begin{equation*}
        \Norm{r_0-h_0}_{L^2(\T^2)} + \Norm{\sqrt{r_0}(\ubf_0-\U_0)}_{L^2(\T^2)} \limit{\gamma}{\infty} 0.
    \end{equation*}

    \noindent Then, according to~\eqref{eq:equivalence_potential_norm}, we have \begin{equation*}
        \ecal_\gamma(r,\ubf|h,\U)(0) \limit{\gamma}{\infty} 0.
    \end{equation*}
    Consequently, for all $\tau \in [0,T]$, we have
    \begin{equation*}
        \ecal_\gamma(r,\ubf|h,\U)(\tau) \limit{\gamma}{\infty} 0.
    \end{equation*}

    \noindent Since $L^2(\T^2) \subset L^{2-\frac{1}{\gamma}}(\T^2)$ for all $\gamma>1$, estimate~\eqref{eq:equivalence_potential_norm} assures the convergence in a space $L^p(\T^2)$ such that $2-1/\gamma \leqslant p$ for all $\gamma$. Since we consider $\gamma>1$, we choose $p=1$ and the convergence towards $0$ of $\ecal_\gamma$ implies~\eqref{eq:convergence_euler_to_sv}:
    \begin{equation*}
        \sup_{\tau \in [0,T]} \Norm{r-h}_{L^1(\T^2)}(\tau) + \Norm{\sqrt{r}(\ubf-\U)}_{L^2(\T^2)}(\tau) \limit{\gamma}{\infty} 0.
    \end{equation*} 
    Moreover if we fix a rate of convergence to $0$ of order $\ocal\left({1}/{\gamma^2}\right)$  for the initial data, then we get the same order of convergence for $\tau \in [0,T]$.

\subsection{Numerical example}\label{sec:numerical_free_surface}

    Using the notation $H=h^{\frac{\gamma}{\gamma-1}}$, the averaged model~\eqref{eq:avg_free_surface_gamma} can be written in the compact form
    using the vector of conservative variables $W=\left(H,Hu\right)^T$ as
    \begin{equation}
        \partial_t W+\partial_x F(W)=S(W),
    \label{eq:compact_form_free_surface}
\end{equation}
where
    \begin{equation*}
        F(W)=\begin{pmatrix} Hu\\ Hu^2+\frac{\gamma-1}{2\gamma-1}gH^{\frac{2\gamma-1}{\gamma}}\end{pmatrix}, \qquad
        S(W)=\begin{pmatrix} 0\\ -\left( \frac{\gamma}{\gamma-1}\right)^{\frac{\gamma}{\gamma-1}}\left( \frac{a}{g}\right)^{\frac{1}{\gamma-1}} \widehat{\kappa}\, u\end{pmatrix}.
    \end{equation*}
    The eigenvalues of the Jacobian matrix $\partial_W F(W)$ are
    \begin{equation}
        \lambda_{\pm}=u\pm c,\qquad c
        =\sqrt{\frac{\gamma-1}{\gamma}\,g\,H^{\frac{\gamma-1}{\gamma}}},
    \label{eq:eigenvalues_free_surface}
    \end{equation}
    in agreement with the analysis of the previous section.

    The system \eqref{eq:compact_form_free_surface} is solved with a finite volume scheme using the HLL numerical flux written as PVM scheme (see e.g. \cite{CastroDiaz2012} for details), with wave speeds estimated from the eigenvalues~\eqref{eq:eigenvalues_free_surface}. The source term is semi-implicitly discretized.

    In the tests below, we take $g=9.81$, $a=1$, friction coefficient $\widehat{\kappa}=10^{-3}$ and $\mathrm{CFL}=0.9$, and the solution of the proposed compressible model (SWC) is compared with that of the classical shallow-water (SW) model.

    \subsubsection{Dam break test}
\begin{figure}[!h]
        \centering
        \includegraphics[width=0.49\linewidth]{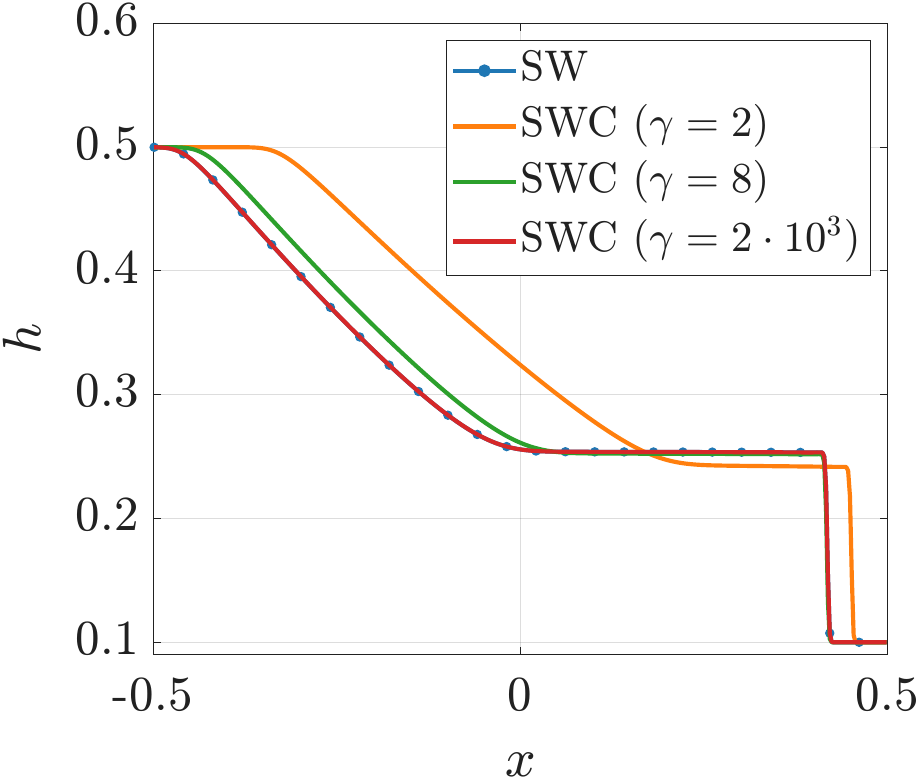}
        \includegraphics[width=0.49\linewidth]{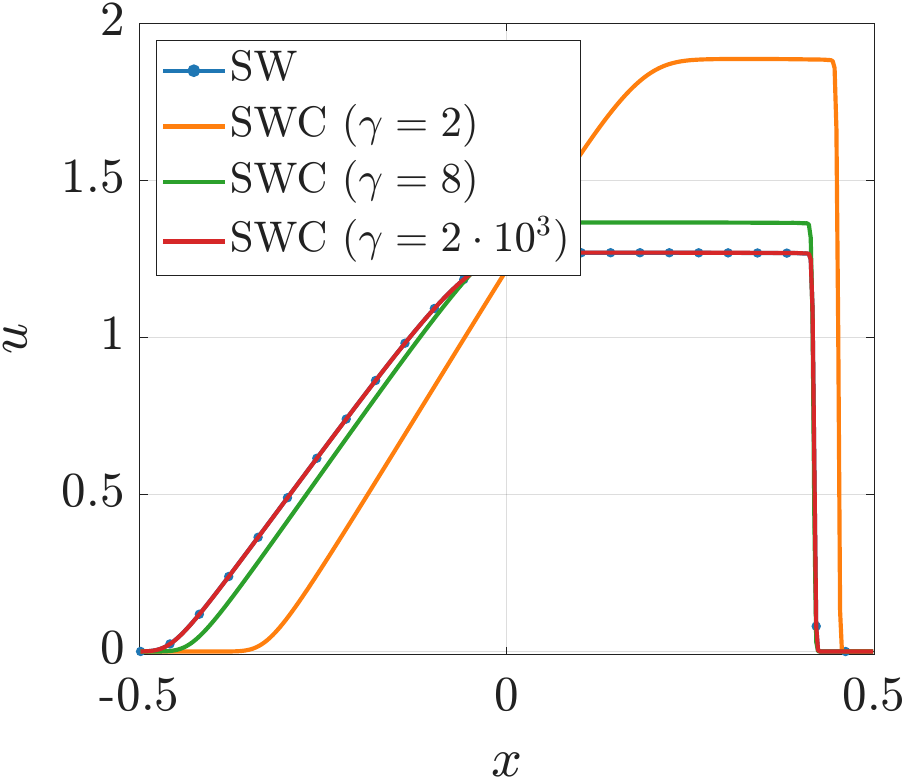}
        \caption{Free surface dam-break problem at $t=0.2\,$s. Left: height $h$. Right: velocity $u$. Comparison between the classical shallow-water model (SW) and the proposed compressible model (SWC) for $\gamma=2$, $\gamma=8$ and $\gamma=2\cdot 10^3$. As $\gamma$ increases, the SWC solution converges to the SW one.}
        \label{fig:FS:test1}
    \end{figure}

        We first consider a dam-break problem on the domain $[-0.5,0.5]$ with the initial condition
    \[ h(0,x)=\begin{cases} 0.5 & \text{if } x<0,\\ 0.1 & \text{if } x>0,\end{cases}\qquad u(0,x)=0.\]
    The computation uses $400$ points, and the results are shown at $t=0.2\,$s in Figure~\ref{fig:FS:test1} for $\gamma=2$, $\gamma=8$ and $\gamma=2\cdot 10^{3}$.
    
    The solution has the usual structure of the dam-break problem: a rarefaction wave traveling to the left, a constant intermediate state and a shock traveling to the right. For $\gamma=2$ the compressible model differs markedly from SW: the intermediate velocity is about $1.89$ instead of about $1.27$, the rarefaction is smoother and propagates more slowly towards the left, and the shock is located further to the right. For $\gamma=8$ the difference is already much smaller, although the intermediate velocity ($\approx 1.37$) and the position of the rarefaction head still differ visibly from SW. For $\gamma=2000$, the solution is indistinguishable from the SW one for both $h$ and $u$. The numerical results therefore confirm that the proposed model converges to the shallow-water model as $\gamma$ increases, in agreement with the analysis presented in the paper.

    \subsubsection{A smooth test: propagation of a Gaussian bump.}
    With the goal of numerically checking the result in Theorem~\ref{thm:convergence_shallow_water},
    which concerns strong solutions, about the relative entropy between both models,
    we consider here a test with smooth initial data. So, we define
    \[ h(0,x)=1+0.1\,e^{-2 x^2},\qquad u(0,x)=0,\]
    on the domain $[-5,5]$, discretized with $2000$ points, and computed up to $t=1\,$s. The solution remains smooth, 
    no shock is formed, and the waves do not reach the boundaries of the domain. 
    The SWC model is computed for $\gamma=2,4,8,16,32,64,128,500$ and $2000$.

\begin{figure}[!h]
        \centering
        \includegraphics[width=0.49\linewidth]{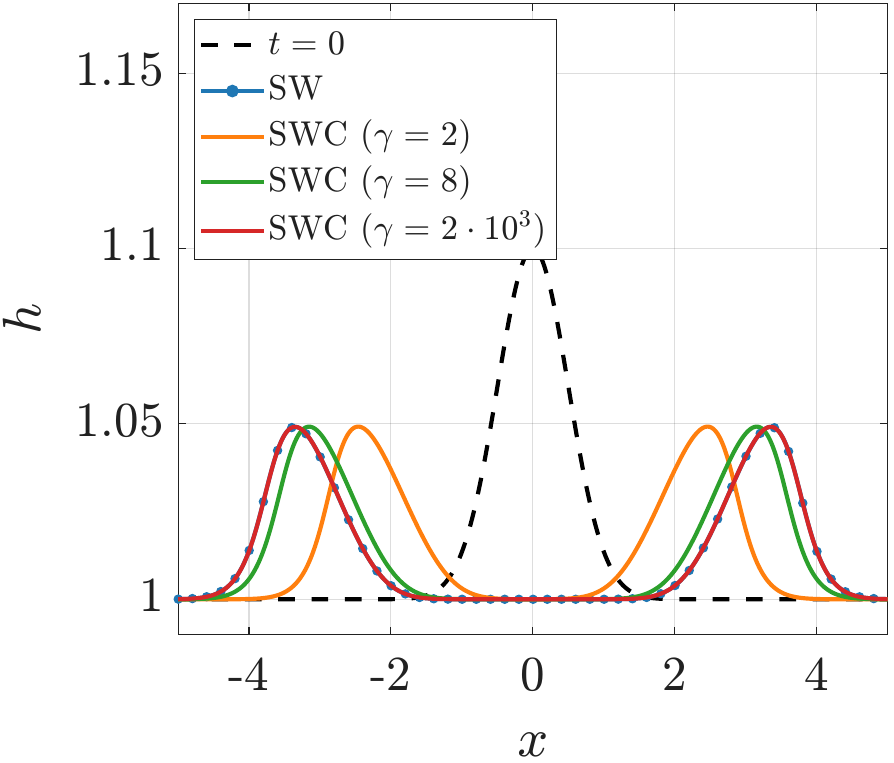}
        \includegraphics[width=0.49\linewidth]{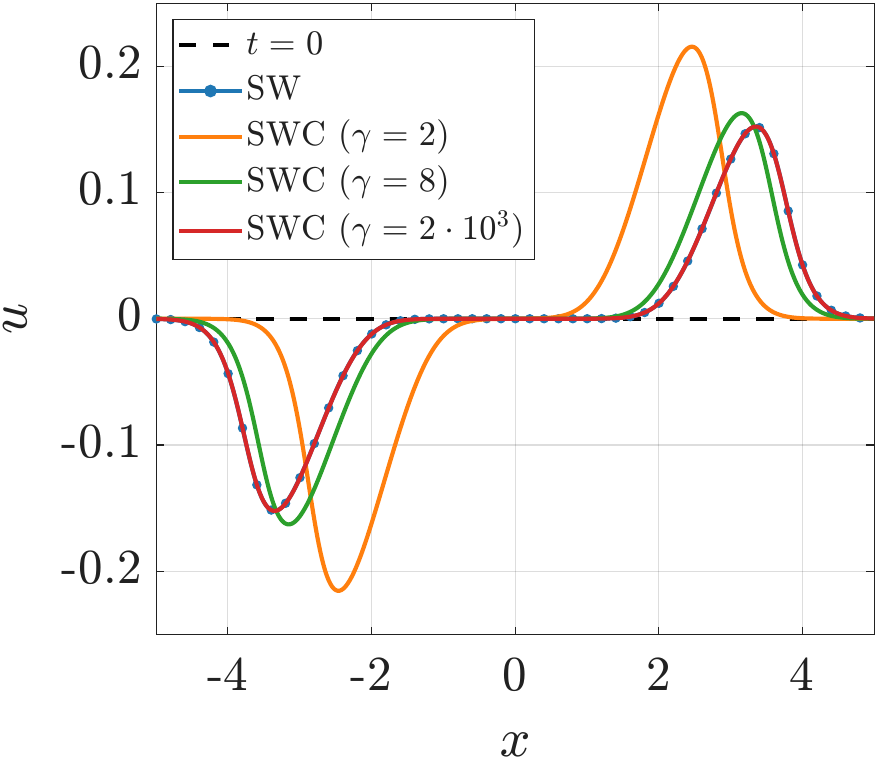}
        \caption{Free surface Gaussian bump. Left: height $h$. Right: velocity $u$. Solution at $t=1\,$s for the classical shallow-water model (SW) and the proposed compressible model (SWC) for $\gamma=2$, $\gamma=8$ and $\gamma=2\cdot 10^3$.}
        \label{fig:FS:gauss_solution}
    \end{figure}

    Figure~\ref{fig:FS:gauss_solution} shows the initial condition and the solution at $t=1\,$s for SW and for SWC with $\gamma=2$, $8$ and $2\cdot 10^3$. 
    The initial bump splits into two waves as expected. Their speed is governed 
    by the eigenvalues~\eqref{eq:eigenvalues_free_surface}: the celerity in the SWC case is $c=\sqrt{\frac{\gamma-1}{\gamma}gh}$,
    which is smaller than $\sqrt{gh}$ (case SW) for small values of $\gamma$, so that the waves of the compressible model are delayed. 
    For $\gamma=2\cdot 10^3$ the solution is indistinguishable from the SW one.

    For each value of $\gamma$, we compute the averaged relative entropy $\ecal_\gamma(t)$ given by \eqref{eq:relative_entropy_integral} 
    at several times, where $(r,\ubf)$ is the solution of the SWC model, with $r=h^{\frac{\gamma}{\gamma-1}}$, 
    and $(h,\U)$ is the solution of the SW model. Notice that $\ecal_\gamma(0)\neq 0$, 
    since $r(0,x)=h(0,x)^{\frac{\gamma}{\gamma-1}}\neq h(0,x)$.

\begin{figure}[!h]
        \centering
        \includegraphics[width=0.49\linewidth]{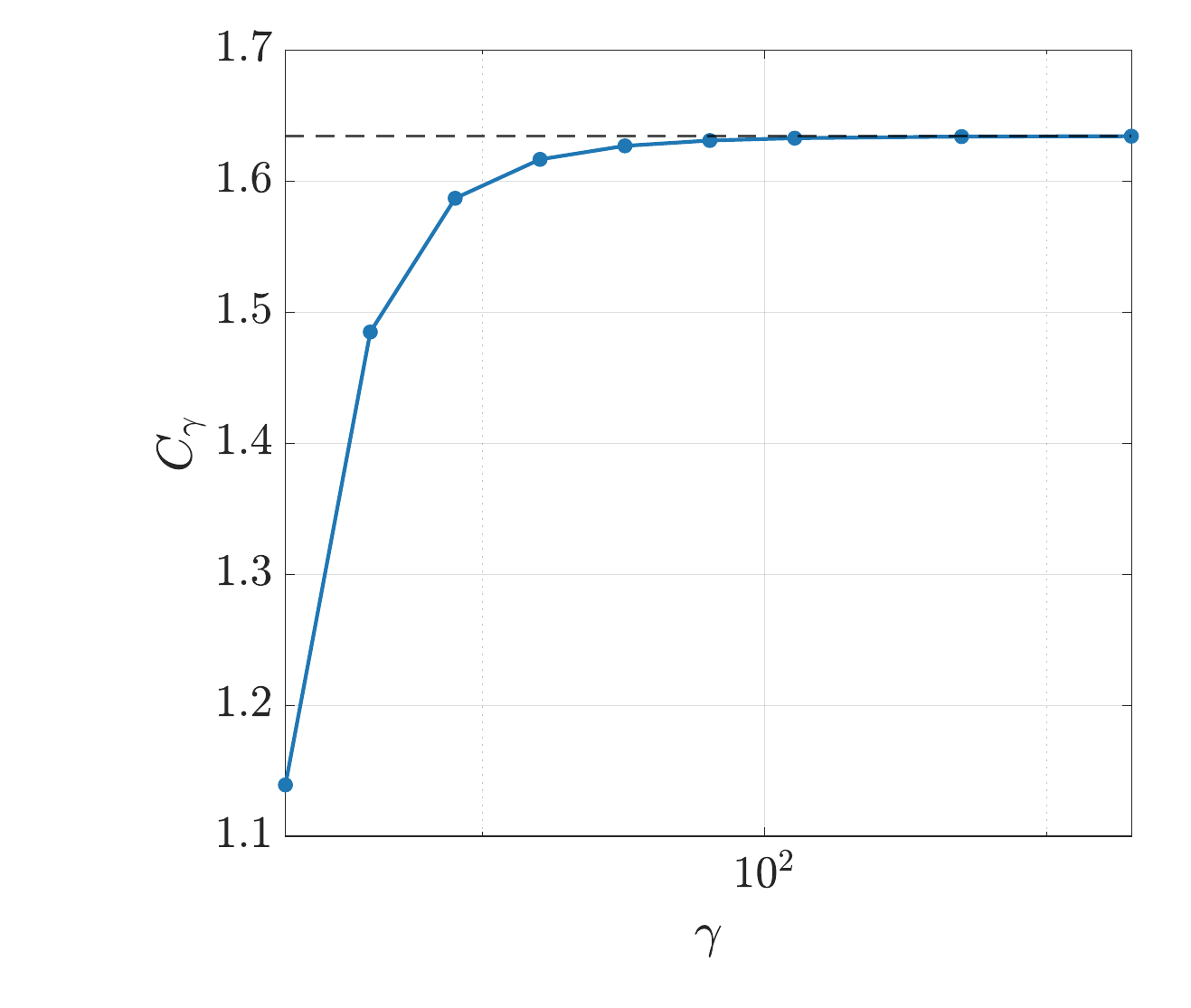}
        \includegraphics[width=0.49\linewidth]{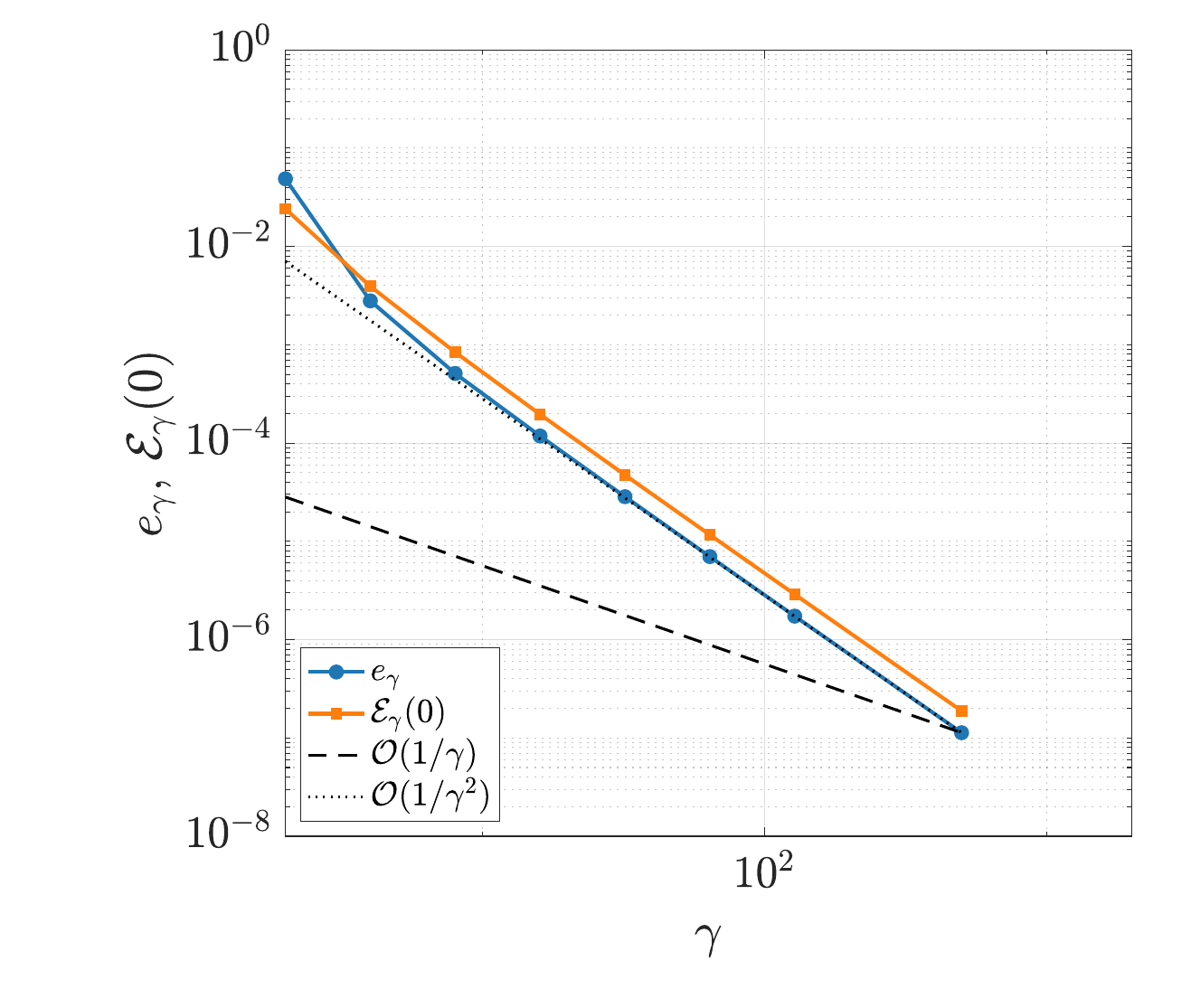}
        \caption{Free surface Gaussian bump. Left: $C_\gamma=\max_{0<t\leqslant 1}\frac{1}{t}\log(\ecal_\gamma(t)/\ecal_\gamma(0))$;
         the dashed line corresponds to $C=\max_\gamma C_\gamma\approx 1.63$. Right: $e_\gamma=\sup_{0\leqslant t\leqslant 1}|\ecal_\gamma(t)-\ecal_\gamma(0)\exp(Ct)|$ and the initial relative entropy $\ecal_\gamma(0)$ as functions of $\gamma$ (log-log scale), compared with the slopes $\ocal(1/\gamma)$ (dashed line) and $\ocal(1/\gamma^2)$ (dotted line).}
        \label{fig:FS:gauss_entropy}
    \end{figure}

    In order to estimate the constant $C$ in~\eqref{eq:entropy_estimate}, we compute
    \[ C_\gamma=\max_{0<t\leqslant 1} C_\gamma(t),\qquad\mbox{with}\qquad C_\gamma(t)=\frac{1}{t}\log\left(\frac{\ecal_\gamma(t)}{\ecal_\gamma(0)}\right).\]
    $C_\gamma(t)$ stabilizes as $t$ approaches $1$ s. The values $C_\gamma$ are plotted 
    in Figure~\ref{fig:FS:gauss_entropy} (left) as a function of $\gamma$. 
    They increase with $\gamma$ and converge to $1.63$ approximately, which is consistent with 
    a constant $C$ independent of $\gamma$, as stated in Theorem~\ref{thm:convergence_shallow_water}. 
    We thus take $C=\max_\gamma C_\gamma\approx 1.63$.

    Finally, with the previous value of $C$, we have $\ecal_\gamma(t)-\ecal_\gamma(0)\exp(Ct)\leqslant 0$
    for all $\gamma$ and $t\in[0,1]$, so that the estimate~\eqref{eq:entropy_estimate} is satisfied. 
    In order to measure the deviation of the relative entropy from the exponential bound in~\eqref{eq:entropy_estimate}, we define
    \[ e_\gamma=\sup_{0\leqslant t\leqslant 1}\left|\ecal_\gamma(t)-\ecal_\gamma(0)\exp(Ct)\right|.\]
    Figure~\ref{fig:FS:gauss_entropy} (right) shows, in logarithmic scale, the convergence rate of $e_\gamma$ and $\ecal_\gamma(0)$ in $\gamma$. 
    Both decrease like $\gamma^{-2}$, that is compatible with $\ocal(1/\gamma^2)$ of~\eqref{eq:entropy_estimate} for this test.

\section{Closing the system in the pressurized case}
\label{sec:closure_pressurized}

    In this section, we analyze the closure of the system \eqref{eq:GeneralDepthAveragedSystem} for the pressurized case. Recall that $h=h(\xbf)$ is prescribed in this case and the unknowns of the system are $\Bar{\rho}$, $\favre{\ubf_\xbf}$, $\Bar{p}$ and $p_S=p|_{z=h}$. We must find a closure of the system using \eqref{eq:barotropy_law} and \eqref{eq:vertical_momentum_asymptotic} to establish relations between the variables to keep just two unknowns, as in the free surface configuration.
 We will see this in the following, first in the quadratic case and then in the general case.

\subsection[The quadratic case]{The averaged model in the pressurized case for $\gamma=2$}

    As in the free surface case, we write the integrated vertical momentum equation \eqref{eq:GeneralDepthAveragedSystem2} as:
        \[ \rho=\rho_S-\frac{g}{2a}(z-h).\]

    \noindent Since we lost the information about the normal component of $\sigma\nbf_h$, we don't know $\rho_S$. It thus becomes a new unknown of the problem. However, $h$ is prescribed now. Our goal in this section is to express $\rho_S$ in terms of $h$ and $\bar{\rho}$ to recover $\bar{p}$ as a function of $\bar{\rho}$ and $h$. Averaging the previous equation, we obtain
    \begin{equation}
        \bar{\rho}=\rho_S - \frac{g}{2a h}\left[ \frac{z^2}{2}-hz\right]_0^h = \rho_S + \frac{gh}{4a}.
    \label{eq:avg_density_pressurized}
    \end{equation}

    \noindent Next, we use the expression of $\rho$ to write the pressure from \eqref{eq:barotropy_law}, 
    \begin{equation*}
        p(\rho)=a \rho^2= a\left(\rho_S + \frac{g}{2a}(h-z)\right)^2.
    \end{equation*}
    Since $\displaystyle \rho_S=\bar{\rho}-\frac{gh}{4a}$ from \eqref{eq:avg_density_pressurized}, we can express the pressure as a function of $\bar{\rho}$ and $h$:
    \begin{equation*}
        p(\rho)=a \left(\bar{\rho}+\frac{gh}{4a} - \frac{gz}{2a}\right)^2=a\left(\bar{\rho}+\frac{gh}{4a}\right)^2 - \left(\bar{\rho}+\frac{gh}{4a}\right)gz+\frac{g^2z^2}{4a}.
    \end{equation*}
    Averaging the previous expression, we get
    \begin{equation}
        \bar{p}=a\left(\bar{\rho}+\frac{gh}{4a}\right)^2 - \left(\bar{\rho}+\frac{gh}{4a}\right)\frac{gh}{2} + \frac{g^2h^2}{12a}=a\bar{\rho}^2+\frac{g^2h^2}{48a}.
    \label{eq:avg_pressure_pressurized}
    \end{equation}
    Finally, evaluating the pressure at $z=h$ we get $p_S$ as
    \begin{equation}
        p_S=p|_{z=h}=a \rho_S^2=a\left(\bar{\rho}-\frac{gh}{4a}\right)^2.
    \end{equation}

    \begin{remark}
        The function $\bar{p}(\bar{\rho},h)=a\bar{\rho}^2+\frac{g^2h^2}{48a}$ coincides with the one obtained in the free surface case in \eqref{eq:p_avg_free_surface_quad} for $\rho_S=0$. In the latter case we just chose to express everything with respect to $h$.
    \hfill $\square$
\end{remark}
\bigskip

    \noindent Therefore, from \eqref{eq:GeneralDepthAveragedSystem}, the averaged model for a pressurized compressible flow for unknowns $\bar{\rho}$ and $\favre{\ubf_\xbf}$ reads
    \begin{subequations}
        \begin{align}
            \partial_t (h\bar{\rho}) + \nabla_\xbf \cdot (h\bar{\rho} \favre{\ubf_\xbf})&=0,
        \label{eq:mass_pressurized}\\
            \partial_t (h\bar{\rho} \favre{\ubf_\xbf}) + \nabla_\xbf \cdot (h\bar{\rho} \favre{\ubf_\xbf} \otimes \favre{\ubf_\xbf}) + \nabla_\xbf \left(a h\bar{\rho}^2 + \frac{g^2h^3}{48a}\right) 
            - a\left(\bar\rho - \frac{g h}{4a}\right)^2 \nabla_\xbf h &= -\widehat{\kappa} \favre{\ubf_\xbf}.
        \label{eq:momentum_pressurized}
        \end{align}
    \label{eq:avg_pressurized}
    \end{subequations}
    
  \begin{remark} Since $\partial_t h=0$, the function $h$ is given by the initial conditions and we have two unknowns and two equations, so the system is closed.
\hfill $\square$
\end{remark}

    \subsubsection{Study of the energy balances}

    \paragraph{Averaging the energy equation.}
        We start from the integrated energy equation associated with the Navier-Stokes system given in \eqref{eq:energyNS_general} that reads
    \begin{equation*}
        \partial_t (h\bar{\rho}\favre{E_\xbf}) + \nabla_\xbf \cdot (h(\bar{\rho} \favre{E_\xbf}+\bar{p}) \favre{\ubf_\xbf})
        =-\widehat{\kappa} \Norm{\favre{\ubf_\xbf}}^2
    \end{equation*}
    \begin{flalign}
        \text{with} && \favre{E_\xbf}=\frac{\Norm{\favre{\ubf_\xbf}^2}}{2}+\favre{e(\rho)+gz}=\frac{\Norm{\favre{\ubf_\xbf}^2}}{2}+\frac{\Bar{p}}{\Bar{\rho}}+\frac{\bar{\rho g z}}{\bar{\rho}}. &&
    \end{flalign}

    \noindent In this case, using the expression of the density, the average of the gravitational term is
    \begin{equation*}
        \bar{\rho g z}=\overline{gz\left(\bar{\rho}+\frac{gh}{4a}\right)-\frac{g^2z^2}{2a}}=\bar{\rho}\frac{gh}{2}-\frac{g^2h^2}{24a}.
    \end{equation*}
    From \eqref{eq:avg_pressure_pressurized}, the second term in $\favre{E_\xbf}$ reads
    $$
    \frac{\bar p}{\bar \rho}=a\bar \rho + \frac{g^2h^2}{48a\bar\rho}.
    $$
    Therefore, the Favre average of the energy reads 
    \begin{equation*}
        \favre{E_\xbf}=\frac{\Norm{\favre{\ubf_\xbf}}^2}{2}+a\bar{\rho}-\frac{g^2h^2}{48a\bar{\rho}}+\frac{gh}{2}.
    \end{equation*}
    Finally, the averaged energy equation reads
    \begin{multline}
        \partial_t \left( h\bar{\rho}\left(\frac{\Norm{\favre{\ubf_\xbf}}^2}{2}+a\bar{\rho}-\frac{g^2h^2}{48a\bar{\rho}}+\frac{gh}{2}\right)\right)\\
        + \nabla_\xbf \cdot \left( h\bar{\rho} \left( \frac{\Norm{\favre{\ubf_\xbf}}^2}{2}+2a\bar{\rho}+\frac{gh}{2}\right) \favre{\ubf_\xbf}\right)
        = -\widehat{\kappa} \Norm{\favre{\ubf_\xbf}}^2.
    \label{eq:averaged_energy_pressurized}
    \end{multline}

\paragraph{The energy balance of the averaged system}

    Let us derive the energy equation linked to the system~\eqref{eq:mass_pressurized}-\eqref{eq:momentum_pressurized}. 
    Note that if we take the scalar product of the momentum equation with $\favre{\ubf_\xbf}$, the part with the
    kinetic energy is well understood using the mass equation on $h\bar\rho$. This gives
    \begin{multline*}
    \partial_t \left( h \bar \rho \frac{\Norm{\favre{\ubf_\xbf}}^2}{2} \right)
   +  \nabla_\xbf \cdot \left( h \bar \rho \frac{\Norm{\favre{\ubf_\xbf}}^2}{2}\favre{\ubf_\xbf} \right)\\
   +\favre{\ubf_\xbf}\cdot  \nabla_\xbf (h\bar p) - a \left(\bar\rho - \frac{gh}{4a}\right)^2 \favre{\ubf_\xbf}\cdot \nabla_x h= -\widehat{\kappa} \Norm{\favre{\ubf_\xbf}}^2.
   \end{multline*}
We use the first term in the second line to write
\begin{multline}\label{general_av_energy0}
    \partial_t \left( h \bar \rho \frac{\Norm{\favre{\ubf_\xbf}}^2}{2} \right)
   +  \nabla_\xbf \cdot \left( h \bar \rho \left(\frac{\Norm{\favre{\ubf_\xbf}}^2}{2} +\frac{\bar p}{\bar\rho}\right)\favre{\ubf_\xbf}\right)\\
   -h\bar p\,  \nabla_\xbf\cdot \favre{\ubf_\xbf} - a \left(\bar\rho - \frac{gh}{4a}\right)^2 \favre{\ubf_\xbf}\cdot \nabla_x h= -\widehat{\kappa} \Norm{\favre{\ubf_\xbf}}^2.
   \end{multline}
We focus on the term of the time derivative in \eqref{eq:averaged_energy_pressurized} and search for the remaining term $\partial_t (h\bar\rho (a\bar{\rho} - \frac{g^2h^2}{48a\bar{\rho}} + \frac{gh}{2}))$. For this aim, we denote $D_t$ the material derivative operator $D_t(\cdot) = \partial_t (\cdot)+ \favre{\ubf_\xbf}\cdot \nabla_\xbf (\cdot)$ and we introduce
   $$I=  h \bar \rho  \, D_t \left(a\bar{\rho} - \frac{g^2h^2}{48a\bar{\rho}} + \frac{gh}{2}\right).$$
Using the mass equation on $h\bar\rho$, \eqref{eq:mass_pressurized}, we equivalently write 
   \begin{equation}\label{eq:Ienergy}
I=\partial_t \left( h \bar \rho \left(a\bar{\rho} - \frac{g^2h^2}{48a\bar{\rho}} + \frac{gh}{2}\right) \right)
   + \nabla_\xbf \cdot \left( h \bar \rho \left(a\bar{\rho} - \frac{g^2h^2}{48a\bar{\rho}} + \frac{gh}{2}\right)\favre{\ubf_\xbf} \right).
   \end{equation}
Let us prove that $I$ coincides with the term appearing in the second line of \eqref{general_av_energy0}, that is,
    $$
       I= - h \bar p\, \nabla_\xbf \cdot \favre{\ubf_\xbf} - a\left(\bar\rho - \frac{gh}{4a}\right)^2 \favre{\ubf_\xbf}\cdot \nabla_\xbf h.$$
From the mass equation \eqref{eq:mass_pressurized} we have
    $$D_t \bar{\rho} = - \bar{\rho} \nabla_\xbf\cdot\favre{\ubf_\xbf} - \frac{\bar{\rho}}{h} D_t h
    $$
    and since $h$ does not depend  on $t$
     $$D_t h = \favre{\ubf_\xbf} \cdot \nabla_\xbf h .
    $$
We compute the second term of $I$,
    $$
    -D_t\left(\frac{g^2 h^2}{48 a\bar\rho}\right) = -\frac{g^2 h}{24 a\bar\rho} D_t h +\frac{g^2h^2}{48 a\bar\rho^2}D_t\bar\rho
           $$
    We merge all the calculations before to write
    \begin{eqnarray*}
    I&=&h \bar \rho  \, D_t \left(a\bar{\rho} - \frac{g^2h^2}{48a\bar{\rho}} + \frac{gh}{2}\right)
    =
    h\bar\rho\left(aD_t \bar\rho-\frac{g^2 h}{24 a\bar\rho} D_t h +\frac{g^2h^2}{48 a\bar\rho^2}D_t\bar\rho+\frac{g}{2}D_t h\right)
    \\&=&
    h\bar\rho\left(\left(a +\frac{g^2h^2}{48 a\bar\rho^2}\right)D_t \bar\rho
    +\left(-\frac{g^2 h}{24 a\bar\rho}+\frac{g}{2}\right) D_t h \right)
    \end{eqnarray*}
Replacing $D_t\bar\rho$ and $D_t h$,

\begin{eqnarray*}
I&=&
-h\bar\rho\left(a\bar{\rho} +\frac{g^2h^2}{48 a\bar\rho}\right)\nabla_\xbf\cdot\favre{\ubf_\xbf} 
-\left(a\bar\rho^2-\bar\rho\frac{gh}{2}+\frac{g^2 h^2}{16 a}\right) \favre{\ubf_\xbf} \cdot \nabla_\xbf h 
\\&=&
-h\bar p \, \nabla_\xbf\cdot\favre{\ubf_\xbf} 
-a\left(\bar\rho -\frac{gh}{4a}\right)^2\favre{\ubf_\xbf} \cdot \nabla_\xbf h 
\end{eqnarray*}
where in the last equality we used \eqref{eq:avg_pressure_pressurized}.
Replacing this term in \eqref{general_av_energy0} with $I$ in \eqref{eq:Ienergy} and using $
    \frac{\bar p}{\bar \rho}=a\bar \rho + \frac{g^2h^2}{48a\bar\rho}
    $, we recover the energy balance \eqref{eq:averaged_energy_pressurized}.

\subsection[General case]{The averaged model in the pressurized case for any $\gamma>1$}\label{se:pressurized_general}

    Following the same procedure as in the free surface case, we write the asymptotic vertical momentum equation as 
    \begin{equation*}
        a \frac{\gamma}{\gamma -1}\partial_z (\rho^{\gamma-1}) = - g.
    \end{equation*}
    Integrating the previous equation from $z$ to $h$, we get
    \begin{equation}
        \rho = \left(\rho_S^{\gamma-1}+\frac{\gamma-1}{a \gamma}g(h-z)\right)^{\frac{1}{\gamma-1}}.
    \label{eq:density_pressurized}
    \end{equation}
    Integrating this expression over $z$ from $0$ to $h$, we obtain
    \begin{equation}
        h\bar{\rho} =  \frac{a}{g} \left(\left(\rho_S^{\gamma-1}+\frac{\gamma-1}{a\gamma}g h\right)^{\frac{\gamma}{\gamma-1}}-\rho_S^\gamma \right)
        = \rho_S^\gamma \frac{a}{g} \left(\left(1+\frac{\gamma-1}{a\gamma}g\frac{h}{\rho_S^{\gamma-1}}\right)^{\frac{\gamma}{\gamma-1}}-1\right).
    \label{eq:avg_density_pressurized_gamma}
    \end{equation}
Next, from \eqref{eq:barotropy_law} the pressure can be expressed as
    \begin{equation*}
        p(\rho) = a \rho_S^\gamma \left(1+\frac{\gamma-1}{a \gamma}g\frac{h-z}{\rho_S^{\gamma-1}}\right)^{\frac{\gamma}{\gamma-1}}.
    \end{equation*}
    Integrating this expression over $z$ from $0$ to $h$, we get
    \begin{equation}
        h\bar{p}=\rho_S^{2\gamma-1} \frac{a^2 \gamma}{g(2\gamma-1)} \left(\left(1+\frac{\gamma-1}{a\gamma}g\frac{h}{\rho_S^{\gamma-1}}\right)^{\frac{2\gamma-1}{\gamma-1}}-1\right).
    \label{eq:avg_pressure_pressurized_gamma}
    \end{equation}
Then, we obtain the system for unknowns $\bar\rho$ and $\favre{\ubf_\xbf}$
	    \begin{equation}\label{eq:pressurizedSystem}
	    	\left\{\begin{array}{l}
	 \partial_t(h\Bar{\rho})+\nabla_\xbf \cdot(h \Bar{\rho} \favre{\ubf_\xbf})=0,\\[2mm]
		\partial_t (h\Bar{\rho}\favre{\ubf_\xbf})+\nabla_\xbf \cdot (h\Bar{\rho}\favre{\ubf_\xbf} \otimes \favre{\ubf_\xbf})+\nabla_\xbf (h\Bar{p})
 - a \rho_S^\gamma\nabla_\xbf h =-\widehat{\kappa} \favre{\ubf_\xbf},
			    	\end{array}\right.
	\end{equation}
	where $h\bar{p}$ is given by \eqref{eq:avg_pressure_pressurized_gamma} as a function of $\rho_S$, and $\rho_S$ is implicitly defined in \eqref{eq:avg_density_pressurized_gamma} as a function of $h$ and $\bar\rho$. Recall that $h$ is given and then the system is closed. In the next lines, we show an equivalent expression of the momentum equation rearranging the pressure terms, and how to obtain $\rho_S$.
    
\begin{remark}[Bottom density and an equivalent form of the pressure terms]\label{rem:pressure_terms_pressurized}
	Let us denote by $\rho_0(t,\xbf)=\rho(t,\xbf,0)$ the density at the bottom. From \eqref{eq:density_pressurized} and \eqref{eq:avg_density_pressurized_gamma} we deduce 
\begin{equation}\label{eq:rho_bottom_pressurized}
		\rho_0^{\gamma-1}=\rho_S^{\gamma-1}+\frac{\gamma-1}{a\gamma}\,g\,h,
\qquad \mbox{and} \qquad 
h\bar{\rho}=\frac{a}{g}\left(\rho_0^{\gamma}-\rho_S^{\gamma}\right)
    \end{equation}
By deriving the first equality with respect to $\xbf$, we obtain
\begin{equation}\label{eq:dx_rho_bottom_pressurized}
\rho_0^{\gamma-2} \nabla_\xbf \rho_0 = \rho_S^{\gamma-2} \nabla_\xbf \rho_S + \frac{g}{a \gamma} \nabla_\xbf h.
\end{equation}
Moreover, from  \eqref{eq:avg_pressure_pressurized_gamma} and \eqref{eq:rho_bottom_pressurized} we obtain 
	\begin{equation} \label{eq:hbarp_rho0}
		h\bar{p}=\frac{a^2\gamma}{g(2\gamma-1)}\left(\rho_0^{2\gamma-1}-\rho_S^{2\gamma-1}\right).
	\end{equation}
Finally, using \eqref{eq:rho_bottom_pressurized},  \eqref{eq:dx_rho_bottom_pressurized}  and \eqref{eq:hbarp_rho0} we get 
\begin{equation}\label{eq:pressure_terms_pressurized}
		\nabla_\xbf(h\bar p)-a\rho_S^\gamma\nabla_\xbf h
		= h\bar\rho\,\nabla_\xbf\left(\frac{a\gamma}{\gamma-1}\rho_0^{\gamma-1}\right)
		= h\bar\rho\,\nabla_\xbf\left(\frac{a\gamma}{\gamma-1}\rho_S^{\gamma-1}+g\,h\right).
	\end{equation}
	Hence, the momentum equation of \eqref{eq:pressurizedSystem} can be equivalently written as
	\begin{equation}\label{eq:pressurizedSystem2}
\partial_t (h\Bar{\rho}\favre{\ubf_\xbf})+\nabla_\xbf \cdot (h\Bar{\rho}\favre{\ubf_\xbf} \otimes \favre{\ubf_\xbf})
		+h\bar\rho\,\nabla_\xbf\left(\frac{a\gamma}{\gamma-1}\rho_S^{\gamma-1}+g\,h\right)=-\widehat{\kappa} \favre{\ubf_\xbf}.	    
	\end{equation}
\hfill $\square$
\end{remark}

The value of $\rho_S$ must be calculated from \eqref{eq:avg_density_pressurized_gamma}. To this end, let us denote $\xi=\rho_S^{\gamma} \geq 0$ and then look for the root of the nonlinear function 
\begin{equation} \label{eq:def_f_rhos}
f(\xi) = \left(\xi^{\frac{\gamma-1}{\gamma}} + \dfrac{\gamma-1}{\gamma}\dfrac{g}{a}h\right)^{\frac{\gamma}{\gamma-1}} - \xi - \dfrac{g}{a}h\bar{\rho}.
\end{equation} 
In fact, there exists a unique solution $\tilde{\xi} \geq 0$ such that $f(\tilde{\xi})=0$ under certain conditions, which are stated in the following result.

\begin{proposition}\label{prop:newton_finding_rhoS}
	Let $g, a, h, \bar{\rho} > 0$ and $\gamma > 1$. The function $f:[0, \infty) \to \mathbb{R}$ defined by \eqref{eq:def_f_rhos} satisfies:
	\begin{enumerate}
        \item If $f(0)=0$ then $\rho_S=0$. Otherwise, we study the case $f(\xi)=0$ for $\xi>0$. 
		\item $f$ is continuous and differentiable,  with $f'(\xi)>0$ and $f''(\xi)<0$ for all $\xi\in(0,\infty)$.
		\item Under condition 
		\begin{equation}\label{eq:cond_f0}
			\bar{\rho} >  \left( \frac{\gamma-1}{\gamma} \right)^{\frac{\gamma}{\gamma-1}} \left( \frac{g}{a} h \right)^{\frac{1}{\gamma-1}},
		\end{equation}
		there exists a unique root $\tilde{\xi} > 0$ such that $f(\tilde{\xi}) = 0$. 

        \item Starting from $\xi_0>0$ verifying $f(\xi_0)<0$,  condition \eqref{eq:cond_f0} ensures that Newton's method converges to the unique root $\tilde{\xi} > 0$.
	\end{enumerate}
\end{proposition}

\begin{proof}
The first derivative of $f$ is 
$$
	f'(\xi) =  \left( 1 + \frac{\gamma-1}{\gamma} \frac{g}{a} \frac{h}{\xi^{\frac{\gamma-1}{\gamma}}} \right)^{\frac{1}{\gamma-1}} - 1 >0, \quad \mbox{for } \xi >0.
$$
In particular, $f'(\xi)$ never vanishes on $(0, \infty)$. For the second derivative, we have
$$	
		f''(\xi) 
		=-\frac{\gamma-1}{\gamma} \frac{g}{a} \frac{h}{\gamma} \, \dfrac{1}{\xi^{\frac{2\gamma-1}{\gamma}}} \left( 1 + \frac{\gamma-1}{\gamma} \frac{g}{a}\dfrac{h}{ \xi^{\frac{\gamma-1}{\gamma}}} \right)^{\frac{2-\gamma}{\gamma-1}} <0 \quad \mbox{for } \xi>0.
$$
In particular, $f''(\xi)$ never vanishes on $(0, \infty)$.\\ Since $f'(\xi) > 0,f''(\xi) < 0 \quad \forall \xi > 0,$ and $f(\xi)\to+\infty$ as $\xi\to+\infty$,  

it is enough to prove that $f(0)<0$. We have $$f(0) = \left( \dfrac{\gamma-1}{\gamma} \dfrac{g}{a} h \right)^{\frac{\gamma}{\gamma-1}} - \dfrac{g}{a} h \bar{\rho},$$ and condition \eqref{eq:cond_f0} implies $f(0)<0$.
	
As a consequence, starting from $\xi_0>0$ satisfying $f(\xi_0)<0$, condition \eqref{eq:cond_f0} ensures that Newton's method converges to the unique root $\tilde{\xi}>0$.
\end{proof}

Note that, in the case $\gamma=2$, condition \eqref{eq:cond_f0} is equivalent to $\rho_S>0$, where $\rho_S$ is given explicitly by \eqref{eq:avg_density_pressurized}.

\subsection{Averaged energy equation versus energy equation of the averaged system.} In this subsection, we compare the averaged energy balance of
the Navier--Stokes for the pressurized configuration to the energy balance of the averaged  system \eqref{eq:pressurizedSystem}.

\paragraph{Averaged energy balance of the Navier--Stokes equation.} It is given in \eqref{eq:energyNS_general},

    \begin{equation*}
        \partial_t (h\bar{\rho}\favre{E_\xbf}) + \nabla_\xbf \cdot (h(\bar{\rho} \favre{E_\xbf}+\bar{p}) \favre{\ubf_\xbf})
        =-\widehat{\kappa} \Norm{\favre{\ubf_\xbf}}^2
    \end{equation*}
    \begin{flalign}
        \text{with} && \favre{E_\xbf}=\frac{\Norm{\favre{\ubf_\xbf}^2}}{2}+\frac{1}{\gamma-1}\frac{\Bar{p}}{\Bar{\rho}}+\frac{\bar{\rho g z}}{\bar{\rho}}. &&
    \end{flalign}

    We use~\eqref{eq:density_pressurized} to compute the average $\bar{\rho g z}$:
    \begin{equation*}
        \bar{\rho g z} = \frac{g}{h}\int_0^h \left( \rho_S^{\gamma-1}+\frac{\gamma-1}{a \gamma}g(h-z)\right)^{\frac{1}{\gamma-1}} z \d z.
    \end{equation*}
    Thanks to integration by parts, we get
    \begin{equation*}
        \bar{\rho g z} =\frac{g}{h} \left( -\frac{ah}{g} \rho_S^{\gamma}
        -\frac{a^2\gamma}{(2\gamma-1)g^2}\Big( \rho_S^{2\gamma-1}-\big(\rho_S^{\gamma-1}+\frac{\gamma-1}{a\gamma}gh\big)^{\frac{2\gamma-1}{\gamma-1}}\Big)\right).  
    \end{equation*}

    \noindent We recognize the expression of $\rho_0$ given in~\eqref{eq:rho_bottom_pressurized}. We can therefore rewrite the previous equation as
    \begin{equation*}
        \bar{\rho g z} = -a\rho_S^{\gamma}+\frac{a^2\gamma}{(2\gamma-1)g h}(\rho_0^{2\gamma-1}-\rho_S^{2\gamma-1}).  
    \end{equation*}
    Thanks to~\eqref{eq:hbarp_rho0}, 
    \begin{equation*}
        \bar{\rho g z} = -a\rho_S^{\gamma}+\bar{p}=\bar{p}-p_S,
    \end{equation*}
and then
$$
\favre{E_\xbf}= \frac{\Norm{\favre{\ubf_\xbf}^2}}{2}+\frac{1}{\gamma-1}\frac{\Bar{p}}{\Bar{\rho}}+\frac{\bar{\rho g z}}{\bar{\rho}} = \frac{\Norm{\favre{\ubf_\xbf}^2}}{2} +\frac{1}{\bar\rho}\left(\frac{\gamma}{\gamma-1}\bar p-p_S\right).
$$

Finally, the averaged energy equation is
    \begin{multline}
        \partial_t \left(h\bar{\rho}\Big(\frac{\Norm{\favre{\ubf_\xbf}}^2}{2}+\frac{1}{\bar{\rho}} \Big(\frac{\gamma}{\gamma-1}\bar{p}-p_S\Big)\Big)\right) \\
        +\nabla_\xbf \cdot \left(h\bar{\rho}\Big(\frac{\Norm{\favre{\ubf_\xbf}}^2}{2}+\frac{1}{\bar{\rho}} \Big(\frac{2\gamma-1}{\gamma-1}\bar{p}-p_S\Big)\Big)\favre{\ubf_\xbf}\right)
        =-\widehat{\kappa} \Norm{\favre{\ubf_\xbf}}^2,
    \label{eq:averaged_energy_equation_pressurized_gamma}
    \end{multline}
    where $\rho_S$ is determined thanks to~\cref{prop:newton_finding_rhoS}.

    \begin{remark}
        The expression for the energy equation computed in the case $\gamma=2$ given in \eqref{eq:averaged_energy_pressurized} is consistent with the previous equation. 
    \hfill $\square$
\end{remark}

\paragraph{Energy balance of the averaged system.}
We start with the averaged system with the momentum equation rewritten in the convenient form given in \eqref{eq:pressurizedSystem2}, namely,
\begin{equation}\label{eq:pressurizedSystem2b}
\begin{array}{l}
\partial_t(h\bar\rho) + \nabla_x\cdot (h\bar\rho \favre{\ubf_\xbf})= 0
\\
\partial_t(h\bar\rho\favre{\ubf_\xbf}) 
+ \nabla_x\cdot (h\bar\rho \favre{\ubf_\xbf}\otimes\favre{\ubf_\xbf})
+ h\bar\rho \nabla_x
  \left(\frac{a\gamma}{\gamma-1} \rho_S^{\gamma-1} 
  + gh \right)= - \widehat{\kappa} \favre{\ubf_\xbf}
\end{array}
\end{equation}
Taking the scalar product of the momentum equation above with $\favre{\ubf_\xbf}$  and using the mass equation related to $h\bar\rho$, we get
\begin{multline}\label{energy_av_gral0}
\partial_t \left( h\bar\rho \frac{\Norm{\favre{\ubf_\xbf}}^2}{2}  \right)
+\nabla_\xbf \cdot \left(h\bar{\rho}\left(\frac{\|\favre{\ubf_\xbf}\|^2}{2}+\frac{a\gamma}{\gamma-1} \rho_S^{\gamma-1} 
  + gh \right) \favre{\ubf_\xbf}
\right)
\\
+ \partial_t (h\bar\rho) \left(\frac{a\gamma}{\gamma-1} \rho_S^{\gamma-1} 
  + gh \right)
= -\widehat{\kappa} \Norm{\favre{\ubf_\xbf}}^2.
\end{multline}
Then, to conclude, we just have to calculate
$$I= \partial_t (h\bar\rho) \left(\frac{a\gamma}{\gamma-1} \rho_S^{\gamma-1} + gh \right)
= \frac{a\gamma}{\gamma-1} \partial_t (h\bar\rho)  
\left(\rho_S^{\gamma-1} + \frac{\gamma-1}{a\gamma} gh \right). $$
Using the formula \eqref{eq:avg_density_pressurized_gamma} we have
 \begin{equation}\nonumber 
        h\bar{\rho} 
        =   \frac{a}{g}
        \left(\left(\rho_S^{\gamma-1} +\frac{\gamma-1}{a\gamma}gh\right)^{\frac{\gamma}{\gamma-1}}-
        \rho_S^{\gamma}\right)
    \end{equation}
We introduce the notation
$$X= \rho_S^{\gamma-1} + \frac{\gamma-1}{a \gamma}gh, \qquad
  Y = \rho_S^{\gamma-1},
$$
that we use to write
$$
h\bar\rho = \frac{a}{g}\left(X^{\frac{\gamma}{\gamma-1}}-Y^{\frac{\gamma}{\gamma-1}}\right)
\quad\text{and}\quad
X=Y+\frac{\gamma-1}{a\gamma}gh.
$$
We then have
$$
I=\frac{a^2\gamma}{g(\gamma-1)} X \partial_t\left( X^{\frac{\gamma}{\gamma-1}} - Y^{\frac{\gamma}{\gamma-1}}\right)
=\frac{a^2\gamma}{g(\gamma-1)} \left(X\partial_t\left( X^{\frac{\gamma}{\gamma-1}}\right) - \left( Y+\frac{\gamma-1}{a\gamma}gh\right)\partial_t\left(Y^{\frac{\gamma}{\gamma-1}}\right) \right)
$$
and we write $I=I_1 + I_2 + I_3$ where
\begin{eqnarray*}
I_1&=&\frac{a^2\gamma}{g(\gamma-1)} \partial_t \left(X^{\frac{\gamma}{(\gamma-1)}}\right) X
     = \frac{\gamma}{\gamma-1} \frac{a^2\gamma}{g(2\gamma-1)} 
     \partial_t \left(X^{\frac{2\gamma-1}{(\gamma-1)}}\right),  \\
I_2 &=& - \frac{a^2\gamma}{g(\gamma-1)} \partial_t \left(Y^{\frac{\gamma}{(\gamma-1)}}\right)Y
      = - \frac{\gamma}{\gamma-1} \frac{a^2\gamma}{g(2\gamma-1)}
      \partial_t \left(Y^{\frac{2\gamma-1}{(\gamma-1)}}\right),
\\
I_3&=& - ah \partial_t \left(Y^{\frac{\gamma}{\gamma-1}}\right).
\end{eqnarray*}
The definition of $\bar p$ in \eqref{eq:avg_pressure_pressurized_gamma} gives 
$$
h\bar p = \frac{a^2\gamma}{g(2\gamma-1)}\left(X^{\frac{2\gamma-1}{\gamma-1}}-Y^{\frac{2\gamma-1}{\gamma-1}}\right).
$$
Then, recalling that $h$ does not depend on time in the pressurized case we write
$$I_1+I_2 = \partial_t\left(h \frac{\gamma}{\gamma-1} \bar p\right)
\quad\text{and }\quad
I_3= - \partial_t (h p_S).$$
Replacing in \eqref{energy_av_gral0} we find
\begin{multline*}
\partial_t \left( h\bar\rho \left(\frac{\Norm{\favre{\ubf_\xbf}}^2}{2} +  \frac{1}{\bar \rho}\left(\frac{\gamma}{\gamma-1}\bar p - p_S\right) \right)\right)
+\nabla_\xbf \cdot \left(h\bar{\rho}\left(\frac{\|\favre{\ubf_\xbf}\|^2}{2}+\frac{a\gamma}{\gamma-1} \rho_S^{\gamma-1} 
  + gh \right) \favre{\ubf_\xbf}
\right)
\\
= -\widehat{\kappa} \Norm{\favre{\ubf_\xbf}}^2.
\end{multline*}
Finally, using \eqref{eq:rho_bottom_pressurized} and \eqref{eq:hbarp_rho0}, after some straightforward calculations, we prove that 
$$ \frac{a\gamma}{\gamma-1} \rho_S^{\gamma-1} 
  + gh=\frac{1}{\bar\rho}\left(\frac{2\gamma-1}{\gamma-1}\bar p-p_S\right) .$$
In conclusion, we get the same energy balance as the averaged energy balance in \eqref{eq:averaged_energy_equation_pressurized_gamma}.

\bigskip

\begin{remark} 
Let us see the resulting limit of the pressurized system \eqref{eq:pressurizedSystem2b} when we formally let $\gamma$ go to $+\infty$. 
We denote $\Pi$ the limit of $a\rho_S^{\gamma-1}$ and $\Pi_0$ the limit of $a\rho_0^{\gamma-1}$. Taking the limit in the first expression in \eqref{eq:rho_bottom_pressurized} gives $\Pi_0=\Pi + gh$, then we get the following equality from the relation between $(h,\bar\rho,\rho_S)$ (second equation in \eqref{eq:rho_bottom_pressurized}):
$$h\bar\rho = h$$
which means that $\bar\rho$ goes to 1 when $\gamma$ goes to $+\infty$.
Therefore, recalling that $h$ does not depend on time, the limit system becomes
$$\nabla_x\cdot (h \favre{\ubf_\xbf})=0$$
$$\partial_t (h \favre{\ubf_\xbf}) + \nabla_x\cdot (h \favre{\ubf_\xbf}\otimes \favre{\ubf_\xbf})
    + h \nabla_x Q =  - \widehat{\kappa} \favre{\ubf_\xbf} 
$$
with $Q= \Pi + gh.$ This corresponds to the lake equations.
\hfill $\square$
\end{remark}

\subsection{Eigenvalues of the System}
Let us introduce the vector of conservative variables
\[
W = \begin{pmatrix} H \\ Q \end{pmatrix}, \qquad \mbox{with} \quad H = h\bar{\rho}, \quad Q = h\bar{\rho} u,
\]
For simplicity in the notation, we write $u$ instead of $\favre{\ubf_\xbf}$ hereinafter since we focus on the 1D case. Recall that, in this pressurized regime, $h=h(x)$ is a prescribed, time-independent datum (the cross-section profile), while $\rho_S=\rho_S(t,x)$ is the actual unknown, recovered from $H$ through the implicit relation \eqref{eq:avg_density_pressurized_gamma}. Consequently, $h\bar{p}$ is a function of the pair $(H,h)$, and its $x$-derivative must be expanded accordingly:
\begin{equation} \label{eq:dxhp}
\partial_x (h\bar{p}) = \frac{\partial (h\bar{p})}{\partial H} \partial_x H + \frac{\partial (h\bar{p})}{\partial h} h'(x).
\end{equation} 
The first (conservative) contribution is the one entering the flux Jacobian, while the second is a geometric source term. With this splitting, the system can be recast in the quasilinear form
\begin{equation}
\partial_t W + A(W) \, \partial_x W = S(W)h'(x) + S_{\textrm{fric}}(W),
\label{eq:quasilinear_system}
\end{equation}
where $A(W)$ is the Jacobian, at fixed $h$, of the flux function
\begin{equation}
    \label{eq:A_F}
F(W) = \begin{pmatrix} Q \\ \dfrac{Q^2}{H} + h\bar{p} \end{pmatrix},
\quad\mbox{with}\quad A(W) = \dfrac{\partial F}{\partial W} = \begin{pmatrix} 0 & 1 \\[4pt] -u^2 + \dfrac{\partial(h\bar{p})}{\partial H} & 2u \end{pmatrix},
\end{equation}
where we have used $u = Q/H$, and where the source terms collect the friction term and the geometric contribution coming from $h'(x)$,
\[
S(W) = \begin{pmatrix} 0 \\[2pt] p(h)- \dfrac{\partial(h\bar{p})}{\partial h}  \end{pmatrix},\qquad S_{\textrm{fric}}(W) = \begin{pmatrix} 0 \\[2pt] -\widehat{\kappa}\, u \end{pmatrix} 
\]
with $p(h)=a\rho_S^\gamma$. 
The eigenvalues of the system matrix $A(W)$ are 
$$
\lambda_1 = u - c, \qquad \lambda_2 = u + c,
$$
where
$$c^2 = \dfrac{\partial(h\bar{p})}{\partial H} = \frac{\partial(h\bar{p})}{\partial\rho_S}\dfrac{\partial\rho_S}{\partial H} =  \frac{\partial(h\bar{p})}{\partial\rho_S}\left(\dfrac{\partial H}{\partial \rho_S}\right)^{-1}.$$
Note that both derivatives are computed at fixed $h$. Using the 
notation of the bottom density $\rho_0^{\gamma-1}=\rho_S^{\gamma-1}+\frac{\gamma-1}{a\gamma} gh$ introduced in Remark~\ref{rem:pressure_terms_pressurized}, the variables
$H=h\bar{\rho}$ and $h\bar{p}$ are given by \eqref{eq:rho_bottom_pressurized} and \eqref{eq:hbarp_rho0}, respectively, and then,
using $d\rho_0/d\rho_S = \rho_0^{2-\gamma}\rho_S^{\gamma -2 }$, we have
\[
\frac{d H}{d\rho_S} = \frac{a\gamma}{g} \rho_S^{\gamma-2} \left( \rho_0 - \rho_S \right),
\quad \mbox{and} \quad 
\frac{d(h\bar{p})}{d\rho_S} = \frac{a^2\gamma}{g} \rho_S^{\gamma-2} \left( \rho_0^\gamma - \rho_S^\gamma \right).\]
Thus, the sound speed of the system is
\begin{equation}
    \label{eq:c}
c^2= a\,\frac{\rho_0^\gamma-\rho_S^\gamma}{\rho_0-\rho_S}=\frac{g H}{\rho_0-\rho_S},
\end{equation}
Since $\rho_0>\rho_S>0$, we have $c^2>0$ and the system is strictly hyperbolic.

\subsection{Steady states}
The steady states at rest of the pressurized system \eqref{eq:pressurizedSystem} are characterized in the following result.
\begin{proposition}\label{prop:steady_pressurized}
	Let $g,a>0$, $\gamma>1$ and a section $h(x)>0$ be given. A state with $\favre{\ubf_\xbf}=0$ is a steady solution of \eqref{eq:pressurizedSystem} if and only if the bottom density $\rho_0$ defined in \eqref{eq:rho_bottom_pressurized} is constant, that is,
	\begin{equation}\label{eq:steady_pressurized}
		\frac{a\gamma}{\gamma-1}\rho_S^{\gamma-1}+g\,h = \frac{a\gamma}{\gamma-1}\rho_0^{\gamma-1}=\mbox{const.}
	\end{equation}
	Given a constant $\rho_0>0$ such that
	\begin{equation}\label{eq:condSteady_rho0}
		\rho_0^{\gamma-1}>\frac{\gamma-1}{a\gamma}\,g\,\max_x h(x),
	\end{equation}
	the corresponding steady state is given explicitly by
	\begin{equation}\label{eq:steady_pressurized_explicit}
		\rho_S(x)=\left(\rho_0^{\gamma-1}-\frac{\gamma-1}{a\gamma}\,g\,h(x)\right)^{\frac{1}{\gamma-1}},\qquad
		h\bar\rho(x)=\frac{a}{g}\left(\rho_0^{\gamma}-\rho_S(x)^{\gamma}\right).
	\end{equation}
\end{proposition}
\begin{proof}
	For $\favre{\ubf_\xbf}=0$, the mass equation is trivially satisfied and the momentum equation reduces to $h\bar\rho\,\nabla_\xbf\left(\frac{a\gamma}{\gamma-1}\rho_0^{\gamma-1}\right)=0$ 
    (see \eqref{eq:pressure_terms_pressurized}). Since $h\bar\rho>0$, this is equivalent to $\rho_0$ being constant. Condition \eqref{eq:condSteady_rho0} ensures $\rho_S>0$ in \eqref{eq:steady_pressurized_explicit}, and $h\bar\rho$ follows from \eqref{eq:rho_bottom_pressurized}.
\end{proof}

\begin{remark}
    Using \eqref{eq:density_pressurized} with the notation using $\rho_0$, one gets
    \[
    \rho(x,z)^{\gamma-1} = \rho_0^{\gamma-1} - g\dfrac{\gamma-1}{a\,\gamma}\,z,
    \]
	then, these steady states, which can be explicitly computed, correspond to a density $\rho(x,z)\equiv\rho(z)$, i.e.,
    it depends only on $z$. For $\gamma=2$, \eqref{eq:steady_pressurized} reduces to $\bar\rho+g h/(4a)=\mbox{const.}$

\hfill $\square$
\end{remark}

\subsection{Well-balanced finite volume method}
The pressurized system~\eqref{eq:quasilinear_system} is solved with a finite volume scheme of HLL type, 
which we write as a PVM (Polynomial Viscosity Matrix) method~\cite{CastroDiaz2012}. To this aim,
first the domain is discretized in uniform control volumes $V_i=[x_{i-1/2},x_{i+1/2}]$
 for $i=1,\dots,M$, with mesh step $\Delta x$, and define the cell averaged state at time $t^n$ as 
\[
    W_i^n = \frac{1}{\Delta x}\int_{x_{i-1/2}}^{x_{i+1/2}} W(x,t^n)dx.
\]
Then, the scheme can be written in the discrete form
\begin{equation}
    W_i^{n+1}=W_i^n-\frac{\Delta t}{\Delta x}\left(\mathcal{F}^n_{i+1/2}-\mathcal{F}^n_{i-1/2}+ \frac{1}{2}\left(\mathcal{B}^n_{i-1/2}+\mathcal{B}^n_{i+1/2}\right) \right) + \Delta t S_{\textrm{fric},i}^{n+1},
\label{eq:discrete_scheme_pressurized}
\end{equation}
where $\mathcal{F}_{i+1/2}$ is the numerical flux,
\[
\mathcal{B}_{i+1/2}=
H_{i+1/2} \left( \dfrac{a\gamma}{\gamma-1} \left( \rho_{S,i+1}^{\gamma-1}-\rho_{S,i}^{\gamma-1}\right)+g\,(h_{i+1}-h_{i})\right)\begin{pmatrix} 0\\[2pt] 1 \end{pmatrix},
\]
with $H_{i+1/2}=\frac{1}{2}(H_i+H_{i+1})$, and $S_{\textrm{fric},i}$ is the source term associated to the friction, which is semi-implicitly discretized.

The numerical flux
is
\begin{equation}
    \mathcal{F}_{i+1/2}=\frac{F_{C,i}+F_{C,i+1}}{2}-\frac{1}{2}\mathcal{D}_{i+1/2},
\label{eq:numerical_flux_pressurized}
\end{equation}
where $F_C$ is the convective flux, 
$$
F_C(W) = \begin{pmatrix} Q \\ \dfrac{Q^2}{H} \end{pmatrix}.
$$
$A_{i+1/2}$ is the Roe's matrix verifying 
\begin{equation}\label{eq:Aroe_prop}
A_{i+1/2}(W_{i+1}-W_i)=F_{C,i+1}-F_{C,i}+c^2_{i+1/2}(H_{i+1}-H_i)\,\begin{pmatrix} 0\\[2pt] 1 \end{pmatrix},
\end{equation} 
where 
$$c_{i+1/2}= \dfrac{gH_{i+1/2}}{\rho_0(h_{i+1/2},\rho_{S,i+1/2})-\rho_{S,i+1/2}}$$ 
is an approximation of the sound speed $c$ \eqref{eq:c} at the interface, and $h_{i+1/2}= \frac{1}{2}(h_i+h_{i+1})$, $\rho_{S,i+1/2}= \frac{1}{2}(\rho_{S,i}+\rho_{S,i+1})$.
The numerical viscosity of the HLL scheme in the PVM framework, using \eqref{eq:Aroe_prop}, is
\begin{equation}\label{eq:difusionEsquema}
\mathcal{D}_{i+1/2}=\alpha_0 \left(W_{i+1}-W_i-A^{-1}_{i+1/2}\mathcal{S}_{i+1/2}\right) 
    + \alpha_1 \left(F_{C,i+1}-F_{C,i}+\mathcal{B}_{i+1/2}\right),
\end{equation}
with the coefficients $\alpha_0$, $\alpha_1$ are defined by
\[ \alpha_0=\frac{S_R|S_L|-S_L|S_R|}{S_R-S_L},\qquad \alpha_1=\frac{|S_R|-|S_L|}{S_R-S_L},\]
where $S_L=\min(\lambda_{1,i},\lambda_{1,i+1})$ and $S_R=\max(\lambda_{2,i},\lambda_{2,i+1})$ are approximations of the smallest and largest wave speeds at the intercell $x_{i+1/2}$ of the finite volume mesh.

Concerning the discrete source term $\mathcal{S}_{i+1/2}$, which is an approximation of $S(W,x)h'(x)$ at the interface, 
we note that, by~\eqref{eq:dxhp},
\[
\frac{\partial (h\bar{p})}{\partial h} h'(x)= \partial_x (h\bar{p}) - \frac{\partial (h\bar{p})}{\partial H} \partial_x H = \partial_x (h\bar{p}) - c^2 \partial_x H.
\]
Then, using the identity~\eqref{eq:pressure_terms_pressurized}, $S(W,x)$ can be written as
\[
S(W,x)h'(x) = \begin{pmatrix} 0 \\[2pt]  p(h)h'-\partial_x(h \bar{p})+\frac{\partial(h \bar{p})}{\partial H}\partial_x H ~
\end{pmatrix}
=  \begin{pmatrix} 0 \\[2pt]  -h\bar\rho\,\partial_x\left(\frac{a\gamma}{\gamma-1}\rho_S^{\gamma-1}+g\,h\right)+c^2 \partial_x H ~
\end{pmatrix}
\]
and we define 
\[
\mathcal{S}_{i+1/2} = \begin{pmatrix} 0 \\[2pt]  
-H_{i+1/2}\, \bigg( \dfrac{a\gamma}{\gamma-1}\left(\rho_{S,i+1}^{\gamma-1}-\rho_{S,i}^{\gamma-1}\right)+g\,\left(h_{i+1}-h_{i}\right)\bigg)+c_{i+1/2}^2 \left(H_{i+1}-H_i\right)
 \end{pmatrix}. 
\]

We focus now on the numerical diffusion of the scheme.
The source term is what makes the scheme well-balanced, namely the term $A^{-1}_{i+1/2}\mathcal{S}_{i+1/2}$ associated with $\alpha_0$.
It is enough to evaluate this term at the stationary solution to be preserved. 
In our case, we focus on the steady states at rest, with  $u=0$. In that case the system matrix is invertible and
\begin{multline}
-A^{-1}_{i+1/2}\mathcal{S}_{i+1/2}  = \\ \dfrac{1}{c_{i+1/2}^2}\begin{pmatrix} H_{i+1/2}\, \bigg( \dfrac{a\gamma}{\gamma-1} \left( \rho_{S,i+1}^{\gamma-1}-\rho_{S,i}^{\gamma-1}\right)+g\,(h_{i+1}-h_{i}) \bigg)-c_{i+1/2}^2 (H_{i+1}-H_i) \\[2pt] 0  \end{pmatrix}.
\end{multline}
and then the contribution of the coefficient $\alpha_0$ to the numerical diffusion \eqref{eq:difusionEsquema} reads
\[
\alpha_0 \left(W_{i+1}-W_i- A^{-1}_{i+1/2}\mathcal{S}_{i+1/2} \right)=
\alpha_0 \begin{pmatrix}
     \dfrac{H_{i+1/2}}{c_{i+1/2}^2 }\, \bigg( \dfrac{a\gamma}{\gamma-1} \left( \rho_{S,i+1}^{\gamma-1}-\rho_{S,i}^{\gamma-1}\right)+g\,(h_{i+1}-h_{i}) \bigg) \\[2pt]
    Q_{i+1}-Q_i
\end{pmatrix},
\]
Using this definition we obtain that the scheme is well-balanced for the stationary solution defined by $u=0$,
$$
\frac{a \gamma}{\gamma-1}\rho_{S}^{\gamma-1}+ gh =\mbox{const.}
$$
that is, for the steady states at rest~\eqref{eq:steady_pressurized} of Proposition~\ref{prop:steady_pressurized}. Indeed, in that case $\mathcal{B}_{i+1/2}=0$ and $Q_i=0$ for all $i$, so that the centred part of~\eqref{eq:discrete_scheme_pressurized}, the term associated with $\alpha_1$ and the term associated with $\alpha_0$ vanish.

\subsection{Numerical example}

    In this section, we study two test cases for the pressurized model: a steady state at rest, which allows us to check the well-balanced property of the numerical scheme, and a dynamic case in which a discontinuity in the density is placed over a variable section $h(x)$ and then evolves in time. In both cases the computational domain is $[-0.6,0.6]$, discretised with $1000$ points, and the CFL number is $0.5$. We take $a=1$, $\gamma=1.4$, $g=9.81$, $\widehat{\kappa}=10^{-3}$ and $h(x)=0.5+0.1\sin(6x)$. The initial velocity is $u=0$ and free outflow boundary conditions are imposed at both ends of the domain.

\subsubsection{Steady case}

\begin{figure}[!h]
	\centering
	\includegraphics[width=0.47\linewidth]{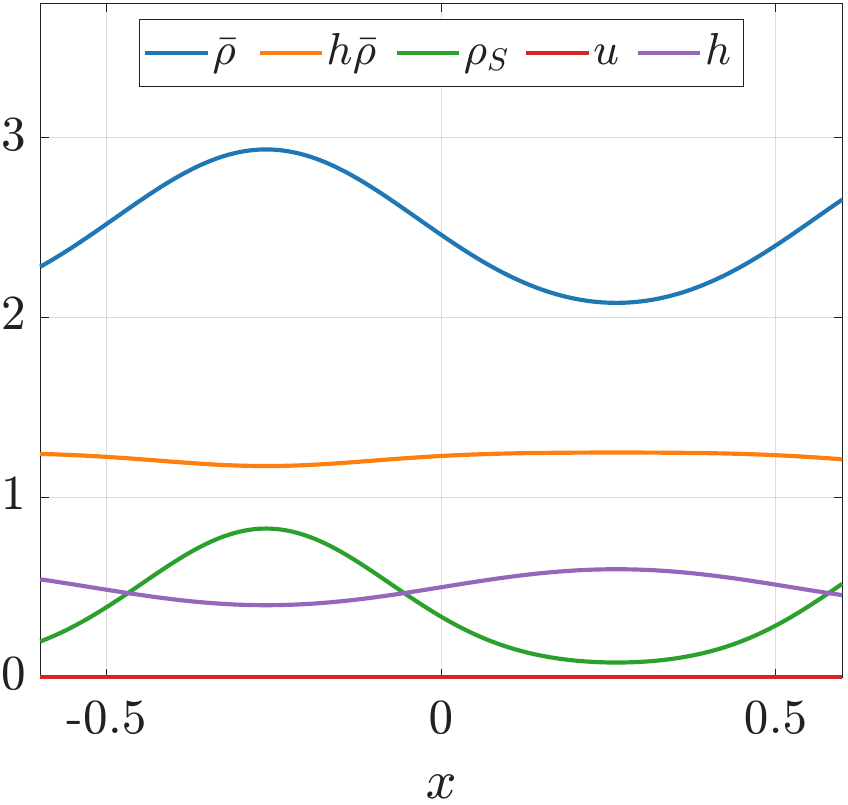}\quad
		\includegraphics[width=0.49\linewidth]{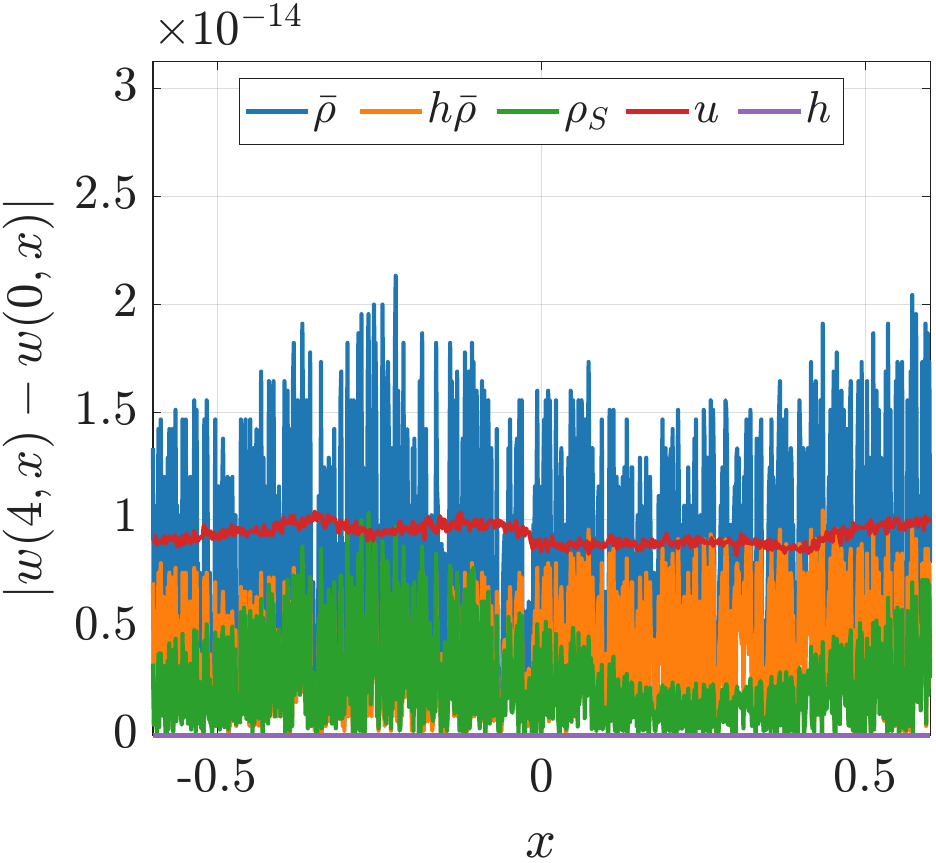}
	\caption{Pressurized case, steady state at rest with $\rho_0=6$. Left: initial condition ($h$, $h\bar{\rho}$, $\bar{\rho}$, $\rho_S$ and $u$). Right: difference between the numerical solution at $t=4$ s and the initial condition.}
	\label{fig:FS:test3}
\end{figure}

    We first consider a steady state at rest given by Proposition~\ref{prop:steady_pressurized}, with constant
     bottom density $\rho_0=6$, which satisfies condition~\eqref{eq:condSteady_rho0}. 
     The initial condition is computed explicitly from~\eqref{eq:steady_pressurized_explicit}. 
     The initial condition (Figure~\ref{fig:FS:test3}, left) shows that $\rho_S$ and $\bar{\rho}$ vary with $x$ 
     to compensate the variation of $h$, with $\bar\rho\in[2.08,2.94]$ and $\rho_S\in[0.08,0.83]$.

    The solution is computed up to $t=4$ s. The difference between the numerical solution at this time and the initial condition is shown in Figure~\ref{fig:FS:test3} (right). It is of the order of $10^{-14}$ for all the variables, that is, at machine precision, so the scheme preserves the steady state.

\subsubsection{Dam break in the density}

\begin{figure}[!ht]
	\centering
	\includegraphics[width=0.48\linewidth]{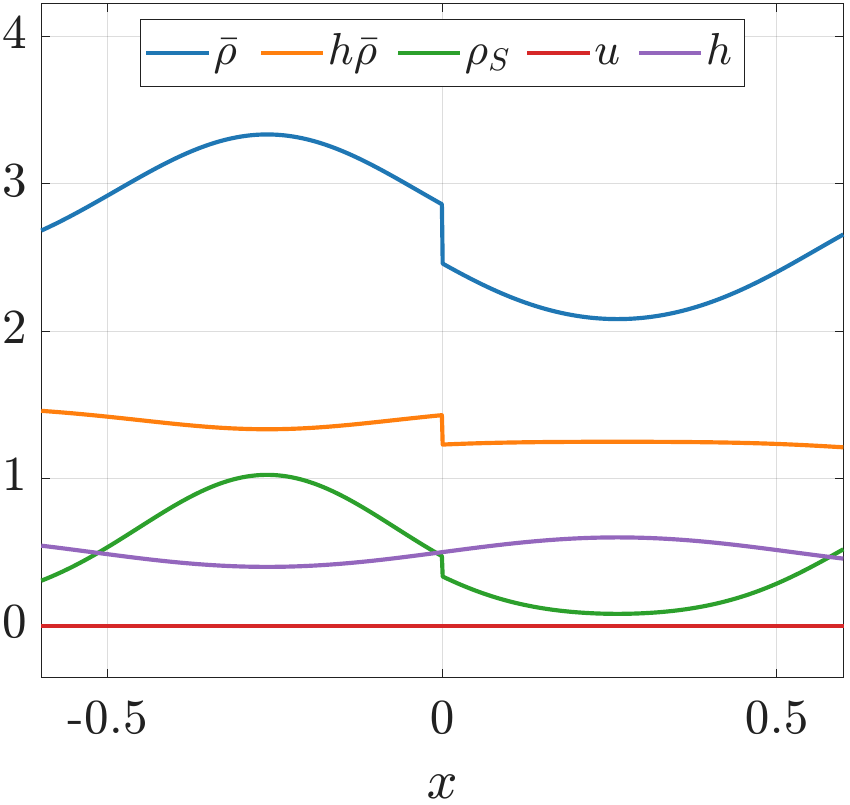}
	\includegraphics[width=0.48\linewidth]{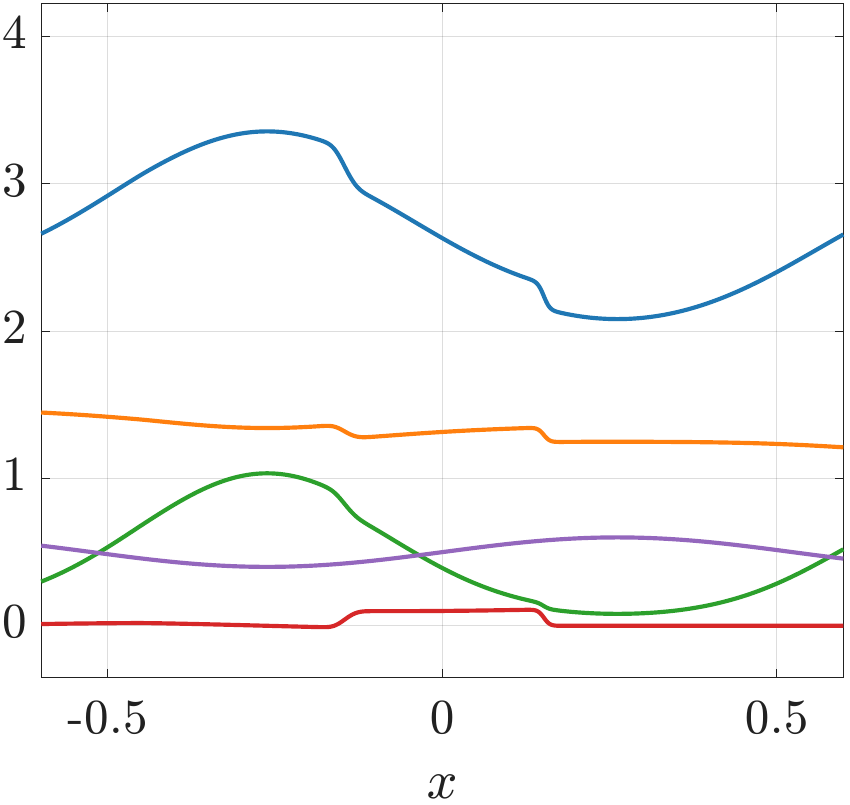}
	\includegraphics[width=0.48\linewidth]{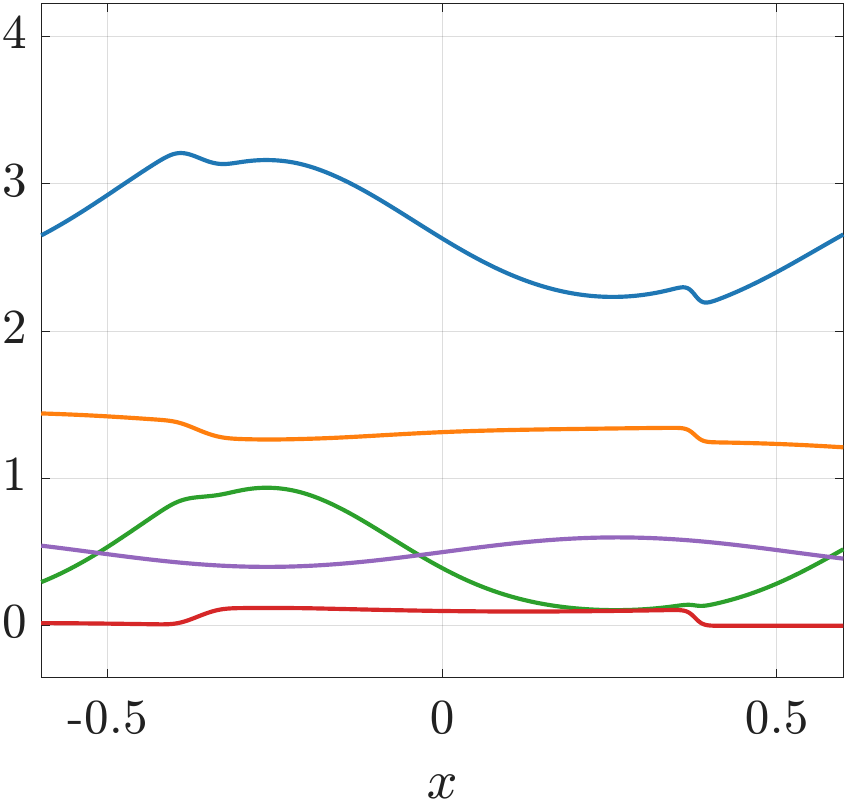}
	\includegraphics[width=0.48\linewidth]{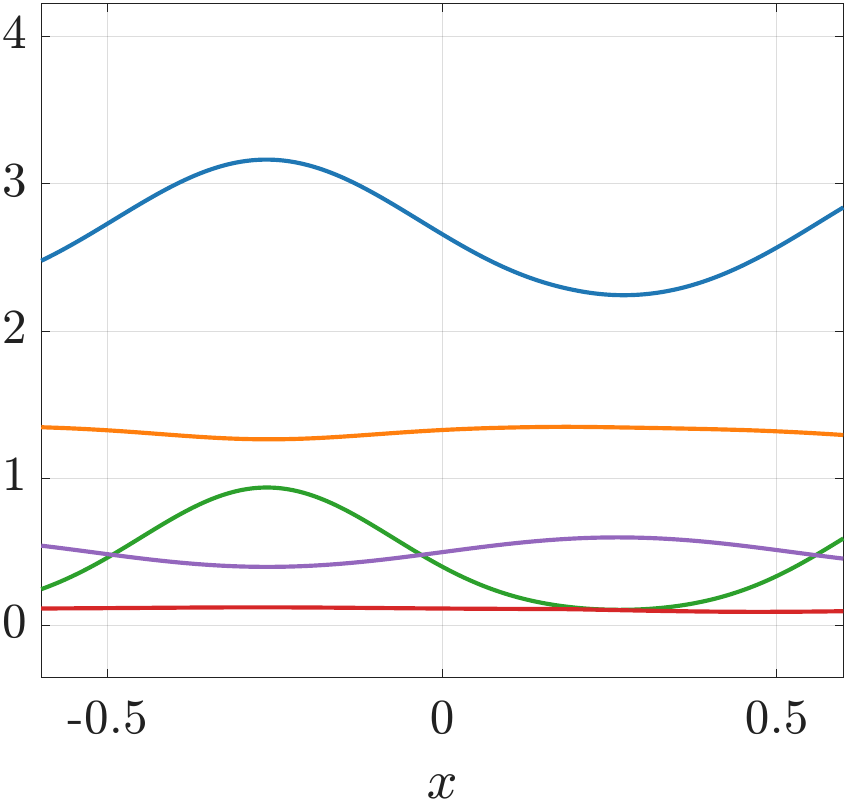}
	\caption{Pressurized case, dam break in the density over a variable section, added to the steady state with $\rho_0=6$. Evolution of $h$, $h\bar{\rho}$, $\bar{\rho}$, $\rho_S$ and $u$ at $t=0$, $0.1$, $0.25$ and $0.5$ s.}
	\label{fig:FS:test2}
\end{figure}

    In the second test, the initial condition is not a steady state. The velocity is $u=0$ and a discontinuity 
    of amplitude $0.4$ is added to the averaged density of the previous steady state,
    \[ \bar{\rho}(0,x)=\begin{cases} \bar\rho_e(x)+0.4 & \text{if } x<0,\\ \bar\rho_e(x) & \text{if } x>0,\end{cases}\]
    where $\bar\rho_e$ denotes the averaged density of the steady state with $\rho_0=6$, 
    over the same height $h(x)=0.5+0.1\sin(6x)$. The value of $\rho_S$ is recovered from $h\bar{\rho}$ by Newton's method, 
    as described in Subsection~\ref{se:pressurized_general}. 
    Figure~\ref{fig:FS:test2} shows the solution at $t=0$, $0.1$, $0.25$ and $0.5$ s. 
    The discontinuity splits into waves that propagate towards both sides of the domain, and which interact with the variations of the section. 
    The density excess on the left generates a flow towards the right. The velocity remains small ($|u|\le 0.13$) and, at the final time, the waves have left the computational domain through the free outflow boundaries.

\bigskip

\noindent {\bf Natural Intelligence Statement.} 
The authors certify that only natural intelligent was used to produce this document.

\bigskip

\noindent {\bf Acknowledgments.} D. Bresch was partially supported by the BOURGEONS project, grant ANR-23-CE40-0014-01 of the French National Research Agency (ANR). This work also benefited of the support of the ANR under France 2030 (PEPR Maths-Vives) bearing the reference ANR-23-EXMA-004 (Complexflows project) and ANR-23-EXMA-001 (Climate project). This work was financially supported by the International Research Project (IRP) PICASSO: https://picasso.math.cnrs.fr

The spanish group want to thank European Union -- NextGenerationEU program and by grants PID2022-137637NB-C22 and PID2025--170651NB--C22 funded by the project named MCIN/AEI/10.13039/501100011033 and “ERDF A way of making Europe”. 

\noindent They also thank IMUS-Maria de Maeztu grant CEX2024-001517-M -- Apoyo a Unidades de Excelencia
Mar\'ia de Maeztu for supporting this research, funded by the project named by MICIU/AEI/10.13039/501100011033 and the applied research and innovation project
”Development of mathematical tools for transfer” (Mathware) at IMUS, co-financed by the EU -- Ministry of Finance and Public Administration -- European Funds --
Junta de Andalucia -- Consejeria de Universidad, Investigacion e Innovacion.

\bibliographystyle{apalike}
\bibliography{biblio.bib}

\end{document}